\documentclass[a4paper,11pt]{article}
\usepackage{graphicx}
\usepackage{jheppub}
\usepackage[T1]{fontenc}
\usepackage{braket}	
\usepackage{dcolumn}
\usepackage{comment}
\usepackage{hyperref}

\newcommand{\be}{\begin{equation}}
\newcommand{\ee}{\end{equation}}

\newcommand{\lt}{\left}
\newcommand{\rt}{\right}
\newcommand{\del}{\partial}

\newcommand{\non}{\nonumber \\}

\newcommand{\Li}{{\rm Li}}

\newcommand{\fn}{\footnote}

\newcommand{\MSb}{\overline{\rm MS}}

\newcommand{\LMS}{\Lambda_{\overline{\rm MS}}}

\title{Formulation for renormalon-free perturbative predictions beyond large-$\beta_0$ approximation}

\author{Hiromasa Takaura}

\affiliation{Department of Physics, Kyushu University, Fukuoka, 819-0395 Japan}

\emailAdd{takaura@phys.kyushu-u.ac.jp}

\abstract{
We present a formulation to give renormalon-free predictions
consistently with fixed order perturbative results.
The formulation has a similarity to Lee's method in that 
the renormalon-free part consists of two parts:
one is given by a series expansion which does not contain renormalons
and the other is the real part of the Borel integral of a singular Borel transform.
The main novel aspect is to evaluate the latter object
using a dispersion relation technique, which was possible only in the large-$\beta_0$ approximation.
Here, we introduce an `` ambiguity function,'' which is defined 
by inverse Mellin transform of the singular Borel transform.
With the ambiguity function, we can rewrite the Borel integral in an alternative formula,
which allows us to obtain the real part using analytic techniques similarly to the case of the large-$\beta_0$ approximation. 
We also present detailed studies of renormalization group properties of the formulation.
As an example, we apply our formulation to the fixed-order result of the static QCD potential,
whose closest renormalon is already visible.
}

\begin{document}

\preprint{KYUSHU--HET--207}

\maketitle
\flushbottom

\section{Introduction}
Perturbation theory is a very basic tool in quantum field theory, 
yet perturbative series are expected to be divergent asymptotic series.
In QCD, due to this property, perturbative predictions have inevitable uncertainties,
and in particular renormalon uncertainties can practically limit accuracies of predictions. 
(See Ref.~\cite{Beneke:1998ui} for a review on renormalon.)
It is generally a non-trivial task to extract an unambiguous part or meaningful prediction
from such a divergent series, particularly when the number of known perturbative coefficients is limited.
Nevertheless, it is necessary to systematically assign a definite value to perturbation theory
in order to go beyond perturbation theory with using the operator product expansion (OPE);
one should systematically add a nonperturbative matrix element to the perturbative contribution for this purpose.

Within the large-$\beta_0$ approximation \cite{Beneke:1994qe}, methods 
to extract an unambiguous part from the series containing renormalons
were developed \cite{Ball:1995ni, Mishima:2016vna}.
In these methods, one can give a renormalon-free (unambiguous) part and renormalon uncertainty
in the form where each is clearly separated.
The renormalon-free part is given in a semi-analytic form
and is useful to gain insight into short-distance behaviors of observables \cite{Mishima:2016vna}.
However, the methods are applicable within the large-$\beta_0$ approximation,
because they rely on the feature that the series is given by the one-loop integral with respect to
the momentum of a dressed gluon propagator.
The large-$\beta_0$ approximation is not sufficient to give accurate predictions,
because, rigorously speaking, it is accurate only at leading order $[\mathcal{O}(\alpha_s)]$, 
and a systematic way to improve this approximation is unclear.
In particular, it is not possible to incorporate exact results of fixed order perturbation theory,
which have been computed currently up to a few to several orders.

In this paper, we devise a general formulation beyond the large-$\beta_0$ approximation
to extract a renormalon-free part from the series containing renormalons, while clearly separating renormalon uncertainties. 
Our formulation has similarities to Lee's method \cite{Lee:2002sn, Lee:2003hh} in the following points.
We consider the Borel transform which is consistent with  fixed order perturbative results
and with the structure of renormalons. Then 
the Borel transform is given by the sum of a regular part [$\delta B(u)$]
and singular part containing the renormalons [$B^{\rm sing}(u)$], i.e.,
$B(u)=\delta B(u)+B^{\rm sing}(u)$.
For this Borel transform, the Borel integral is considered.
This is the same procedure as Refs.~\cite{Lee:2002sn, Lee:2003hh}.
We evaluate the Borel integral of the regular Borel transform by a series expansion in $\alpha_s$, 
as it does not contain renormalons. A novel point of the present paper is to devise
a procedure to evaluate the Borel integral of the singular Borel transform.
We introduce an ``ambiguity function'', which is defined by inverse Mellin transform of the singular Borel transform.
With the use of the ambiguity function, we obtain a resummation formula alternative to the Borel integral.
This resummation formula is given by the one-dimensional integral which
has similar features to the resummation formula in the large-$\beta_0$ approximation.
Then, it is possible to use a dispersion relation technique to obtain the real part 
of the quantity (an unambiguous part of the Borel integral)
in a parallel manner to the case of the large-$\beta_0$ approximation \cite{Ball:1995ni,Mishima:2016xuj,Mishima:2016vna}.
This work can be regarded as an extension of the preceding studies 
\cite{Neubert:1994vb,Ball:1995ni,Sumino:2005cq,Mishima:2016vna},
developed mainly within the large-$\beta_0$ approximation. 
As a result, we obtain the unambiguous prediction in a closed form
from the resummation formula.
The final renormalon-free result is consistent with fixed order perturbation theory
and does not suffer from renormalon uncertainties similarly to Refs.~\cite{Lee:2002sn, Lee:2003hh}. 
We also study renormalization group (RG) properties of the formulation in detail.

The method using the Borel resummation, 
as done in Refs.~\cite{Lee:2002sn, Lee:2003hh} and in the present paper,
has the following advantages.
First, one can (in principle) define the perturbative contribution in a renormalization group (RG) invariant way.
This feature is assumed in the OPE argument to discuss renormalon structure
and the Borel resummation respects this property.
Secondly, the renormalon uncertainty is given in the form such that it can be canceled against 
a nonperturbative matrix element in the OPE.
These features are obvious in our construction
and quite useful to go beyond perturbation theory with using the OPE.
On the other hand, in minimal term truncation methods (where perturbative series is truncated around 
the order where the term of series gets minimal), these features are not obvious.
See the recent paper Ref.~\cite{Ayala:2019uaw} for possible improvement in
these issues using truncation.

As a practical application, we use our method to give a renormalon-free prediction
for the static QCD potential starting from the currently known fixed-order result \cite{Anzai:2009tm, Smirnov:2009fh, Lee:2016cgz}. 
Then we can obtain an accurate prediction which is consistent with the fixed-order result\fn{
This means that our renormalon-free part reproduces the original perturbative expansion
{\it once} it is expanded in $\alpha_s$. As long as we do not expand it, we have a finite and unambiguous prediction.}
and does not suffer from a renormalon uncertainty.
Although our definition of a renormalon-free part itself reduces to a quite similar one to Ref.~\cite{Lee:2002sn},
the original point in this paper is that
we present a systematic and analytic method to extract a renormalon-free part
from the Borel integral of a singular Borel transform
and describe how it is related to an ambiguous part of the Borel integral.
We also add an insight into a short-distance behavior of the observable.

This paper is organized as follows.
In Sec.~\ref{sec:2}, we present a general formulation to extract
a renormalon-free part from a given all-order perturbative series while clearly
separating renormalon uncertainties. We give detailed RG arguments as well.
Then, we explain how to use the formulation in practical situations,
where perturbative series is known to finite orders.
In Sec.~\ref{sec:3}, we test our formulation by using all-order perturbative series 
obtained with certain approximations.
We study the Adler function in the large-$\beta_0$ approximation,
and the static QCD potential with using the RG method in Ref.~\cite{Sumino:2005cq} at leading log (LL) and next-to-LL (NLL).
In Sec.~\ref{sec:4}, we apply our formulation to the static QCD potential
starting from the available fixed-order perturbative series.
Sec.~\ref{sec:5} is devoted to the conclusions and discussion.
In App.~\ref{app:A}, we show RG invariance of the Borel integral 
(this issue has been discussed in Ref.~\cite{Ayala:2019uaw}
and we give App.~\ref{app:A} for a self-contained explanation).
In App.~\ref{app:B}, we present convenient formulae for numerical evaluation of the 
renormalon-free part. 

\section{Formulation}
\label{sec:2}
 
Let us first clarify the notation used in this paper.
We consider a general dimensionless observable $X(Q)$ depending on a single scale $Q$.
We denote its perturbative series as
\be
X(Q^2)=\sum_{n=0}^{\infty} d_n(Q,\mu) \alpha_s(\mu)^{n+1} \, ,
\ee
where $\mu$ is a renormalization scale.
Corresponding to this perturbative series, we define the Borel transform as
\be
B_X(u; Q, \mu)=\sum_{n=0}^{\infty} \frac{1}{n!} \frac{d_n(Q,\mu)}{b_0^n} u^n \, .
\ee
Here, $b_0$ is the first coefficient of the beta function, which is defined as
\be
\beta(\alpha_s)=\mu^2 \frac{d \alpha_s}{d \mu^2} =-\sum_{i=0}^{\infty} b_i \alpha_s^{i+2} \, .
\ee
Explicitly the first two coefficients are given by
\be
b_0=\frac{1}{4 \pi} \lt(11-\frac{2}{3} n_f \rt),~b_1=\frac{1}{(4 \pi)^2} \lt(102-\frac{38}{3} n_f \rt),
\, 
\ee
for QCD, where $n_f$ is the number of quark flavors.
The $\Lambda$ parameter in the $\MSb$ scheme is given by
\be
\LMS^2=\mu^2 \exp \lt[-\lt\{\frac{1}{b_0 \alpha_s(\mu)}+\frac{b_1}{b_0^2} \log(b_0 \alpha_s(\mu)) 
+\int_0^{\alpha_s(\mu)} dx \lt(\frac{1}{\beta(x)}+\frac{1}{b_0 x^2}-\frac{b_1}{b_0^2 x} \rt)   \rt\} \rt] \, . \label{Lambda}
\ee
The resummation of the perturbative series is given by the Borel integral (or Borel sum):
\be
X(Q^2)=\frac{1}{b_0} \int_0^{\infty} du \, B_X(u; Q, \mu) e^{-u/(b_0 \alpha_s(\mu))} \, . \label{Borelint}
\ee
In the presence of IR renormalons [which refer to singularities of $B_X(u; Q, \mu)$ on the real $u$-axis], 
we can regularize the Borel integral \eqref{Borelint} by contour deformation
$\int_0^{\infty} \to \int_{0 \pm i \epsilon}^{\infty \pm i \epsilon} \equiv \int_{C_{\pm}}$, 
\be
X_{\pm}(Q^2)=\frac{1}{b_0} \int_{C_{\pm}} du \, B_X(u; Q, \mu) e^{-u/(b_0 \alpha_s(\mu))} \, . \label{Xpm}
\ee
In this case, the Borel sum possesses an imaginary part, whose sign is dependent on which contour is chosen. 
This imaginary part is regarded as a renormalon uncertainty.
The real part is an unambiguous part, which we call a renormalon-free part.
It is equal to the principal value prescription of the integral, i.e.,
the average of the integral along $C_{+}$ and that along $C_{-}$.

The subsequent contents in this Section are as follows.
In Sec.~\ref{sec:2.1}, we decompose the Borel transform into two parts,
a regular part and singular part, as in Lee's method.
For the Borel integral of the singular Borel transform,
we give an alternative resummation formula
by introducing an ``ambiguity function.''
In Sec.~\ref{sec:2.2}, we show some formulae and examples of the ambiguity function.
In Sec.~\ref{sec:2.3}, we define a ``preweight,''
which is obtained by the dispersive integral of the ambiguity function.
The preweight plays a central role in extracting an unambiguous part from
the Borel integral of the singular Borel transform.
In Secs.~\ref{sec:2.4} and \ref{sec:2.5}, 
we explain methods to extract an unambiguous part from the resummation formula 
given in Sec.~\ref{sec:2.1}.
This is done in two different regularizations:
cutoff regularization in Sec.~\ref{sec:2.4}
and contour regularization in Sec.~\ref{sec:2.5}.
The unambiguous (renormalon-free) parts reduce to the same result in both regularizations.
As we shall see, regarding the ambiguous part, the method in Sec.~\ref{sec:2.5} is superior
in the sense that the ambiguous part is compatible with the OPE.
In Sec.~\ref{sec:2.6}, we discuss renormalization group properties of the formulation.
Here we assume that there are only IR renormalons.
The former contents in this subsection can be regarded as a new insight into Lee's method.
Also it clarifies how we should change the domain of the ambiguity function (corresponding to
an IR renormalon) when varying a renormalization scale.
In Sec.~\ref{sec:2.7}, we explain a practical way to use the formula,
although the contents up to Sec.~\ref{sec:2.6} are formal in the sense that
we assume that all necessary information (for instance an all-order perturbative series) is known. 
We also give a general discussion on error size in practical situations.

\subsection{Resummation formula with ambiguity function}
\label{sec:2.1}
For a given Borel transform $B_X(u)$,
we decompose it into a singular part and regular part
similarly to Refs.~\cite{Lee:2002sn, Lee:2003hh}:
\be
B_X(u)=B_X^{\rm sing}(u)+\delta B_X(u)
\ee
such that $\delta B_X(u)$ does not possess renormalons,
and all renormalons are contained in $B_X^{\rm sing}(u)$.
(This decomposition is not unique.)
We denote perturbative coefficients involved in $\delta B_X(u)$ by $\delta_n$,
\be
\delta B_X(u)=\sum_{n=0}^{\infty} \frac{1}{n!} \frac{\delta_n(Q,\mu)}{b_0^n} u^n \, ,
\ee
and those in $B_X^{\rm sing}(u)$ by $d_n^{{\rm ren}}$,
\be
B_X^{\rm sing}(u)=\sum_{n=0}^{\infty} \frac{1}{n!} \frac{d^{{\rm ren}}_n(Q,\mu)}{b_0^n} u^n \, ,
\ee
and thus $d_n=\delta_n+d_n^{{\rm ren}}$.
Since the perturbative series
\be
\sum_{n=0}^{\infty} \delta_n(Q,\mu) \alpha_s(\mu)^{n+1} \label{deltapart}
\ee
does not contain renormalon divergences, this part shows a more convergent behavior than 
the original series and is free from the renormalon uncertainties.
We refer to this series as a $\delta$ part hereafter.
On the other hand, we have to apply the Borel sum to the series including renormalons,
\be
\sum_{n=0}^{\infty} d_n^{{\rm ren}}(Q,\mu) \alpha_s(\mu)^{n+1} \, . \label{dnren}
\ee
In other words, we adopt the Borel sum \eqref{Xpm} 
to define the perturbative calculation and it can be  decomposed as 
\be
X_{\pm}(Q^2)=\sum_{n=0}^{\infty} \delta_n(Q,\mu) \alpha_s(\mu)^{n+1}+X^{{\rm ren}}_{\pm}(Q) \label{XdeltaandBR}
\ee
with
\be
X^{{\rm ren}}_{\pm}(Q)=\frac{1}{b_0} \int_{C_{\pm}} du \, B^{\rm sing}_X(u;Q,\mu) e^{-u/(b_0 \alpha_s(\mu))} \, . \label{Xrenpm}
\ee
The $\delta$ part is regarded as an unambiguous part as it is free from renormalons.
The main purpose of this paper is to develop a method to evaluate the unambiguous part (or real part) of $X^{\rm ren}_{\pm}(Q)$.

Now we derive a resummation formula alternative to the Borel integral \eqref{Xrenpm}.
We introduce an ambiguity function $Am_X$, which specifies 
the imaginary ambiguity in the Borel integral.
We define it by inverse Mellin transform of the singular Borel transfrom:
\begin{align}
Am_X(x; Q, \mu):=\frac{1}{2 i} \int_{- i \infty}^{i \infty} du \, B^{\rm sing}_X(u;Q,\mu) x^u \, . \label{BoreltoAmb}
\end{align}
Such a function was first introduced in Ref.~\cite{Neubert:1994vb}
in the context of the large-$\beta_0$ approximation.
In the large-$\beta_0$ approximation, $x$ corresponds to the loop momentum of a dressed gluon. 
Beyond the large-$\beta_0$ approximation  
we do not have such a diagrammatic correspondence.
This function gives the renormalon uncertainty when we take $x=e^{-1/(b_0 \alpha_s(\mu))} (\ll 1)$:
\begin{align}
Am_X(x=e^{-1/(b_0 \alpha_s(\mu))};Q,\mu)
&=\frac{1}{2 i} \lt( \int_{C_{+}} -\int_{C_{-}} \rt) du \, B^{\rm sing}_X(u; Q, \mu) e^{-u/(b_0 \alpha_s)} \non
&=\pm b_0 {\rm Im} X_{\pm}(Q^2) \,  \label{AmbtoAmbiguity}
\end{align}
as seen from Eq.~\eqref{Xpm}.
Here, we assumed that for small $x \ll 1$ the integration contour in the ambiguity function 
can be deformed  as above due to $x^u=e^{u \log{x}}$ with $\log{x} <0$.
We note that the subtraction of the regular part does not change the renormalon uncertainty.

We have the inverse formula of Eq.~\eqref{BoreltoAmb}, i.e., 
we obtain $B_X^{\rm sing}$ from the ambiguity function as
\be
B^{\rm sing}_X(u;Q, \mu)=\frac{1}{\pi} \int_0^{\infty} d x \, Am_X(x; Q, \mu) x^{-u-1} \, . \label{BfromAm}
\ee
One can show the equality, for instance, for pure imaginary $u$.
Then, if both are analytic functions the equality can be enlarged to the whole complex $u$-plane.
(In practical applications below, we rather use this relation to {\it define} $B_X^{\rm sing}(u)$
from an ambiguity function.)\fn{
Here we make comments on the case where a regular Borel transform is considered 
in Eq.~\eqref{BoreltoAmb} instead of (or additionally to) $B_X^{\rm sing}(u)$.
As an example, let us consider $1+u$.
The corresponding ambiguity function is given by a hyperfunction as $\pi \delta(\log{x})+\pi \frac{\del}{\del \log{x}}  \delta(\log{x})$.
One can confirm that this ambiguity function gives the perturbative series in Eq.~\eqref{resumAmb} as $\alpha_s+b_0 \alpha_s^2$
consistently with the considered Borel transform.}

Using the above inverse formula, 
we can rewrite the Borel integral in terms of the ambiguity function:\fn{This calculation is just formal.
We present a calculation with regularization in the subsequent subsections (Secs.~\ref{sec:2.4} and \ref{sec:2.5}).}
\begin{align}
X^{{\rm ren}}(Q^2)
&=\frac{1}{b_0} \int_0^{\infty} \frac{dx}{\pi x} \, Am_X(x;Q,\mu) \int_0^{\infty} du \, x^{-u} e^{-u/(b_0 \alpha_s(\mu))} \non
&=\frac{1}{b_0}  \int_0^{\infty} \frac{dx}{\pi x} \, Am_X(x;Q,\mu) \frac{1}{\log{x}+\frac{1}{b_0 \alpha_s(\mu)}} \, . \label{resumAmb}
\end{align}
This is an alternative formula to the Borel integral, which allows us to resum the perturbative series. 
In this paper, we mainly adopt this resummation formula (with necessary regularization).

Let us make comments on the resummation formula \eqref{resumAmb}.
In Eq.~\eqref{resumAmb}, the singularity on the integration path (positive real $x$-axis)
is solely given by the simple pole at $x=e^{-1/(b_0 \alpha_s(\mu))}$. 
This singularity structure is much simpler than the integrand of the Borel integral,
which generally has an infinite number of cut singularities on the positive $u$-axis.
(It is well known that cut singularities in an integrand can be rewritten in terms of a pole singularity.)
This feature makes it easy to handle the all-order resummed series.
In the resummation formula \eqref{resumAmb}, the imaginary ambiguity is correctly obtained from the contribution around this simple pole
as\fn{The correspondence between $X^{{\rm ren}}_{\pm}$ and how to deform the contour in the $x$-plane is explained in Sec.~\ref{sec:2.5}.} 
\begin{align}
X^{{\rm ren}}_{\pm}(Q^2)
&=\frac{1}{b_0} \int_{C_{\mp}} \frac{dx}{\pi x}  Am_X(x;Q,\mu) \frac{1}{\log{x}+\frac{1}{b_0 \alpha_s(\mu)}} \non
&\sim \frac{1}{b_0} \int_{C_{\mp}} \frac{dx}{\pi x}  Am_X(x;Q,\mu) \frac{e^{-1/(b_0 \alpha_s(\mu))}}{x-e^{-1/(b_0 \alpha_s(\mu))}} \non
&\sim \pm  \frac{i}{b_0} Am_X(e^{-1/(b_0 \alpha_s(\mu))};Q,\mu) \, ,
\end{align}
where we show only the imaginary part when the symbol $\sim$ is used.\fn{
We note that although the pole position is $\mu$ dependent,
the resulting uncertainty is $\mu$ independent.
This is because the Borel integral is RG invariant, 
and hence so does its imaginary part, identified as the renormalon ambiguity.
}
The singularity structure is similar to the case of the large-$\beta_0$ approximation,
where the resummation formula is given by a single integral with respect to the momentum of a dressed gluon~\cite{Neubert:1994vb}.
Hence, it is possible to make use of the techniques developed in 
the large-$\beta_0$ approximation \cite{Ball:1995ni, Mishima:2016vna} by adopting 
the resummation formula with the ambiguity function.

The relation between $d_n^{{\rm ren}}$ and the ambiguity function
is simply given by expanding the integrand in $\alpha_s(\mu)$ before integration in Eq.~\eqref{resumAmb}:
\be
X^{{\rm ren}}(Q^2)_{\rm pert.}=\sum_{n=0}^{\infty} \int_0^{\infty} \frac{dx}{\pi x} \, Am_X(x;Q,\mu) \alpha_s(\mu) (-b_0 \alpha_s(\mu) \log{x})^n \, .
\ee
That is, we have [cf. Eq.~\eqref{dnren}]
\be
\frac{d^{{\rm ren}}_n}{b_0^n}=\int_0^{\infty} \frac{dx}{\pi x} \, Am_X(x;Q,\mu) (-\log{x})^n \, . \label{FOfromAmb}
\ee
We can obtain this relation also from Eq.~\eqref{BfromAm}, by taking derivatives with
respect to $u$ and sending $u \to 0$.

\subsection{Explicit form of ambiguity function}
\label{sec:2.2}
In this subsection, we present explicit forms of ambiguity functions in some examples.
Here and hereafter, we set $\mu=Q$ unless otherwise stated and omit the arguments of $\mu$ and $Q$.
(We will discuss $\mu$ dependence in Sec.~\ref{sec:2.6}.)
Assuming that a Borel transform $B^{\rm sing}_X(u=R e^{i \theta})$ exhibits a good convergence at $R \to \infty$,
the ambiguity function is obtained as
\be
Am_{X}(x)=\frac{1}{2 i} \lt( \int_{0+i \epsilon}^{\infty+i\epsilon}-\int_{0-i\epsilon}^{\infty-i \epsilon} \rt) du \, B^{\rm sing}_X(u) x^u \quad \text{for $x \ll 1$} \, , \label{Ambright}
\ee
and
\be
Am_{X}(x)=\frac{1}{2 i} \lt( \int_{0+i \epsilon}^{-\infty+i\epsilon}-\int_{0-i\epsilon}^{-\infty-i \epsilon} \rt) du \, B^{\rm sing}_X(u) x^u \quad \text{for $x \gg 1$} \, ,\label{Ambleft}
 \ee
where we note that $x^u=e^{u \log{x}}$ can be a suppression factor in right or left side of the complex $u$-plane 
depending on the sign of $\log{x}$.
Since one expects that the Borel transform is expanded around its singularity at $u=u_i$ as
\begin{align}
B_X(u;\mu,Q) \simeq \frac{K_{u_i}\Gamma(1+\nu)}{\lt(1-u/u_i\rt)^{1+\nu}} \sum_{k=0}^{\infty} c_k \lt(1-u/u_i \rt)^{k} \, ,
\label{Borelexp}
\end{align}
the following formulae are convenient to obtain the ambiguity function:
\begin{align}
&\frac{1}{2 i} \lt( \int_{0+i \epsilon}^{\infty+i\epsilon}-\int_{0-i\epsilon}^{\infty-i \epsilon} \rt) du \,  \frac{\Gamma(1+\nu) }{\lt(1-\frac{u}{u_i}\rt)^{1+\nu-k}} x^u \non
&=
\begin{cases}
\pi u_i  x^{u_i} \lt[\frac{1}{u_i \log(1/x)} \rt]^{-\nu}  & \text{for $k=0$}\\
\pi u_i \nu (\nu-1) \cdot \cdots \cdot (\nu-k+1) x^{u_i} \lt[\frac{1}{u_i \log(1/x)} \rt]^{k-\nu} & \text{for $k\geq1$}
\end{cases}  \quad \text{for $u_i>0$ and $x<1$}  \label{ambformula1}
\end{align}
and
\begin{align}
&\frac{1}{2 i} \lt( \int_{0+i \epsilon}^{-\infty+i\epsilon}-\int_{0-i\epsilon}^{-\infty-i \epsilon} \rt) du \,  \frac{\Gamma(1+\nu) }{\lt(1-\frac{u}{u_i}\rt)^{1+\nu-k}} x^u \non
&=
\begin{cases}
- \pi u_i  x^{u_i} \lt[\frac{-1}{u_i \log{x}} \rt]^{-\nu} & \text{for $k=0$} \\
- \pi u_i \nu (\nu-1) \cdot \cdots \cdot (\nu-k+1) x^{u_i} \lt[\frac{-1}{u_i \log{x}} \rt]^{k-\nu} & \text{for $k \geq1$}
\end{cases} \quad \text{for $u_i<0$ and $x>1$} \, . \label{ambformula2}
\end{align}
One can see that the IR renormalons determine the small-$x$ behavior of the ambiguity function,
whereas the UV renormalons do the large-$x$ behavior.

\begin{figure}[h!]
\begin{center}
\includegraphics[width=7cm]{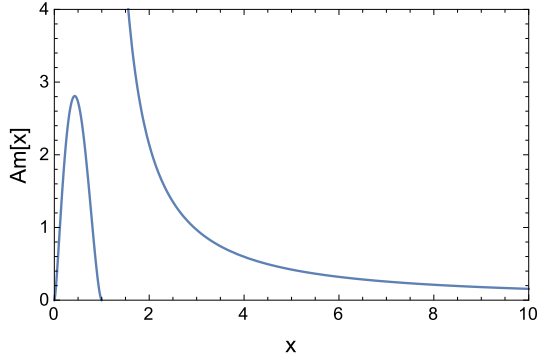}
\end{center}
\caption{Ambiguity function of Eq.~\eqref{AmbToy}. The parameters are taken as
$b_1/b_0^2=102/11^2$, $a=1$, and $b=1$.}
\label{fig:AmbToy}
\end{figure}

As an example, we consider the Borel transform as
\be
B^{\rm sing}(u)=\frac{a \Gamma(1-\frac{b_1}{b_0^2})}{(1+u)^{1-\frac{b_1}{b_0^2}}}+\frac{b \Gamma(1+\frac{2 b_1}{b_0^2})}{(1- u/2)^{1+\frac{2 b_1}{b_0^2}}} \, , \label{Bex1}
\ee
which possesses renormalons at $u=-1$ and $2$.
Then the ambiguity function is given  by
\be
Am(x)=
\begin{cases}
2 \pi b x^2 (-2\log{x})^{\frac{2 b_1}{b_0^2}} \quad \text{for}~x<1 \\
\pi a \frac{1}{x} \lt(\frac{1}{\log{x}} \rt)^{\frac{b_1}{b_0^2}} \quad \text{for}~x>1 \, . \label{AmbToy}
\end{cases}
\ee
One can check that the above ambiguity function indeed gives
the Borel transform of Eq.~\eqref{Bex1} through Eq.~\eqref{BfromAm} and thus gives the same perturbative coefficients 
via Eq.~\eqref{FOfromAmb} as the ones from the Borel transform.
We show the behavior of the ambiguity function in Fig.~\ref{fig:AmbToy}.\fn{
Although the ambiguity function is divergent as $x \to 1+0$, the integral of the ambiguity function 
$\int_1^{R} dx \, Am(x) x^{-u-1}$ is convergent (where $R>1$).}

As a second example, we consider the Borel transform which possesses a singularity only at $u=u_i$
and gives the renormalon uncertainty as
\be
\frac{1}{2 b_0 i} \lt( \int_{0+i \epsilon}^{\infty+i \epsilon} -  \int_{0-i \epsilon}^{\infty-i \epsilon}\rt) du \, B^{\rm sing}(u) e^{-u/(b_0 \alpha_s(Q))}
=N_i \lt(\frac{\LMS^2}{Q^2} \rt)^{u_i}  \label{ex3} \, .
\ee
Since the ambiguity function can be obtained by the replacement of $\alpha_s \to -\frac{1}{b_0} \frac{1}{\log{x}}$
in the Borel integral (and multiplying $b_0$ as the overall factor) [cf. Eqs.~\eqref{BoreltoAmb} and \eqref{AmbtoAmbiguity}],  
one sees that the corresponding ambiguity function is given by [cf. Eq.~\eqref{Lambda}]
\begin{align}
Am(x)=N_i b_0
\lt[x (\log{(1/x)})^{b_1/b_0^2} e^{-\int_0^{-\frac{1}{b_0 \log{x}}} dt \lt(\frac{1}{\beta(t)}+\frac{1}{b_0 t^2}-\frac{b_1}{b_0^2 t} \rt)} \rt]^{u_i} \, . \label{AmbLambda}
\end{align}
In this way, we can directly obtain the ambiguity function from the renormalon uncertainty
and often avoid an explicit calculation of the Borel transform.

For instance, at the two-loop level (where we set $b_2=b_3=\cdots=0$) 
we explicitly have 
\be
Am(x)=
\begin{cases}
& N_i b_0 \lt[ x \lt(-\log{x}+\frac{b_1}{b_0^2} \rt)^{b_1/b_0^2} \rt]^{u_i} \quad \text{for $0 < x<e^{b_1/b_0^2}$} \\
& 0  \qquad \text{for $x>e^{b_1/b_0^2}$} 
\end{cases} \, .  \label{TwoloopAmb}
\ee
In this case, the explicit form of the Borel transform is actually inferred as
\be
B^{\rm sing}(u)=\frac{b_0 N_i}{\pi u_i} e^{u_i b_1/b_0^2} u_i^{-u_i b_1/b_0^2} 
\Gamma(1+u_i b_1/b_0^2) \frac{e^{-u b_1/b_0^2}}{(1-u/u_i)^{1+u_i \frac{b_1}{b_0^2}}} \label{BorelNLL}
\ee
by noting Eq.~\eqref{ambformula1} and $-\log{x}+b_1/b_0^2=-\log{(x e^{-b_1/b_0^2})}$.
From this expression, in particular from the factor $e^{-u b_1/b_0^2}$, 
one sees that the integral to obtain the ambiguity function for small $x$ [cf. Eq.~\eqref{ambformula1}]
is convergent for $\log{x}-b_1/b_0^2<0$.
This is a clear exposition why the expression of the above ambiguity
function is restricted to the region $0 < x<e^{b_1/b_0^2}$.
We can also confirm 
\be
\frac{1}{\pi} \int_0^{e^{b_1/b_0^2}} dx \, Am(x) x^{-u-1}=B^{\rm sing}(u) \, .
\ee

\subsection{Preweight}
\label{sec:2.3}
As a preparation for extracting the unambiguous part from $X^{{\rm ren}}(Q^2)$ of Eq.~\eqref{resumAmb},
we introduce a new function, given by the dispersive integral of the ambiguity function,
\be
W_X(z):=\int_{0}^{\infty} \frac{dx}{\pi} \frac{Am_X(x)}{x-z} \, . \label{preweight}
\ee
We refer to this function as a preweight.
This function is defined in the complex $z$-plane, and 
satisfies
\be
{\rm Im} W_X(z+i0)=-{\rm Im} W_X(z-i0)= Am_X(z) \quad \text{for}~z \in \mathbb{R}_{\geq 0} \, . \label{Wproperty}
\ee
This function also has a real part. 
As we shall see below, the real part gives (part of) an unambiguous part of the perturbative prediction.
Namely the preweight plays an important role in reviving an unambiguous part,
while the preweight itself is obtained from the renormalon ambiguity.
 
For later convenience, we also define
\be
W_{X+}(z):=W_{X}(-z)=\int_{0}^{\infty} \frac{dx}{\pi} \frac{Am_X(x)}{x+z} \, . \label{WX+}
\ee
This function is regular for positive $z$.

From the preweight we define {\it extended} Borel transforms as [cf. Eq.~\eqref{BfromAm}]
\be
C_{X}(u) \equiv \frac{1}{\pi} \int_0^{\infty} dz \, W_{X}(z+i0) z^{-u-1} \, ,
\ee
\be
C_{X+}(u) \equiv \frac{1}{\pi} \int_0^{\infty} dz \, W_{X+}(z) z^{-u-1} \, .
\ee 
They are in fact related to the Borel transform $B_X^{\rm sing}$ as
\begin{align}
C_{X+}(u)
&=\frac{1}{\pi} \int_0^{\infty} \frac{dx}{\pi} Am_X(x) \int_0^{\infty} dz \frac{z^{-u-1}}{x+z} \non
&=-\frac{1}{\sin(\pi u)} \frac{1}{\pi} \int_0^{\infty} dx \,  Am_X(x) x^{-u-1} \non
&=-\frac{1}{\sin(\pi u)} B^{\rm sing}_X(u) \, ,
\end{align}
and\fn{To obtain Eq.~\eqref{CX}, we use 
\be
\int_0^{\infty} dz \frac{z^{-u-1}}{x-z-i0}=-\int_0^{\infty}  dz \frac{z^{-u-1}}{z+xe^{i \pi}}  \label{CX+}
\ee
and then use the previous result of the $z$-integration.}
\begin{align}
C_{X}(u)
&=\frac{1}{\pi} \int_0^{\infty} \frac{dx}{\pi} Am_X(x) \int_0^{\infty} dz \frac{z^{-u-1}}{x-z-i0} \non
&=-\frac{e^{-i \pi u}}{\sin(\pi u)}B^{\rm sing}_X(u) \, . \label{CX}
\end{align}
Here, we used Eq.~\eqref{BfromAm}.
(These functions were considered in Refs.~\cite{Ball:1995ni,Mishima:2016vna}
in the context of the large-$\beta_0$ approximation.)
These formulae are convenient to know asymptotic behaviors of the preweight at $z \sim 0$ and $z \sim \infty$,
since with the inverse formulae,
\be
W_X(z+i0)=\frac{1}{2 i} \int_{-i \infty}^{i \infty} du \, C_X(u) x^u \, ,
\ee
\be
W_{X+}(z)=\frac{1}{2 i} \int_{-i \infty}^{i \infty} du \, C_{X+}(u) x^u \, ,
\ee
one can calculate the small- or large-$z$ behavior by deforming the contour as in Eqs.~\eqref{Ambright}
and \eqref{Ambleft}
and picking up the contributions from singularities.

\subsection{Renormalon-free part in cutoff regularization}
\label{sec:2.4}
Now we explain how to extract the unambiguous part from the resummation formula \eqref{resumAmb}.
In this subsection, we use cutoff regularization. 
The present formulation is an extension of Ref.~\cite{Mishima:2016vna}, whose study 
is performed within the large-$\beta_0$ approximation.
In this regularization, we consider
\be
B^{\rm sing}_X(u;\hat{\mu}_f):=\frac{1}{\pi} \int_{\hat{\mu}_f}^{\infty} d x \, Am_X(x) x^{-u-1} \, ,
\ee
with $\hat{\mu}_f \gg e^{-1/(b_0 \alpha_s(Q))} (\sim \LMS^2/Q^2)$,
instead of $B^{\rm sing}_X(u)$ in Eq.~\eqref{BfromAm}.
$B^{\rm sing}_X(u;\hat{\mu}_f)$ is free from IR renormalons (singularities on the positive $u$-axis),
because they stem from the integral around $x \sim 0$.
Then, the regularized Borel integral and corresponding expression in terms of the ambiguity function
are given by
\begin{align}
X^{{\rm ren}}(Q;\hat{\mu}_f)
&=\frac{1}{b_0} \int _0^{\infty} du\, B^{\rm sing}_X(u;\hat{\mu}_f) e^{-u/(b_0 \alpha_s(Q))} \non
&=\frac{1}{b_0} \int_{\hat{\mu}_f}^{\infty} \frac{d x}{\pi x} \, Am_X(x) \int _0^{\infty} du\,  e^{-u/(b_0 \alpha_s(Q))} x^{-u}  \non
&=\frac{1}{b_0} \int_{\hat{\mu}_f}^{\infty} \frac{d x}{\pi x} \, Am_X(x) \frac{1}{\log{x}+\frac{1}{b_0 \alpha_s(Q)}} \, . \label{Xrencut}
\end{align}
In the last equality, we note $\log{x}+1/(b_0 \alpha_s(Q))>0$ due to the cutoff,
which ensures convergence of the $u$-integral.
The defined quantity shows dependence on the artificial parameter $\hat{\mu}_f$,
and this dependence is regarded as an uncertainty of this quantity.

\begin{figure}[t!]
\begin{minipage}{0.5\hsize}
\begin{center}
\includegraphics[width=6cm]{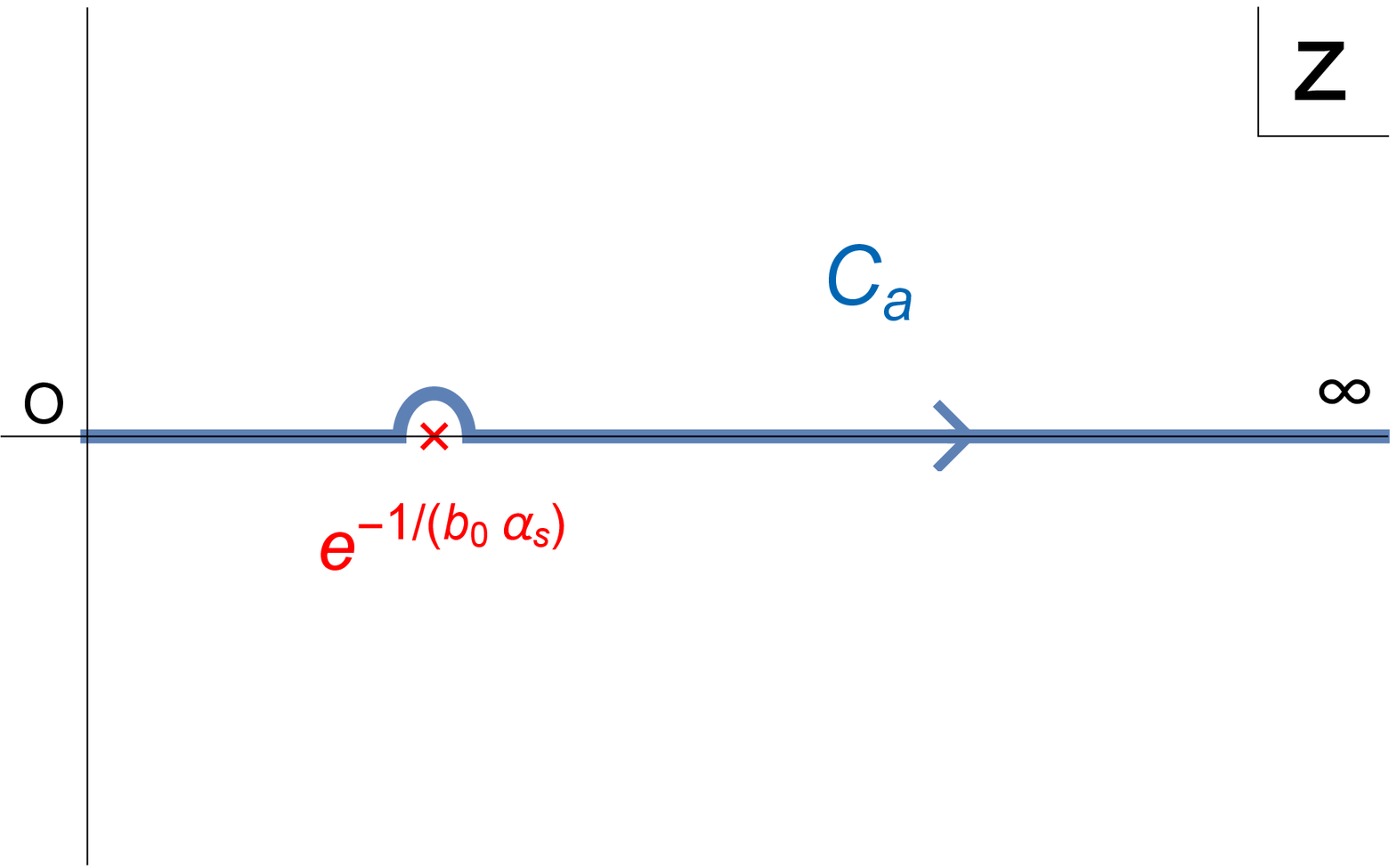}
\end{center}
\end{minipage}
\begin{minipage}{0.5\hsize}
\begin{center}
\includegraphics[width=6cm]{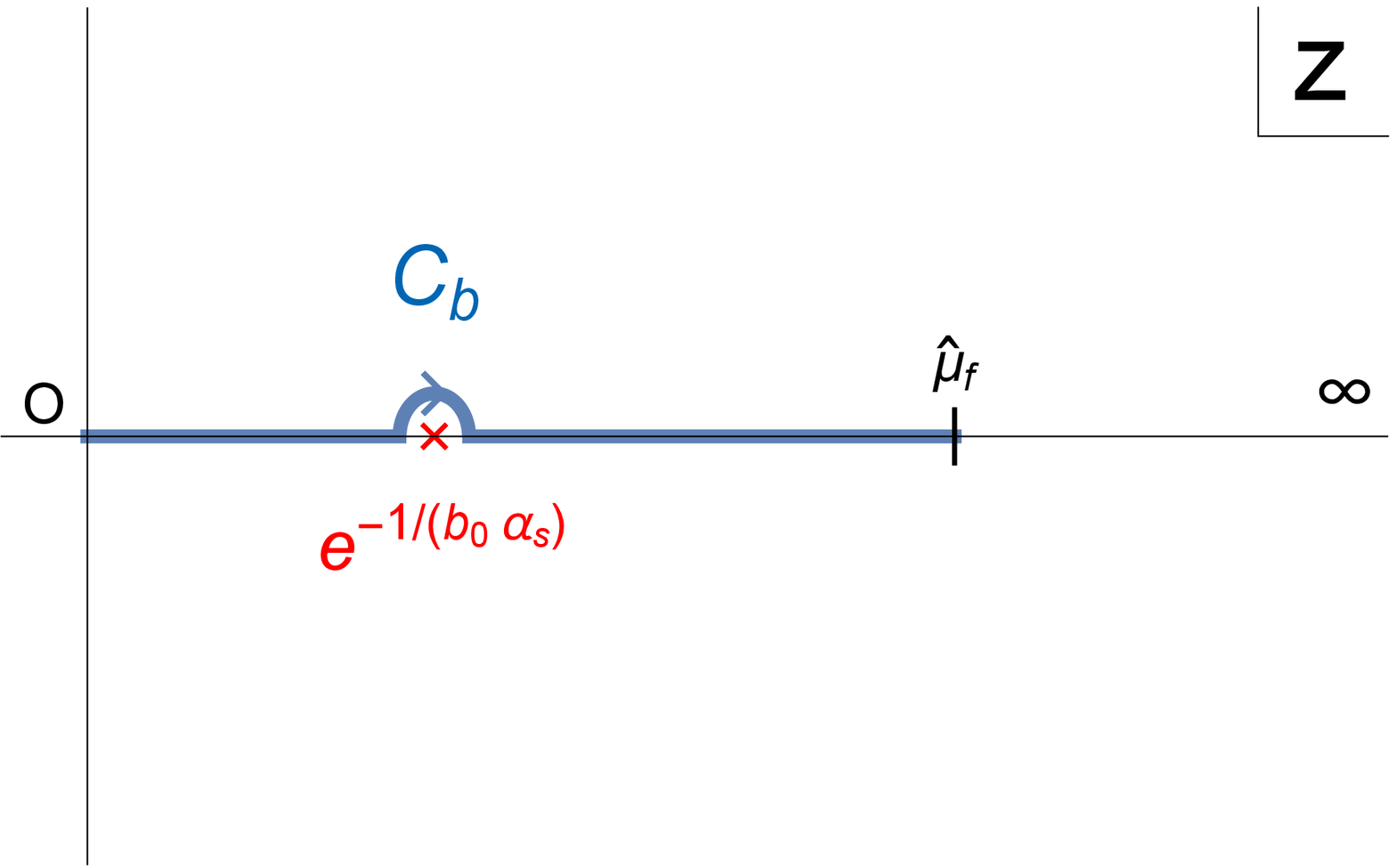}
\end{center}
\end{minipage}
\caption{Contours $C_a$ and $C_b$ in Eq.~\eqref{eqCaCb}.}
\label{fig:CaCb}
\end{figure}

We extract a regularization parameter ($\hat{\mu}_f$) independent part,
which we can identify as the unambiguous part.
Using Eq.~\eqref{Wproperty}, we can rewrite Eq.~\eqref{Xrencut} as
\begin{align}
X^{{\rm ren}}(Q^2;\hat{\mu}_f)
&=\frac{1}{b_0}  {\rm Im}   \int_{\hat{\mu}_f}^{\infty} \frac{dz}{\pi z} W_X(z+i0) \frac{1}{\log{z}+\frac{1}{b_0 \alpha_s(Q)}} \non
&=\frac{1}{b_0} {\rm Im} \lt( \int_{C_a} -\int_{C_b} \rt) \frac{dz}{\pi z} W_X(z+i0) \frac{1}{\log{z}+\frac{1}{b_0 \alpha_s(Q)}} \, . \label{eqCaCb}
\end{align}
Here the contour $C_a$ connects the origin $z=0$ to $z=\infty$ in the upper half plane avoiding the pole at $z=e^{-1/(b_0 \alpha_s(Q))}$,
and $C_b$ connects the origin $z=0$ to $z=\hat{\mu}_f$ in the upper half plane; see Fig.~\ref{fig:CaCb}.
The integral along $C_a$ is evaluated as
\begin{align}
\frac{1}{b_0} {\rm Im} \int_{C_a} \frac{dz}{\pi z} W_X(z+i0) \frac{1}{\log{z}+\frac{1}{b_0 \alpha_s(Q)}}
&=\frac{1}{b_0} \int_0^{\infty} \frac{dz}{\pi z} W_{X+}(z)   {\rm Im} \frac{1}{\log{z}+i\pi+\frac{1}{b_0 \alpha_s(Q)}} \non
&=\frac{1}{b_0} \int_0^{\infty} \frac{dz}{\pi z} W_{X+}(z) \frac{- \pi}{\lt(\log{z}+\frac{1}{b_0 \alpha_s(Q)} \rt)^2+\pi^2} \, ,
\end{align}
where we rotate the integration path to the negative real axis. [$W_{X+}(z)$ is defined in Eq.~\eqref{WX+}.]
On the other hand, the integral along $C_b$ is evaluated as follows.
Since we can decompose the preweight as
\be
W_X(z+i0)={\rm Re} \, W_X(z)+i Am_X(z) \, ,
\ee
for $z >0$, noting the contribution from the pole at $z=e^{-1/(b_0 \alpha_s(Q))}$,
we have 
\begin{align}
&-\frac{1}{b_0} {\rm Im} \int_{C_b} \frac{dz}{\pi z} W_X(z) \frac{1}{\log{z}+\frac{1}{b_0 \alpha_s(Q)}} \non
&=\frac{1}{b_0} {\rm Re} W_X(e^{-1/(b_0 \alpha_s(Q))})
-\frac{1}{b_0} {\rm Im} \int_{C_b} \frac{dz}{\pi z} i Am_X(z) \frac{1}{\log{z}+\frac{1}{b_0 \alpha_s(Q)}} \, .
\end{align}
As a result, we obtain
\begin{align}
X^{{\rm ren}}(Q^2;\hat{\mu}_f)
&=\frac{1}{b_0} \int_0^{\infty} \frac{dz}{\pi z} W_{X+}(z) \frac{- \pi}{\lt(\log{z}+\frac{1}{b_0 \alpha_s(Q)} \rt)^2+\pi^2} 
+\frac{1}{b_0} {\rm Re} W_X(e^{-1/(b_0 \alpha_s(Q))})\non
&\quad-\frac{1}{b_0} {\rm Im} \int_{C_b} \frac{dz}{\pi z} i Am_X(z) \frac{1}{\log{z}+\frac{1}{b_0 \alpha_s(Q)}} \, . \label{Xrencutoff}
\end{align}
Thus, we obtain the cutoff-independent part, i.e., unambiguous part as
\be
X_{\rm disp}^{\rm RF}(Q^2)
\equiv \frac{1}{b_0} \int_0^{\infty} \frac{dz}{\pi z} W_{X+}(z) \frac{- \pi}{\lt(\log{z}+\frac{1}{b_0 \alpha_s(Q)} \rt)^2+\pi^2}
+\frac{1}{b_0} {\rm Re} W_X(e^{-1/(b_0 \alpha_s(Q))}) \, . \label{XRFdisp1}
\ee
We refer to this part as $X_{\rm disp}^{\rm RF}(Q^2)$, which is obtained based on the dispersive integral of the ambiguity function.
Interestingly, we have the unambiguous part while starting from the ambiguity function.

As a whole, we have the following perturbative prediction:
\begin{align}
X(Q^2;\hat{\mu}_f)
&\equiv \sum_{n=0}^{\infty} \delta_n \alpha_s(Q)^{n+1} +X^{{\rm ren}}(Q^2;\hat{\mu}_f) \non
&=\sum_{n=0}^{\infty} \delta_n \alpha_s(Q)^{n+1}+X^{\rm RF}_{\rm disp}(Q^2)  \non
&-\frac{1}{b_0} {\rm Im} \int_{C_b} \frac{dz}{\pi z} i Am_X(z) \frac{1}{\log{z}+\frac{1}{b_0 \alpha_s(Q)}} \label{wholepre1} \, .
\end{align}
We obtain a renormalon-free part by the first line:
\be
X^{\rm RF}(Q^2)=\sum_{n=0}^{\infty} \delta_n \alpha_s(Q)^{n+1}+X^{\rm RF}_{\rm disp}(Q^2) \, . \label{XRF1}
\ee 
The renormalon-free part consists of the $\delta$ part and $X^{\rm RF}_{\rm disp}$.
On the other hand, the second line of Eq.~\eqref{wholepre1}  remains cutoff dependent,
and is regarded as the ambiguity in this regularization.

\subsection{Renormalon-free part in contour regularization}
\label{sec:2.5}
In this subsection, we extract the unambiguous part 
by starting from the contour regularization as in Eq.~\eqref{Xrenpm}.
This is an extension of the work within the large-$\beta_0$ approximation \cite{Ball:1995ni}.
Let us consider $X^{{\rm ren}}_{+}(Q^2)$:
\begin{align}
X^{{\rm ren}}_+(Q^2)
&=\frac{1}{b_0}\int_{C_+} du \, B^{\rm sing}_X(u) e^{-u/(b_0 \alpha_s(Q))} \, .
\end{align}
In this case it is convenient to use the following relation to give the Borel transform from the ambiguity function [cf. Eq.~\eqref{BfromAm}],
\be
B^{\rm sing}_X(u)=\frac{1}{\pi}  \int_0^{\infty-i 0} dx \, Am_X(x) x^{-u-1} \, .
\ee
This slight deformation of the integration contour does not change the result of the integral
(and thus correctly gives the Borel transform)
as long as the ambiguity function has a good convergence property at infinity.
Then, we deform the integration contour in the complex $u$-plane and make use of this relation:
\begin{align}
X^{{\rm ren}}_+(Q^2)
&=\frac{1}{b_0} \int_0^{i \infty} du \, B^{\rm sing}_X(u) e^{-u/(b_0 \alpha_s(Q))} \non
&=\frac{1}{b_0}  \int_{C_{-}} \frac{dx}{\pi x} Am_X(x)  \int_0^{\infty} i dv \, e^{-iv (\log{x}+\frac{1}{b_0 \alpha_s(Q)})} \non
&=\frac{1}{b_0}  \int_{C_{-}} \frac{dx}{\pi x} Am_X(x) \frac{1}{\log{x}+\frac{1}{b_0 \alpha_s(Q)}} \, .
\end{align}
In the final step, we use ${\rm Im} [\log{x}+1/(b_0 \alpha_s(Q))]<0$, which 
ensures convergence of the $v$-integral (where $u=iv$).
In a parallel manner, one sees that $X_-^{\rm ren}$ corresponds to the $x$-integration along $C_{+}$.
Thus, we obtain\fn{
We note that a naive idea that the choice of the $u$-integration contour $C_{+}$
corresponds to complexifying $\alpha_s$ as $\alpha_s \to \alpha_s-i \epsilon$
does not work and leads to a wrong (opposite) result.
One can see that such a shift of $\alpha_s$ does not serve for convergence of the $u$-integral.}
\be
X_{\pm}^{{\rm ren}}(Q^2)=\frac{1}{b_0} \int_{C_{\mp}} \frac{d x}{\pi x} Am_X(x) \frac{1}{\log{x}+\frac{1}{b_0 \alpha_s(Q)}} \, . \label{XpmAmb}
\ee

Now we rewrite Eq.~\eqref{XpmAmb} in the form where its real (unambiguous) part
and imaginary (ambiguous) part are clearly separated.
By using a preweight in Eq.~\eqref{preweight} and its property \eqref{Wproperty}, we obtain
\begin{align}
&\int_{C_{-}} \frac{dz}{\pi z} W_X(z) \frac{1}{\log{z}+\frac{1}{b_0 \alpha_s(Q)}} 
-\int_{C_{+}} \frac{dz}{\pi z} W_X(z) \frac{1}{\log{z}+\frac{1}{b_0 \alpha_s(Q)}} \non
&=(-2i) \lt[\int_{C_{\mp}} \frac{dz}{\pi z} Am_X(z) \frac{1}{\log{z}+\frac{1}{b_0 \alpha_s(Q)}} \mp i Am_X(e^{-\frac{1}{b_0 \alpha_s(Q)}})   \rt]
+2i {\rm Re} \, W_X(e^{-\frac{1}{b_0 \alpha_s(Q)}}) \, ,
\end{align}
by noting the existence of the pole at $z=e^{-1/(b_0 \alpha_s(Q))}$.
From this equation, we have
\begin{align}
X^{{\rm ren}}_{\pm}(Q^2)
&=\frac{1}{b_0}\int_{C_{\mp}} \frac{dz}{\pi z} Am_X(z) \frac{1}{\log{z}+\frac{1}{b_0 \alpha_s(Q)}} \non
&=-\frac{1}{2 i}  \frac{1}{b_0}\lt[ \int_{C_{-}} \frac{dz}{\pi z} W_X(z) \frac{1}{\log{z}+\frac{1}{b_0 \alpha_s(Q)}} 
-\int_{C_{+}} \frac{dz}{\pi z} W_X(z) \frac{1}{\log{z}+\frac{1}{b_0 \alpha_s(Q)}}  \rt] \non
&\quad{}+\frac{1}{b_0}{\rm Re} \, W_X(e^{-\frac{1}{b_0 \alpha_s(Q)}}) \pm \frac{1}{b_0} i Am_X(e^{-\frac{1}{b_0 \alpha_s(Q)}}) \, .
\end{align}
By rotating the integration paths $C_{\pm}$ to the negative real axis without hitting the pole,
we obtain
\begin{align}
X^{{\rm ren}}_{\pm}(Q^2)
&=\frac{1}{b_0}\int_0^{\infty} \frac{d z}{\pi z} W_{X+}(z)  \frac{-\pi}{\lt(\log{z}+\frac{1}{b_0 \alpha_s(Q)} \rt)^2+\pi^2} 
+\frac{1}{b_0} {\rm Re} \, W_X(e^{-\frac{1}{b_0 \alpha_s(Q)}}) \non
&\quad{}\pm\frac{1}{b_0} i Am_X(e^{-\frac{1}{b_0 \alpha_s(Q)}}) \, . \label{Xrendecom}
\end{align}
This is one of the main results in this paper.
The first line is a real part and corresponds to the unambiguous part.
We denote it by $X^{\rm RF}_{\rm disp}(Q^2)$, 
\begin{align}
X^{\rm RF}_{\rm disp}(Q^2)
&\equiv\frac{1}{b_0}\int_0^{\infty} \frac{d z}{\pi z} W_{X+}(z)  \frac{-\pi}{\lt(\log{z}+\frac{1}{b_0 \alpha_s(Q)} \rt)^2+\pi^2} +\frac{1}{b_0}{\rm Re} \, W_X(e^{-\frac{1}{b_0 \alpha_s(Q)}}) \non
&=\frac{1}{b_0} \int_0^{\infty} \frac{d z}{\pi z} W_{X+}(z)  \frac{-\pi}{\lt(\log{z}+\frac{1}{b_0 \alpha_s(Q)} \rt)^2+\pi^2} +\frac{1}{b_0}{\rm Re} \, W_X(0) \non
&\quad{}+\frac{1}{b_0} [{\rm Re} \, W_X(e^{-\frac{1}{b_0 \alpha_s(Q)}})-{\rm Re} \, W_X(0) ] \, , \label{XRFdisp2}
\end{align}
which is completely the same as Eq.~\eqref{XRFdisp1}.
In the last equality, we decompose $X^{\rm RF}_{\rm disp}$
into two parts. 
The first and second lines give qualitatively different behaviors
and such a decomposition is useful to understand the short-distance behavior of an observable, 
as we shall see.
In particular, the second line of Eq.~\eqref{XRFdisp2} gives a non-trivial power-like behavior to an observable.

We have a pure imaginary part in the second line of Eq.~\eqref{Xrendecom},
which indeed coincides with the renormalon uncertainty appearing in the Borel integral.
In this way, we can obtain an explicit result where the unambiguous part and ambiguous part
are clearly separated.

As a whole, we have the following perturbative prediction:
\begin{align}
X_{\pm}(Q^2)
&=\sum_{n=0}^{\infty} \delta_n \alpha_s(Q)^{n+1} +X^{{\rm ren}}_{\pm}(Q^2) \non
&=\sum_{n=0}^{\infty} \delta_n \alpha_s(Q)^{n+1}+X^{\rm RF}_{\rm disp}(Q^2) \non
&\quad{}\pm \frac{1}{b_0} i Am_X(e^{-\frac{1}{b_0 \alpha_s(Q)}}) \label{wholepre2}
\end{align}
The first line is real, which we call a renormalon-free part:
\be
X^{\rm RF}(Q^2)
=\sum_{n=0}^{\infty} \delta_n \alpha_s(Q)^{n+1} +X_{\rm disp}^{\rm RF}(Q^2) \, .
\ee
This is again the same as Eq.~\eqref{XRF1}.
The renormalon-free part gives a net and reliable part of
the originally divergent asymptotic series.
We note that the last term in Eq.~\eqref{wholepre1} and that in Eq.~\eqref{wholepre2}
(which are regarded as ambiguous parts) differ.
This is because we adopt different regularizations.
The method in the present subsection is superior in the sense that
the ambiguity coincides with the renormalon uncertainty of the Borel integral,
which can be canceled against an uncertainty of a nonperturbative matrix element in the OPE.
Thus, this method to calculate the perturbative contribution seems to be optimal to 
be systematically combined with the OPE framework.

\subsection{Renormalization group properties}
\label{sec:2.6}

So far, we have assumed that $\mu=Q$. 
Here we reveal some aspects of the formulation 
from RG analyses by considering general $\mu$.
First, the sum of the $\delta$ part and $X^{\rm ren}_{\pm}$, which is the 
final result we give, is RG invariant. Namely, it is independent of
the choice of the renormalization scale $\mu$.
This is because the sum coincides with the Borel integral of 
the original series, which is RG invariant as shown in App.~\ref{app:A}.
This conclusion that the Borel integral is RG invariant
agrees with the previous study~\cite{Ayala:2019uaw} but disagrees with Ref.~\cite{Chyla:1990na}.
Appendix~\ref{app:A} can be regarded as an explicit extension 
of Ref.~\cite{Ayala:2019uaw} to all order with respect to the beta function.

Now we focus on the $\mu$ dependence of each object in the separation,
i.e. the $\mu$ dependence of $\delta$ part or that of $X^{\rm ren}_{\pm}$.
(We adopt contour regularization here.) 
Since we know that the sum of them is $\mu$ independent,
it is sufficient to study one of them.
Then, we study $\mu$ dependence of $X^{\rm ren}_{\pm}$, 
\be
X^{\rm ren}_{\pm}=\int_{C_{\pm}}dt \,  \tilde{B}_X^{\rm sing}(t;Q,\mu) e^{-t/\alpha_s(\mu)} \, . \label{XrenBsing}
\ee
[In this subsection we adopt the same convention for the Borel transform as in App.~\ref{app:A};
we add a tilde to distinguish it from the one we have used so far.
See Eq.~\eqref{Boreltconv} for the definition of $\tilde{B}_X(t;Q,\mu)$.]
In the present discussion, we assume that there are only IR renormalons.\fn{
A study of the case with UV renormalons is left for future work.} 
We first study the RG property of the singular part of the Borel transform 
$\tilde{B}_X^{\rm sing}(t;Q,\mu)$. This RG property can be revealed from
the fact that a renormalon ambiguity is RG invariant.
(We note that a renormalon ambiguity is an imaginary part of the RG invariant quantity,  
$\int_{C_{\pm}} dt\, \tilde{B}_X(t;Q,\mu) e^{-t/\alpha_s(\mu)}$.) 
A renormalon ambiguity is given in a form  \cite{Beneke:1998ui}\fn{
It is known in the OPE argument that the parameters $a$, $\nu$, and $s_k$ can be parameterized by the coefficients of 
the beta function and the anomalous dimension of the operator responsible for the cancellation of the renormalon, 
and the perturbative coefficients of the Wilson coefficient of the operator.
However, this information is not needed to show that $\tilde{B}^{\rm sing}(t;Q,\mu)$ satisfies a desired RG equation.}
\be
{\text{(Renormalon amb.)}}=N(Q,\mu) e^{-a/\alpha_s(\mu)} [b_0 \alpha_s(\mu)]^{-\nu} \sum_{k=0}^{\infty} s_k(Q,\mu) \alpha_s(\mu)^k \, .
\ee
From the RG invariance of this quantity, we obtain
\be
\lt( \mu^2 \frac{\del}{\del \mu^2}+\beta(\alpha_s) \frac{\del}{\del \alpha_s} \rt) {\text{(Renormalon amb.)}}
=0 \, .
\ee 
Here we approximate the beta function by $\{b_0, \dots , b_n\}$: $\beta(\alpha_s)=-\sum_{i=0}^n b_i \alpha_s^{i+2}$.
Then, we obtain the following equations for $N(Q,\mu)$ and $s_k(Q,\mu)$:
\be
\frac{\del \log{N(Q,\mu)}}{\del \log{\mu^2}}=a b_0  \label{RGforN}
\ee
and
\begin{align}
\mu^2 \frac{\del}{\del \mu^2} s_k
&=b_0 (k-\nu-1) s_{k-1}+\sum_{i=1}^{{\rm min}\{n,k-1\}} b_i (k-i-1-\nu) s_{k-i-1}
+\sum_{i=1}^{{\rm min}\{n,k\}}b_i a  s_{k-i}   \label{RGfors}
\end{align}
for $k \geq 1$.
We assumed $\mu^2 \del s_0/\del \mu^2=0$ as the $\mu$ dependence can always be absorbed by $N(Q,\mu)$.
The first equation tells us that $N(Q,\mu) \propto (\mu^2/Q^2)^{a b_0}$.
The singular part of the Borel transform corresponding to the above renormalon ambiguity
is obtained by
\be
\tilde{B}_X^{\rm sing}(t;Q,\mu)= c \frac{N(Q,\mu)\Gamma(1+\nu)}{(1-t/a)^{1+\nu}} \sum_{k=0}^{\infty} s_k'(Q,\mu) (1-t/a)^k \, , \label{Btildesing}
\ee
where $s_k'$ is related to $s_k$ as
\be
s_0=s'_0
\ee
\be
s'_k=a^k \frac{1}{\nu \cdot \cdots \cdot (\nu-k+1)} s_k \quad {\text{for $k \geq 1$}} \, ,
\ee
and $c$ is a constant, $c=1/(a^{1+\nu} b_0^{\nu} \pi)$.
Then, $\mu^2$ dependence of $s'_k$ is controlled by Eqs.~\eqref{RGforN} and \eqref{RGfors} as
\begin{align}
\mu^2 \frac{\del}{\del \mu^2} s'_k
&=-b_0 a s'_{k-1} \non
&\quad{}-\sum_{i=1}^{{\rm min}\{n,k-1\}} b_i \frac{a^{i+1}}{(\nu-k+i) \cdots (\nu-k+1)} s'_{k-i-1}
+\sum_{i=1}^{{\rm min}\{n,k\}} b_i \frac{a^{i+1}}{(\nu-k+i) \cdots (\nu-k+1)} s'_{k-i} \, .
\end{align}
Now we can show that the singular part of the Borel transform satisfies 
the RG equation 
$\frac{\del^n}{\del t^n} \mu^2 \frac{\del}{\del \mu^2} \tilde{B}_X^{\rm sing}(t;Q,\mu)
=\sum_{i=0}^n b_i \frac{\del^{n-i}}{\del t^{n-i}} [t \tilde{B}_X^{\rm sing}(t)]$, which is the same RG equation
as the one that the total Borel transform satisfies [see Eqs.~\eqref{eq:(A5)}--\eqref{eq:(A7)}].
Using
\begin{align}
\frac{\del^m}{\del t^m} [t \tilde{B}_X^{\rm sing}(t)]
&=a \frac{\del^m}{\del t^m} [1-(1-t/a) \tilde{B}_X^{\rm sing}(t)] \non
&=a c N(Q,\mu) \Gamma(1+\nu)
\bigg[\sum_{k=0}^{\infty}  (1+\nu-k) \cdots (m+\nu-k)a^{-m} s'_k (1-t/a)^{-1-\nu+k-m}  \non
&\qquad{}\qquad{}\qquad{} -\sum_{k=0}^{\infty}  (\nu-k) \cdots (m-1+\nu-k) a^{-m} s'_k (1-t/a)^{-\nu+k-m} \bigg]
\end{align}
we obtain
\begin{align}
&[c a N(Q,\mu) \Gamma(1+\nu) (1-t/a)^{1+\nu+n} ]^{-1} \lt[\frac{\del^n}{\del t^n}\mu^2\frac{\del}{\del \mu^2}  \tilde{B}_X^{\rm sing}(t)
-\sum_{i=0}^n b_i \frac{\del^{n-i}}{\del t^{n-i}} [t \tilde{B}_X^{\rm sing}(t)] \rt]\non
&=\sum_{k=0}^{\infty} (1+\nu-k) \cdots (n+\nu-k) a^{-n} \lt[b_0 s'_k+\frac{1}{a}\mu^2 \del s'_k/\del \mu^2 \rt] (1-t/a)^k \non
&\quad{}-b_0  \sum_{k=0}^{\infty} (1+\nu-k) \cdots (n+\nu-k) a^{-n}  s'_k (1-t/a)^k \non
&\quad{}+b_0 \sum_{k=0}^{\infty}  (\nu-k) \cdots (n-1+\nu-k) a^{-n} s'_k (1-t/a)^{k+1} \non
&\quad{}-\sum_{i=1}^n b_i 
\bigg[\sum_{k=0}^{\infty}  (1+\nu-k) \cdots (n-i+\nu-k)a^{-(n-i)} s'_k (1-t/a)^{k+i}  \non
&\qquad{}\qquad{} -\sum_{k=0}^{\infty}  (\nu-k) \cdots (n-i-1+\nu-k) a^{-(n-i)} s'_k (1-t/a)^{k+i+1} \bigg] \non
&=(1+\nu-k) \cdots (n+\nu-k) a^{-n-1}  \non
&\times \bigg\{ \sum_{k=1}^{n} \bigg[\mu^2 \frac{\del s'_k}{\del \mu^2}+b_0 a s'_{k-1}-\sum_{i=1}^{k} b_i  \frac{1}{(1+\nu-k) \cdots (\nu-k+i)} a^{i+1} s'_{k-i} \non
&\qquad{}\qquad{}+\sum_{i=1}^{k-1} b_i \frac{1}{(1+\nu-k) \cdots (\nu-k+i)} a^{i+1} s'_{k-i-1} \bigg] (1-t/a)^k \non
&\quad{}+\sum_{k=n+1}^{\infty} \bigg[\mu^2 \frac{\del s'_k}{\del \mu^2}+b_0 a s'_{k-1} 
-\sum_{i=1}^{n} b_i  \frac{1}{(1+\nu-k) \cdots (\nu-k+i)} a^{i+1} s'_{k-i} \non
&\qquad{}\qquad{}\qquad{}+\sum_{i=1}^{n} b_i \frac{1}{(1+\nu-k) \cdots (\nu-k+i)} a^{i+1} s'_{k-i-1} \bigg] (1-t/a)^k \bigg\}  \non
&=0 \, . \label{BsingRG}
\end{align}
Now we can see the RG property of $X^{\rm ren}_{\pm}$. From
\begin{align}
\mu^2 \frac{\del}{\del \mu^2} X^{\rm ren}_{\pm}
&=\int_{0 \pm i0}^{\infty \pm i0} dt \, \mu^2 \frac{\del}{\del \mu^2} \tilde{B}_X^{\rm sing}(t) e^{-t/\alpha_s(\mu)} \non
&=\sum_{k=0}^{n-1} \alpha_s^{k+1} \mu^2 \frac{\del}{\del \mu^2} \frac{\del^k \tilde{B}_X^{\rm sing}}{\del t^k}(t=0) \non
&\quad{}+\alpha_s^n \int_{0 \pm i0}^{\infty \pm i0}
 dt \, \mu^2 \frac{\del}{\del \mu^2} \frac{\del^n}{\del t^n} \tilde{B}_X^{\rm sing}(t) \cdot e^{-t/\alpha_s}
\end{align}
and
\begin{align}
\beta(\alpha_s) \frac{\del}{\del \alpha_s} X^{\rm ren}_{\pm}
&=\frac{\beta(\alpha_s)}{\alpha_s^2} \int_{0 \pm i0}^{\infty \pm i0} dt \, t \tilde{B}_X^{\rm sing}(t) e^{-t/\alpha_s(\mu)} \non
&=-\sum_{i=0}^n b_i \sum_{k=1}^{n-i-1} \alpha_s^{i+k+1} \frac{\del^{k} (t \tilde{B}_X^{\rm sing})}{\del t^{k}} (t=0) \non
&\quad{}-\sum_{i=0}^n b_i \alpha_s^n \int_{0\pm i0}^{\infty \pm i0} dt \, \frac{\del^{n-i}}{\del t^{n-i}} [t \tilde{B}_X^{\rm sing}(t)] \cdot e^{-t/\alpha_s(\mu)} \, ,
\end{align}
where we have used integration by parts repeatedly and have surface terms, 
we obtain
\begin{align}
\mu^2 \frac{d}{d \mu^2} X^{\rm ren}_{\pm}
&=
\sum_{k=0}^{n-1} \alpha_s^{k+1} \mu^2 \frac{\del}{\del \mu^2} \frac{\del^k \tilde{B}_X^{\rm sing}}{\del t^k}(t=0) \non
&\quad{}-\sum_{i=0}^n b_i \sum_{k=1}^{n-i-1} \alpha_s^{i+k+1} \frac{\del^{k} (t \tilde{B}_X^{\rm sing})}{\del t^{k}} (t=0)  \, . \label{RGofBintofBsing}
\end{align}
It turned out that $X^{\rm ren}_{\pm}$ is not $\mu$ independent.
However, as can be seen from the above derivation, 
due to the RG equation \eqref{BsingRG}, 
the RG invariance of $X^{\rm ren}_{\pm}$ is broken only by the surface terms.
As a result, the breaking of the RG invariance~\eqref{RGofBintofBsing} is given by a finite series (when
we approximate the beta function up to $b_n$) despite the fact that 
the Borel integral itself contains an infinite series.

We study the same issue also in the alternative representation
using an ambiguity function, where our formulation is developed.
We define the ambiguity function with a tilde by
\be
\Tilde{Am}_X(x;Q,\mu)=\frac{1}{2 i} \int_{-i \infty}^{i \infty} dt \,  \tilde{B}_X^{\rm sing}(t;Q,\mu) x^t \, .
\ee
We note that we assume that only IR renormalons exist.
The ambiguity function corresponding to an IR renormalon
is expected to have a finite domain, as indicated in Eq.~\eqref{Ambright}.
Then, we consider the domain of the ambiguity function to be $0 \leq x \leq \tilde{\rho}$
with a parameter $\tilde{\rho}>0$.
(The domain of the ambiguity function without a tilde is $0 \leq x \leq \rho=\tilde{\rho}^{1/b_0}$.)
As we shall see, the RG argument here tells us how $\tilde{\rho}$ should depend on $\mu$.

Corresponding to the above domain, $\tilde{B}_X^{\rm sing}(t)$ is defined by
\be
\tilde{B}_X^{\rm sing}(t;Q,\mu)= \int_0^{\tilde{\rho}} \frac{dx}{\pi x} \, \tilde{Am}_X(x;Q,\mu) x^{-t} \label{Bsingamb}
\ee
and the resummation formula is given by
\be
X_{\pm}^{\rm ren}=\int_{0 \mp i0}^{\tilde{\rho} \mp i0} \frac{dx}{\pi x} \, \tilde{Am}_X(x;Q,\mu) \frac{1}{\log{x}+\frac{1}{\alpha_s(\mu)}} \, . \label{Xrenamb}
\ee
(We remark that $X_{\pm}^{\rm ren}$ here is not always identical to the above one \eqref{XrenBsing},
which is defined with the Borel transform of Eq.~\eqref{Btildesing}.
This is because we can choose any small $\tilde{\rho}$ in Eqs.~\eqref{Xrenamb} and \eqref{Bsingamb},
and singular Borel transforms with different $\tilde{\rho}$ differ by a regular function.)
Since an IR renormalon ambiguity $\Tilde{Am}_X(x=e^{-1/\alpha_s(\mu)};Q,\mu)$ is RG invariant,
we have the RG equation for the ambiguity function:
\be
\mu^2 \frac{\del}{\del \mu^2} \Tilde{Am}_X(x;Q,\mu)+\frac{\beta\lt(-\frac{1}{\log{x}} \rt)}{\lt(-\frac{1}{\log{x}} \rt)^2} x \frac{\del}{\del x} \Tilde{Am}(x;Q,\mu)=0 \, . \label{RGforAm}
\ee
Noting that 
\be
\frac{1}{\log{x}+\frac{1}{\alpha_s}}
=\sum_{k=0}^{m-1} \alpha_s^{k+1} (-\log{x})^k+ (-\alpha_s \log{x})^m \frac{1}{\log{x}+\frac{1}{\alpha_s}}
\ee
for an arbitrary positive integer $m$, we obtain
\begin{align}
&\mu^2 \frac{d}{d \mu^2} X_{\pm}^{\rm ren} \non
&=\mu^2 \frac{\del}{\del \mu^2} \int_{0 \mp i0}^{\tilde{\rho} \mp i0}\frac{dx}{\pi x} \, \tilde{Am}_X(x;Q,\mu) \frac{1}{\log{x}+\frac{1}{\alpha_s(\mu)}} \non
&\quad{}-\frac{\beta(\alpha_s)}{\alpha_s^2} \int_{0 \mp i0}^{\tilde{\rho} \mp i0} \frac{dx}{\pi x} \, \tilde{Am}_X(x;Q,\mu) x \frac{\del}{\del x} \frac{1}{\log{x}+\frac{1}{\alpha_s(\mu)}} \non
&= \mu^2 \frac{\del}{\del \mu^2}  \int_{0 \mp i0}^{\tilde{\rho} \mp i0}\frac{dx}{\pi x} \,\tilde{Am}_X(x;Q,\mu) 
\lt[\sum_{k=0}^{n-1} \alpha_s^{k+1} (-\log{x})^k+ (-\alpha_s \log{x})^n \frac{1}{\log{x}+\frac{1}{\alpha_s}} \rt] \non
&\quad{}+\sum_{i=0}^n b_i \alpha_s^i  \int_{0 \mp i0}^{\tilde{\rho} \mp i0} \frac{dx}{\pi x} \, \tilde{Am}_X(x;Q,\mu) \cdot 
x \frac{\del}{\del x} \lt[ \sum_{k=0}^{n-i-1} \alpha_s^{k+1} (-\log{x})^k+ (-\alpha_s \log{x})^{n-i} \frac{1}{\log{x}+\frac{1}{\alpha_s(\mu)}} \rt] \non
&=\sum_{k=0}^{n-1} \alpha_s^{k+1} \mu^2 \frac{\del}{\del \mu^2}  \int_{0 \mp i0}^{\tilde{\rho} \mp i0}\frac{dx}{\pi x} \, \tilde{Am}_X(x;Q,\mu)   (-\log{x})^k \non
&\quad{}+\sum_{i=0}^n b_i \sum_{k=0}^{n-i-1} \alpha_s^{i+k+1}  \int_{0 \mp i0}^{\tilde{\rho} \mp i0} \frac{dx}{\pi x} \, \tilde{Am}_X(x;Q,\mu) x \frac{\del}{\del x} (-\log{x})^k \non
&\quad{}+(-\alpha_s \log{\tilde{\rho}})^n \lt[\frac{1}{\tilde{\rho}} \mu^2\frac{\del \tilde{\rho}}{\del \mu^2}+\sum_{i=0}^n b_i \frac{1}{(-\log{\tilde{\rho}})^i}  \rt] \frac{1}{\pi} \tilde{Am}_X(\tilde{\rho}) \frac{1}{\log{\tilde{\rho}}+\frac{1}{\alpha_s(\mu)}} \, , \label{RGofXrenwithAmfn}
\end{align}
where we have used integration by parts and the RG equation \eqref{RGforAm}.
The first two lines of the right hand side show the same structure as Eq.~\eqref{RGofBintofBsing};
they are equal to
$\sum_{k=0}^{n-1} \alpha_s^{k+1} \mu^2 \frac{\del}{\del \mu^2} \frac{\del^k \tilde{B}_X^{\rm sing}}{\del t^k}(t=0)
-\sum_{i=0}^n b_i \sum_{k=1}^{n-i-1} \alpha_s^{i+k+1} \frac{\del^k (t \tilde{B}_X^{\rm sing})}{\del t^k}(t=0)$,
where $\tilde{B}_X^{\rm sing}$ is defined in Eq.~\eqref{Bsingamb}.
The third line represents an extra contribution.
It would be natural to eliminate the extra contribution
so that we can keep the good property that the breaking of the RG invariance
is represented merely by a finite series.
We can realize this property by considering running $\tilde{\rho}(\mu)$ satisfiying the RG equation, 
\be
\frac{1}{\tilde{\rho}} \mu^2 \frac{\del \tilde{\rho}}{\del \mu^2}=-\sum_{i=0}^n b_i \lt(-\frac{1}{\log{\tilde{\rho}} }\rt)^i \, .
\ee
If we define $f(\tilde{\rho}):=-\frac{1}{\log{\tilde{\rho}}}$, we can see that $f(\tilde{\rho})$ satisfies the same RG equation as
the running coupling,
\be
\mu^2 \frac{d f(\tilde{\rho})}{d \mu^2}=-\sum_{i=0}^n b_i f(\tilde{\rho})^{i+2} \, .
\ee
Hence, we have a relation
\be
\frac{\mu_0^2}{\mu^2}=\exp \lt[-\int_{f(\tilde{\rho}(\mu_0))}^{f(\tilde{\rho}(\mu))} \frac{dx}{\beta(x)} \rt] \, , \label{relationforrho}
\ee
where $\mu_0$ is an RG invariant scale.
Instead of considering the running of $\tilde{\rho}$,
we may choose $\tilde{\rho}$ such that 
\be
\tilde{Am}_X(\tilde{\rho})=0 \, . \label{secondop}
\ee
It is worth noting that, once one takes $\tilde{\rho}$ in an above way,
it is easy to show that $\tilde{B}_X^{\rm sing}$ of Eq.~\eqref{Bsingamb} satsfies the RG equation 
$\frac{\del^n}{\del t^n} \mu^2 \frac{\del}{\del \mu^2} \tilde{B}_X^{\rm sing}(t;Q,\mu)
=\sum_{i=0}^n b_i \frac{\del^{n-i}}{\del t^{n-i}} [t \tilde{B}_X^{\rm sing}(t)]$.
In the subsequent analyses, we adopt the first option, i.e., running $\tilde{\rho}(\mu)$, in Sec.~\ref{sec:4}, 
while we adopt the second option, i.e., $\tilde{\rho}$ is taken as a zero of an ambiguity function, in Sec.~\ref{sec:3.3}.
When we adopt the first option, $\tilde{\rho}_0$ is not fixed a priori
and we have to choose some value.

We make a comment on differences between 
the $b_0$ level analysis ($n=0$) and beyond it ($n \geq 1$).
For $n \geq 1$, $X^{\rm ren}$ is not RG invariant.
Then, the $\delta$ part is required to exist so that it cancels the $\mu$ dependence of $X^{\rm ren}_{\pm}$.
This argument shows necessity of the $\delta$ part for $n \geq 1$ from the viewpoint of the RG property.
On the other hand, for $n=0$ Eq.~\eqref{RGofBintofBsing} is zero and the Borel integral of the singular part alone is RG invariant
(as long as it satisfies the RG equation $\mu^2 \del \tilde{B}^{\rm sing}/\del \mu^2=b_0 t \tilde{B}^{\rm sing}$).
This indicates that the $\delta$ part may not be necessary at the $b_0$ level analysis.
Indeed in the formulations in the large-$\beta_0$ approximation, 
we do not have the $\delta$ part \cite{Ball:1995ni, Mishima:2016vna}.
In this way, we can naturally understand the origin of the $\delta$ part
in analyses beyond the large-$\beta_0$ approximation.
A more explicit explanation on the necessity of the $\delta$ part is as follows. 
At the $b_0$ level, from the differential equation
$\tilde{B}^{\rm sing}$ should be a form $\tilde{B}^{\rm sing}(t;Q,\mu)=(\mu^2/Q^2)^{b_0 t} \times ({\text{function of $t$}})$,
and this form is consistent with the complete Borel transform in the large-$\beta_0$ approximation
[cf. Eqs.~\eqref{BorelD} and \eqref{BorelVLL}].
On the other hand, beyond the $b_0$ level, although the singular Borel transform \eqref{Btildesing}
satisfies the RG equation, this Borel transform is not consistent with the complete Borel transform.
One can see this from the fact that perturbative coefficients corresponding to the singular Borel transform 
are not polynomials of $\log{(\mu^2/Q^2)}$ unlike original perturbative coefficients
due to the overall factor $N(Q,\mu) \propto (\mu^2/Q^2)^{ab_0}$.
These explain why we do not need to split the Borel transform in the large-$\beta_0$ approximation
but we need to split the Borel transform beyond the $b_0$ level to
deal with renormalon divergences.

\subsection{Practical use of the formulation and discussion on error}
\label{sec:2.7}

The argument so far is formal in the sense that we assumed, for instance, 
that we know perturbative series to all orders.
Here we discuss some practical issues.
Before this, we clarify a role of the formulation in the context of
the OPE.
The OPE of an observable $X(Q^2)$ is given by
\be
X^{\rm OPE}(Q^2)=C_1(Q^2/\mu^2,\alpha_s(\mu))
+C_{\mathcal{O}}(Q^2/\mu^2,\alpha_s(\mu)) \frac{\langle 0| \mathcal{O}(\mu) |0 \rangle}{Q^d}+\cdots \, ,
\ee
where $\mathcal{O}$ is a renormalized local operator of mass dimension $d$,
and $C_1$ and $C_{\mathcal{O}}$ are Wilson coefficients.
Since the local condensate $\langle 0| \mathcal{O}(\mu) |0 \rangle$ is a nonperturbative effect,
the perturbative expansion of the observable is identified with that of $C_1$.
Due to renormalon divergences in this series, $C_1$ is regularized in the Borel procedure
and after that it is given by the sum of a real and imaginary part.
The imaginary part, which is a renormalon uncertainty, is expected to cancel with
the imaginary ambiguity in the nonperturbative condensate.
Hence, we formally have
\be
X^{\rm OPE}(Q^2)={\rm Re} \, C_1(Q^2/\mu^2,\alpha_s(\mu))
+C_{\mathcal{O}}(Q^2/\mu^2,\alpha_s(\mu)) \frac{{\rm Re} \, \langle 0| \mathcal{O}(\mu) |0 \rangle}{Q^d}+\cdots . \label{realOPE}
\ee
In this discussion, we just focus on the first IR renormalon and 
ignore renormalons in $C_{\mathcal{O}}(Q^2/\mu^2,\alpha_s(\mu))$.
Also we assume that there are no UV renormalons.
The real part of the nonperturbative condensate is treated as a parameter.
What we have studied in the present paper is 
how to obtain ${\rm Re} \, C_1(Q^2/\mu^2,\alpha_s(\mu))$,
which we call a renormalon-free part.
To explicitly show that what we treat is the Wilson coefficient $C_1$,
we call that renormalon-free part ${\rm Re} \, C_1$ here, instead of $X^{\rm RF}$.

Although we have assumed so far that we know 
the all-order perturbative series and the complete form of the ambiguity function 
(or equivalently the complete form of the renormalon ambiguity), the knowledge on them is practically limited.
Consider the situation where we know the perturbative series to the $n_1$th order,
and the form of the renormalon ambiguity to the $n_2$th order:
\be
X(Q^2)|_{\rm pert}=\sum_{k=0}^{n_1} d_k \alpha_s^{k+1} \, , \label{trunX}
\ee
\be
{\text{(Renormalon amb.)}}=N e^{-\frac{u_0}{b_0 \alpha_s}} (b_0 \alpha_s)^{-\nu} \sum_{k=0}^{n_2-1} s_k \alpha_s^{k}
\ee
where $u_0(=d/2)$ is the first IR renormalon.\fn{
We regard that the renormalon ambiguity at $n_2=0$ 
is the one where $\nu$ is set to zero. 
This corresponds to regarding the form of the renormalon ambiguity obtained
in the large-$\beta_0$ approximation is consistent with the $n_2=0$ result.}
(Here we mean that the parameters $u_0$, $\nu$, $s_0, \dots s_{n_2-1}$ are known
but the parameter $N$ is not known.)
We note that the order $n_1$ and $n_2$ are independent.
Now, we explain the practical procedure to obtain an approximated ${\rm Re} \, C_1(Q^2/\mu^2,\alpha_s(\mu))$
in this situation.
First, we construct the ambiguity function approximately:
\be
Am^{(n_1,n_2)}(x)=b_0 N^{(n_1)} x^{u_0} (-\log{x})^{\nu} \sum_{k=0}^{n_2} s_k \lt(-\frac{1}{b_0 \log{x}} \rt)^k \, .
\ee
Here, $N^{(n_1)}$ is the normalization constant estimated from the $n_1$th order perturbative series for $C_1$.
For the estimate, we use the method proposed in Ref.~\cite{Lee:1996yk},
where $N^{(n_1)} \to N$ as $n_1 \to \infty$ is ensured.
From this ambiguity function, we can calculate ${C_1}^{\rm RF}_{\rm disp}$ (which means $X^{\rm RF}_{\rm disp}$) approximately;
we obtain an approximated preweight $W_{X}^{(n_1,n_2)}$ using Eq.~\eqref{preweight} and 
then ${C_1}^{{\rm RF}(n_1,n_2)}_{\rm disp}$ using Eq.~\eqref{XRFdisp2}.
With the above ambiguity function, 
we can obtain approximated $d_k^{\rm ren}$ from Eq.~\eqref{FOfromAmb},
and then construct the $\delta$ part,
\be
\delta_k^{(n_1,n_2)}=d_k-d_k^{{\rm ren} (n_1,n_2)} \, ,
\ee
for $k=0, \dots, n_1$.
In this way, we obtain the approximated ${\rm Re} \, C_1$ as
\be
{\rm Re} \, C_1^{(n_1,n_2)}={C_1}^{{\rm RF}(n_1,n_2)}_{\rm disp}+\sum_{k=0}^{n_1} \delta_k^{(n_1,n_2)} \alpha_s^{k+1} \, .
\ee
In Sec.~\ref{sec:4}, we study the QCD potential with the accuracy $(n_1,n_2)=(3,3)$.

Secondly, we discuss the error of ${\rm Re} \, C_1^{(n_1,n_2)}$. 
Although ${\rm Re} \, C_1^{(n_1,n_2)}$ will converge to ${\rm Re} \, C_1$ as $n_1$ and $n_2$ are large enough, 
they can be different for finite $n_1$ and $n_2$.
To estimate the size of the difference we consider the asymptotic expansion of ${\rm Re} \, C_1^{(n_1,n_2)}$:
\be
{\rm Re} \, C_1^{(n_1,n_2)}=\sum_{k=0}^{n_1} d_k \alpha_s^{k+1}+\sum_{k=n_1+1}^{\infty} d_k^{{\rm ren}(n_1,n_2)} \alpha_s^{k+1} \, .
\ee
Here we note that $X^{\rm RF}_{\rm disp}$ contains an all-order perturbative series.
On the other hand, the asymptotic expansion of ${\rm Re} \, C_1$ is of course given by
\be
{\rm Re} \, C_1=\sum_{k=0}^{\infty} d_k \alpha_s^{k+1}=\sum_{k=0}^{\infty} (d_k^{\rm ren}+\delta_k) \alpha_s^{k+1} \, .
\ee
Corresponding to the singular Borel transform of Eq.~\eqref{Btildesing},
$d^{\rm ren}_k$ is obtained as
\be
d^{\rm ren}_k=c N \Gamma(k+\nu+1) \lt(\frac{b_0}{u_0} \rt)^k 
\sum_{m=0}^{\infty} s'_m \frac{(\nu+1-m) \cdots \nu}{(k+\nu) \cdots (k+\nu-m+1)}
\ee
and, on the other hand, $d^{{\rm ren}(n_1,n_2)}_k$ is obtained as
\be
d^{{\rm ren}(n_1,n_2)}_k=c N^{(n_1)} \Gamma(k+\nu+1) \lt(\frac{b_0}{u_0} \rt)^k 
\sum_{m=0}^{n_2-1} s'_m \frac{(\nu+1-m) \cdots \nu}{(k+\nu) \cdots (k+\nu-m+1)}
\ee
for $a=u_0/b_0$. 
Then, in terms of the asymptotic expansion the difference is given by
\begin{align}
&{\rm Re} \, C_1 - {\rm Re} \, C_1^{(n_1,n_2)} \non
&=[d_{n_1+1}^{\rm ren}+\delta_{n_1+1}-d_{n_1+1}^{{\rm ren}(n_1,n_2)}] \alpha_s^{n_1+2}+\mathcal{O}(\alpha_s^{n_1+3})  \non
&\simeq \lt[\lt(1-\frac{N^{(n_1)}}{N}\rt)+s'_{n_2} \frac{(\nu+1-n_2) \cdots \nu}{(n_1+1+\nu) \cdots (n_1+1+\nu-n_2+1)} \rt]
c N \Gamma(n_1+\nu+2) \lt(\frac{b_0}{u_0} \rt)^{n_1+1} \alpha_s^{n_1+2} \non
&\quad{}+\delta_{n_1+1} \alpha_s^{n_1+2} \, . \label{difference}
\end{align}
If one uses a usual truncated perturbative series \eqref{trunX} instead of ${\rm Re} \, C_1^{(n_1,n_2)}$,
the difference between ${\rm Re} \, C_1$ and $X|_{\rm pert}$ is given
by $\sim d_{n_1+1} \alpha_s^{n_1+2}$, which means that 
the factor in the square brackets in Eq.~\eqref{difference} is replaced with $1$.
Hence, we can largely reduce the error size compared to that case.
Here, we assume that $n_1$ is large enough such that $N^{(n_1)}$ 
is obtained with a good accuracy (and hence
the first term insides the square brackets is much smaller than one).
(We will use our formulation in this situation.
When $n_1$ is small, usual perturbation theory is still useful 
because the accuracy is improved by just going to higher order.)
The second term insides the square brackets is typically suppressed as $\mathcal{O}(1/n_1^{n_2+1})$.\fn{
Usual situations would be $n_1 \gg n_2$.}

A possible and important application of the formulation is to determine nonperturbative condensates,
${\rm Re} \, \langle{0| \mathcal{O} |0 \rangle}$, in particular the gluon condensate,
which is a universal nonpertubative input in the OPE.
The determination can be done by comparing the exact measurement of an observable (for instance using lattice QCD)
and the Wilson coefficient ${\rm Re} \, C_1$.
We note that a reasonable determination is possible only when the error size of an approximated 
$C_1$, $|{\rm Re} \, C_1-{\rm Re}\, C_1^{\rm approx.}|$,  
is smaller than the size of the second term of the OPE \eqref{realOPE}.
If one truncates the series at an optimal order $N_*$,
the remaining error $\sim d_{N_*+1} \alpha_s^{N_*+2}$ 
is comparable to the size of the second term, at least parametrically \cite{Ayala:2019uaw}.
Since  in our formulation the error size is suppressed compared to $d_{n_1+1} \alpha_s^{n_1+2}$ 
as shown in Eq.~\eqref{difference},
we expect that we can perform an accurate determination of the gluon condensate 
around the order $n_1 \lesssim N_*$.
We finally note that in Eq.~\eqref{difference} the expansion coefficient is much smaller than $d_{n_1+1}$
and hence the order at which the divergence of the asymptotic expansion starts is relatively late.
Then, the estimate of the error based on the asymptotic expansion would 
be reasonable up to relatively higher order.

When we practically estimate the error of ${\rm Re} \, C_1^{(n_1,n_2)}$, 
we vary the orders of $n_1$ and $n_2$ 
and examine the differences. This will be done in Sec.~\ref{sec:4}.

\section{Test of formulation}
\label{sec:3}
In this section, we test our formulation by using all-order perturbative series
obtained in certain methods. In particular, we check behaviors of renormalon subtracted
coefficient $\delta_n$ 
explicitly, and check validity of our renormalon-free predictions.
In Sec.~\ref{sec:3.1}, we consider the Adler function in the large-$\beta_0$ approximation
and briefly explain how the previous result in Refs.~\cite{Ball:1995ni,Mishima:2016xuj,Mishima:2016vna}
is reproduced with the method in Sec.~\ref{sec:2}.
In Secs.~\ref{sec:3.2} and \ref{sec:3.3}, we study the static QCD potential 
with the RG method in Ref.~\cite{Sumino:2005cq}, 
which allows us to obtain approximated all-order perturbative series containing renormalons.
We note that although a method to extract a renormalon-free prediction
was developed in Ref.~\cite{Sumino:2005cq},
we do not adopt it here.
We only use the perturbative series obtained with the method of Ref.~\cite{Sumino:2005cq},
and apply the formulation in Sec.~\ref{sec:2} to extract a renormalon-free part. 

\subsection{Adler function in the large-$\beta_0$ approximation}
\label{sec:3.1}
The Adler function $D(Q^2)$ is defined from the correlator of the electro-magnetic 
quark current $J^{\mu}(x)=\bar{q}(x) \gamma^{\mu} q(x)$ as
\be
D(Q^2)=4 \pi^2 Q^2 \frac{d \Pi(Q^2)}{d Q^2}-1\, ,
\ee
where
\be
(q^{\mu} q^{\nu} -g^{\mu \nu} q^2) \Pi(Q^2)=-i \int d^4 x \, e^{-i q\cdot x} \langle0|J^{\mu}(x) J^{\nu}(0) |0\rangle  \, ,
\ee
with $Q^2=-q^2>0$.
The Borel transform in the large-$\beta_0$ approximation is given by \cite{Beneke:1992ch, Broadhurst:1992si}
\be
B_D(u)= \frac{8 N_C C_F}{\pi} \lt(\frac{e^{5/3} \mu^2 }{Q^2} \rt)^u 
\frac{1}{2-u}  \sum_{k=2}^{\infty} \frac{(-1)^k k}{[k^2-(1-u)^2]^2} \, , \label{BorelD}
\ee
with $N_c=3$ and $C_F=4/3$. We take $\mu=Q$.
We identify this Borel transform as $B_D^{\rm sing}(u)$ and set $\delta B_D(u)=0$.
Thus, we do not have a $\delta$ part in this case.
The Borel transform has both UV and IR renormalons; 
the singularities are located at $u=\cdots$, $-2$, $-1$, $2$, $3$, $\cdots$.
Then, from Eq.~\eqref{BoreltoAmb}, one can obtain
the ambiguity function as \cite{Neubert:1994vb} 
\begin{align}
Am_D(e^{-5/3} x) &=\frac{C_F}{2} \times \non
\quad{}&\begin{cases}
& (7-4 \log{x})x^2+4x(1+x) [\Li_2(-x)+\log{x} \log{(1+x)}]  \quad \text{for $x<1 $} \\
& 3+2 \log{x}+4x (1+\log{x}) +4x(1+x) [\Li_2(-1/x)-\log{x} \log{(1+1/x)} \quad \text{for $x>1$}
\end{cases} \, .
\end{align}
The behavior of $Am_D(x)$ for $x<e^{-5/3}$ is determined from the IR renormalons while that for $x>e^{-5/3}$
from the UV renormalons.
Corresponding to the first IR renormalon at $u=2$ and first UV renormalon at $u=-1$,
the ambiguity function behaves as $Am_D(x) \sim x^2$ for small $x$
and $Am_D(x) \sim x^{-1}$ for large $x$.
A preweight $W_D$ can be analytically calculated and the result is presented in Eq.~(38) of \cite{Mishima:2016vna}.
Then, one can extract a renormalon-free part according to Eq.~\eqref{XRFdisp2}.
This is the same result as that in Refs.~\cite{Ball:1995ni,Mishima:2016xuj,Mishima:2016vna}.

\subsection{Static QCD potential with RG method at LL}
\label{sec:3.2}

We consider the static QCD potential in this and next subsections. 
The static QCD potential is extracted from an expectation value of a rectangular Wilson loop.
It can be written as
\be
V_{\rm QCD}(r)=-\frac{2C_F}{\pi r} \int_0^{\infty} \frac{dq}{q} \sin(qr) \alpha_V(q) 
\ee
with the $V$-scheme coupling $\alpha_V(q)$.
From this expression, according to the method in Ref.~\cite{Sumino:2005cq} using RG estimate, 
one can obtain approximated all-order perturbative series.
In this method one first considers RG improved $\alpha_V(q)$;
at N$^k$LL one considers $\log^{k} {(\mu/q)} \alpha_s(\mu)^{n+k+1}$ terms for arbitrary $n \geq 0$ in $\alpha_V(q)$.
Then performing the $q$-integral, 
one obtains all-order perturbative series for $V_{\rm QCD}(r)$, which contains renormalon divergences.
There are only IR renormalons in this observable (with this treatment) 
and this is a difference from the Adler function.

\begin{figure}[t!]
\begin{center}
\includegraphics[width=7cm]{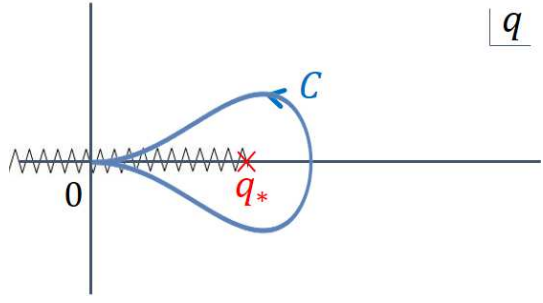}
\end{center}
\caption{Contour $C$. $q_*$ represents a singularity of the running coupling.
At LL, it is a simple pole. Beyond LL, the singularity is given as a cut singularity.}
\label{fig:C}
\end{figure}

In this subsection, we work at LL. 
The renormalon uncertainty in this method has been revealed in Ref.~\cite{Sumino:2020mxk}
at general order of the RG improvement.
The renormalon uncertainty for the dimensionless potential $v(r)=r V_{\rm QCD}(r)$
at LL is given by \cite{Sumino:2020mxk}
\begin{align}
{\rm Im} \, v_{\pm}
&=\frac{1}{b_0} {\rm Im} \int_{C_{\pm}} du \, B_v(u) e^{-u/(b_0 \alpha_s(1/r))} \non
&=\mp\frac{2 C_F}{\pi} \frac{1}{2i} \int_C \frac{d q}{q} \sin(qr) [\alpha_V(q)]_{\text{LL}} \non
&=\mp\frac{2 C_F}{\pi} \frac{1}{2i} \int_C \frac{d q}{q} \sin(qr) \frac{1}{b_0} \frac{1}{\log(q^2/\LMS^2)} \non
&=\mp\frac{2 C_F}{\pi} \frac{1}{2i} \int_C \frac{d q}{q} \sin(qr) \frac{1}{2 b_0} \frac{\LMS}{q-\LMS} \non
&=\mp\frac{C_F}{b_0}  \sin(\LMS r) \, ,
\end{align}
where the integration contour $C$ is shown in Fig.~\ref{fig:C}.\fn{
At LL, without relying on the formula in Ref.~\cite{Sumino:2020mxk}
one can easily obtain the renormalon uncertainty by a
calculation of the Borel transform.}
Then, the ambiguity function is obtained as \cite{Mishima:2016vna} [cf.~\eqref{AmbtoAmbiguity}]
\be
Am_v(x)=-C_F \sin(x^{1/2}) \, .  \label{AmbvLLform}
\ee
Note that $\LMS^2 r^2=e^{-1/(b_0 \alpha_s(r^{-1}))}$ at LL.
We adopt this form for $x<1$:\fn{
It is possible to adopt this form for all $x$, $0<x<\infty$.
In this case, we do not have a $\delta$ part,
as the correct Borel transform [Eq.~\eqref{BorelVLL}] is reproduced by $\frac{1}{\pi}\int_0^{\infty} dx \, Am_v(x) x^{-u-1}$.
However, in the analyses below beyond LL, we cannot adopt a non-trivial form of the ambiguity function
for whole $x$ (Secs.~\ref{sec:3.3} and \ref{sec:4}). 
Then, as a test, we limit the range of the ambiguity function.}
\be
Am_v(x)=
\begin{cases}
-C_F \sin(x^{1/2}) \quad &\text{for $x<1$} \\
0 \quad &\text{for $x>1$} 
\end{cases} \, . \label{AmbvLL}
\ee
Then $B^{\rm sing}_v(u)$ is obtained as 
\begin{align}
B_v^{\rm sing}(u)
&=-\frac{C_F}{\pi} \int_0^1dx  \sin(x^{1/2}) x^{-u-1}  \, . \label{BsingvLL}
\end{align}
In this case, we {\it define} $B^{\rm sing}_v(u)$
from  the ambiguity function of Eq.~\eqref{AmbvLL} through Eq.~\eqref{BfromAm}. 
The Borel transform itself is given by \cite{Aglietti:1995tg}
\begin{align}
B_v(u)
&=-\frac{2 C_F}{\pi} \int_0^{\infty} \frac{dq}{q} \, \sin(qr) \lt(\frac{\mu^2}{q^2} \rt)^u \non
&=-\frac{1}{\sqrt{\pi}} C_F \lt(\frac{\mu r}{2} \rt)^{2 u} \frac{\Gamma(\frac{1}{2}-u)}{\Gamma(u+1)} \label{BorelVLL}
\end{align}
and one can confirm that Eq.~\eqref{BsingvLL} gives the singular parts correctly (with $\mu=1/r$):
\begin{align}
B_v^{\rm sing}(u)
&=-\frac{C_F}{\pi} \int_0^1dx \lt(x^{1/2}-\frac{1}{6}x^{3/2}+\cdots \rt) x^{-u-1} \non
&=\frac{C_F}{\pi} \frac{1}{u-\frac{1}{2}}-\frac{C_F}{6 \pi} \frac{1}{u-\frac{3}{2}}+\cdots \,
\end{align}
If one changes the range of $x$ to adopt the form \eqref{AmbvLLform} in Eq.~\eqref{AmbvLL},
it corresponds to a change of the definition of $B^{\rm sing}_v(u)$ and $\delta B_v(u)$. 
However, it is important to note that
even in this case, the IR renormalons are correctly encoded in a new $B^{\rm sing}_v(u)$
because they stem from the integral around $x \sim 0$ in Eq.~\eqref{BsingvLL} [or generally in Eq.~\eqref{BfromAm}].
This means that regardless of details of the range of $x$, 
the IR renormalons are always removed from $\delta B_X(u)$.

\begin{figure}[tbp]
\begin{minipage}{0.5\hsize}
\begin{center}
\includegraphics[width=7cm]{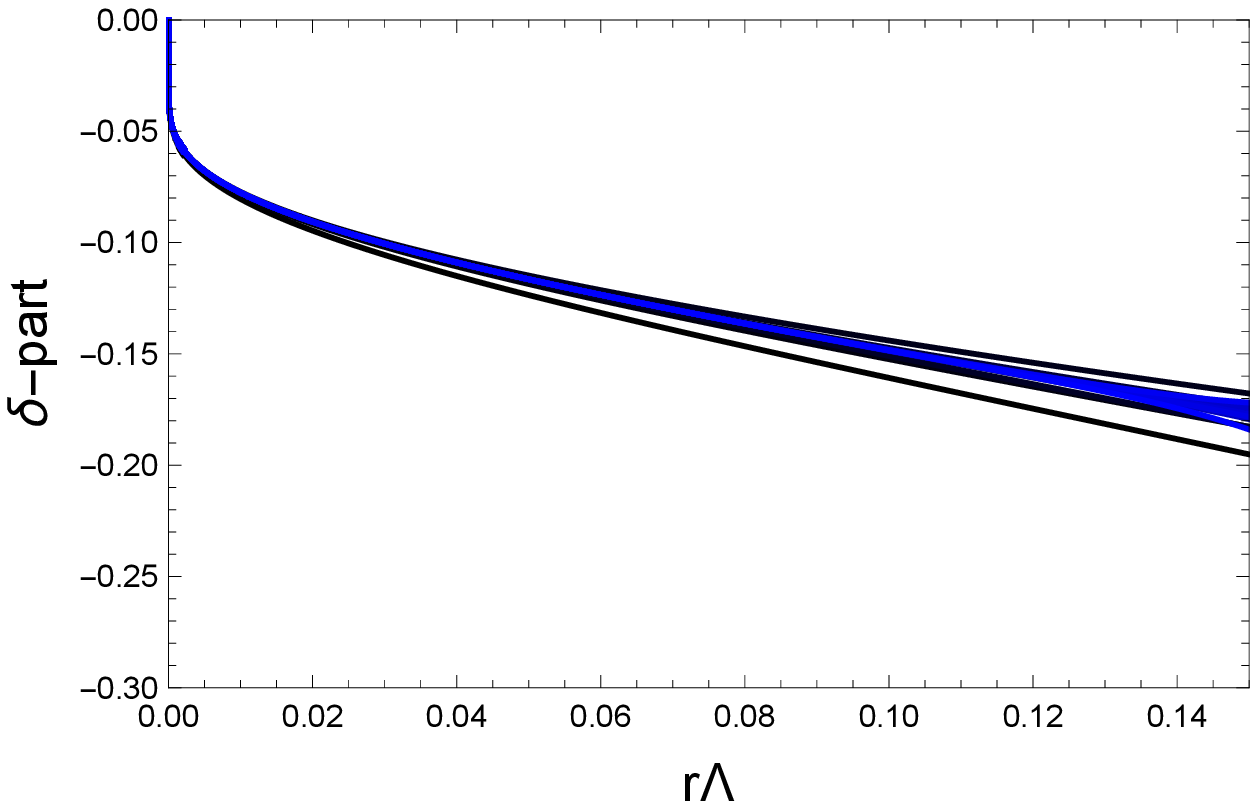}
\end{center}
\end{minipage}
\begin{minipage}{0.5\hsize}
\begin{center}
\includegraphics[width=7cm]{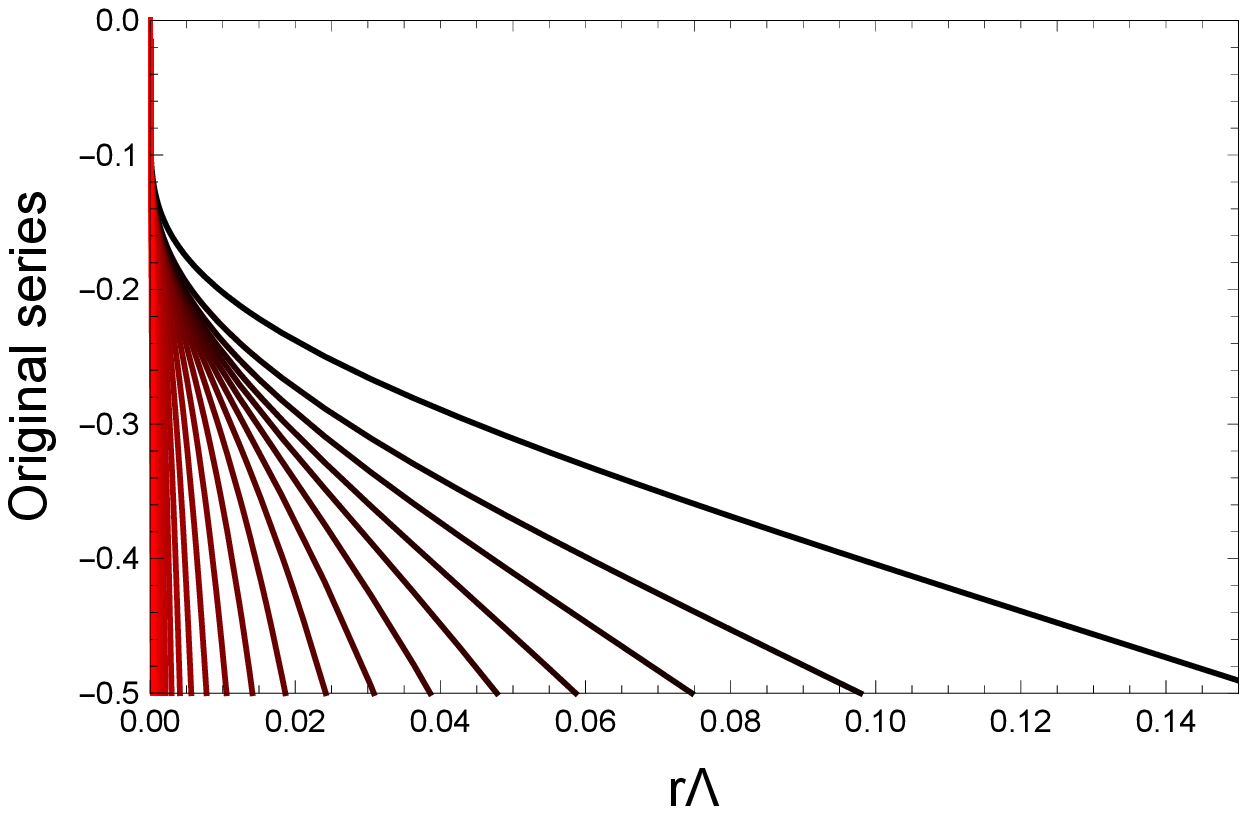}
\end{center}
\end{minipage}
\caption{Perturbative series of the $\delta$ part for the dimensionless potential $r V_{\rm QCD}(r)$ (left).
This is compared to the original series containing renormalons (right).
Deeper blue (red) line corresponds to higher order result. 
The highest order is $\mathcal{O}(\alpha_s^{21})$.
}
\label{fig:deltaoriginal}
\end{figure}

Now we examine a $\delta$ part.
Namely, we evaluate
\be
\delta_n := d_n-d_n^{{\rm ren}} \, ,
\ee
where $d_n^{{\rm ren}}$ are obtained by
\be
\frac{d_n^{{\rm ren}}}{b_0^n}=\int_0^1 \frac{dx}{\pi x} (-C_F \sin{(x^{1/2})}) (-\log{x})^n \, .
\ee
We can obtain $d_n$ at an arbitrary order
by performing the $q$-integral of the LL result of 
$\alpha_V(q)|_{\text{LL}}=\alpha_s(q)=\alpha_s(\mu)+\alpha_s(\mu)^2 b_0 \log{(\mu^2/q^2)}+\cdots$.
The results for $d_n/b_0^n$ and $\delta_n/b_0^{n}$ are given in Table~\ref{tab1}.
We can confirm that the perturbative coefficients $\delta_n$ are significantly smaller than $d_n$
as a consequence of the renormalon subtraction.
\begin{table}[t]
\begin{minipage}{0.5\hsize}
\begin{center}
\begin{tabular}{c|D{.}{.}{10}D{.}{.}{10}}
\hline
\multicolumn{1}{c|}{$n$} & \multicolumn{1}{c}{$d_n/b_0^n$} & \multicolumn{1}{c}{$\delta_n/b_0^n$}  \\ \hline
0 & -1.33333 & -0.530273    \\
1 & -1.53924 & 0.127532   \\
2 & -6.16344  & 0.585703  \\
3 & -4.2887 \times 10^1& -2.22658 \\
4 & -3.20373 \times 10^2 & 5.35337 \\
5 & -3.26704 \times 10^3 & -8.28977 \\
10 & -3.15415 \times 10^9 & -4.90861 \times 10^3 \\
20 & -2.16543 \times 10^{24} & -1.25656 \times 10^{10} \\
30 & -2.41757 \times 10^{41}   & 8.69321 \times 10^{16} \\
\hline
\end{tabular}
\end{center}
\end{minipage}
\begin{minipage}{0.5\hsize}
\begin{center}
\begin{tabular}{c|D{.}{.}{10}D{.}{.}{10}}
\hline
\multicolumn{1}{c|}{$n$} & \multicolumn{1}{c}{$\frac{d_n}{b_0^n} (2^n n!)^{-1}$} & \multicolumn{1}{c}{$\frac{\delta_n}{b_0^n} ( n!)^{-1}$}  \\ \hline
0 & -1.33333 & -0.530273    \\
1 & -0.769621 & 0.127532   \\
2 & -0.770430  & 0.292851  \\
3 & -0.893478 & -0.371097 \\
4 & -0.834305  & 0.223057 \\
5 & -0.850792 & -6.90814 \times 10^{-2} \\
10 & -0.848827 & -1.35268 \times 10^{-3} \\
20 & -0.848826 & -5.16486 \times 10^{-9} \\
30 & -0.848826  & 3.27733 \times 10^{-16} \\
\hline
\end{tabular}
\end{center}
\end{minipage}
\caption{\small
Original perturbative coefficient $d_n$ and renormalon subtracted perturbative coefficient $\delta_n$.
In the right panel, we divide $d_n$ by the large order behavior expected from the $u=1/2$ renormalon,
and divide $\delta_n$ by $n!$.
We take $n_f=3$.
\label{tab1}
}
\end{table}
In Fig.~\ref{fig:deltaoriginal}, we show the $\delta$ part,
\be
\sum_{k=0}^{n} \delta_k \alpha_s^{k+1}(1/r) \, ,
\ee
where $\alpha_s(1/r)$ is the running coupling at LL.
One sees that the $\delta$ part exhibits much better convergence than the original series, as expected.

\begin{figure}
\begin{center}
\includegraphics[width=9cm]{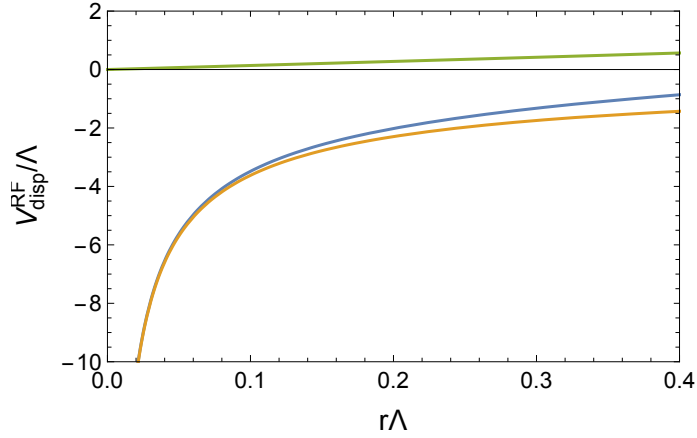}
\end{center}
\caption{$v_{\rm disp}^{\rm RF}/(r \Lambda)=V_{\rm disp}^{\rm RF}/\Lambda$ as a function of $r \Lambda$ (blue).
Orange line corresponds to the first line of Eq.~\eqref{vRFdispLL} and green one to the second line of Eq.~\eqref{vRFdispLL}.
}
\label{fig:RFdispLL}
\end{figure}

Now we study the renormalon-free part obtained via a preweight.
The preweight is given by
\be
W_{v+}(z)=-C_F\int_0^1 \frac{dx}{\pi} \frac{\sin{\sqrt{x}}}{x+z} 
\ee
from the ambiguity function of Eq.~\eqref{AmbvLL}.
It is possible to give an analytic expression for the preweight.
Using this function, the renormalon-free part corresponding to $X^{\rm RF}_{\rm disp}$ is given by
\begin{align}
v^{\rm RF}_{\rm disp}(r)
&=\frac{1}{b_0} \int_0^{\infty} \frac{dz}{\pi z} 
W_{v+} (z) \frac{-\pi}{\lt(\log{z}+\frac{1}{b_0 \alpha_s(r^{-1})} \rt)^2+\pi^2}
+\frac{1}{b_0} {\rm Re} \, W_{v}(e^{-1/(b_0 \alpha_s(r^{-1}))}) \non
&=\frac{1}{b_0} \int_0^{\infty} \frac{dz}{\pi z} 
W_{v+} (z) \frac{-\pi}{\lt(\log{z}+\frac{1}{b_0 \alpha_s(r^{-1})} \rt)^2+\pi^2}+\frac{1}{b_0} {\rm Re} \, W_{v}(0) \non
&\quad{}+\frac{1}{b_0} \lt[{\rm Re}\, W_{v}(e^{-1/(b_0 \alpha_s(r^{-1}))})- {\rm Re}\, W_{v}(0) \rt] \, . \label{vRFdispLL}
\end{align}
The result for $v_{\rm disp}^{\rm RF}/(r \LMS)=V_{\rm disp}^{\rm RF}/\LMS$ is shown as a function of $r \LMS$  in Fig.~\ref{fig:RFdispLL}.
We evaluate  the integral with respect to $z$ in the first line of Eq.~\eqref{vRFdispLL} numerically.
The first line gives a Coulomb-like potential and the second line gives a linear-like potential.
(See Ref.~\cite{Sumino:2003yp} for the first observation of such a behavior.)

\begin{figure}[tbp]
\begin{minipage}{0.5\hsize}
\begin{center}
\includegraphics[width=7cm]{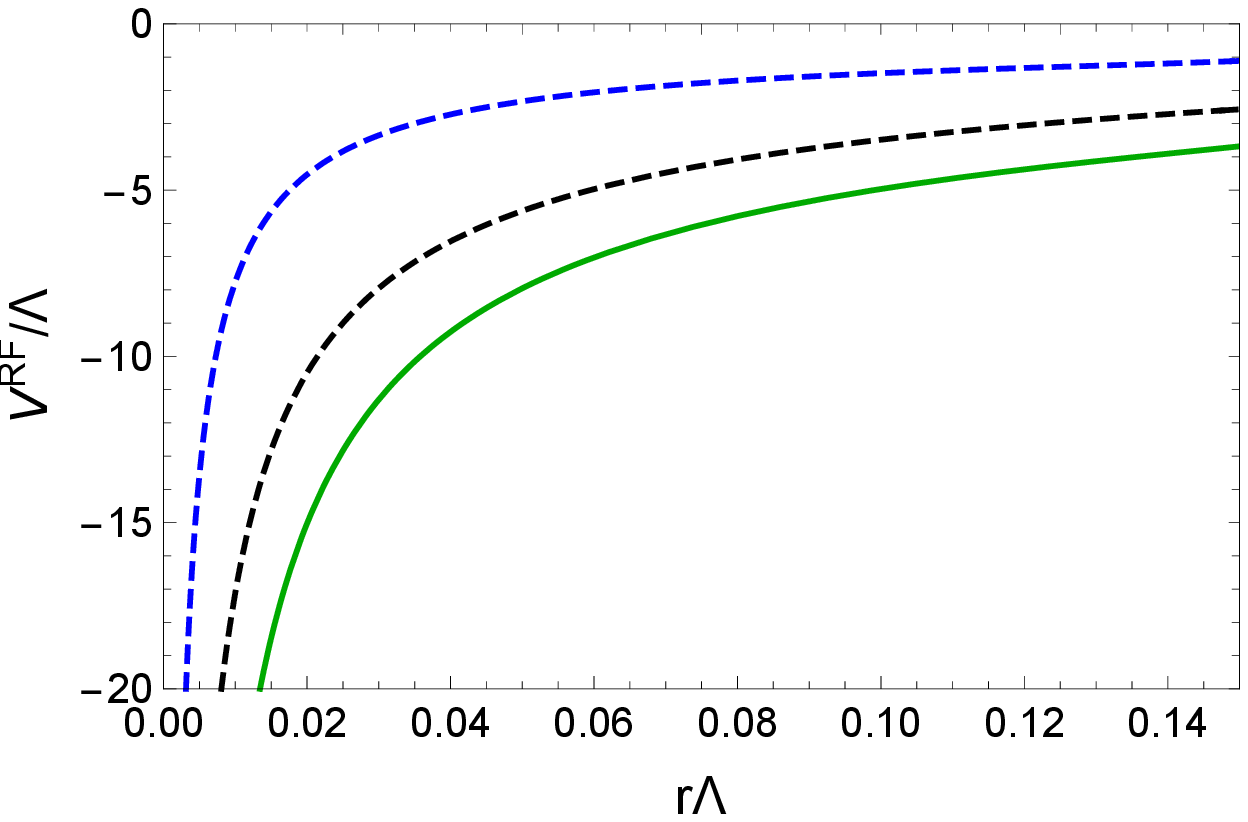}
\end{center}
\end{minipage}
\begin{minipage}{0.5\hsize}
\begin{center}
\includegraphics[width=7cm]{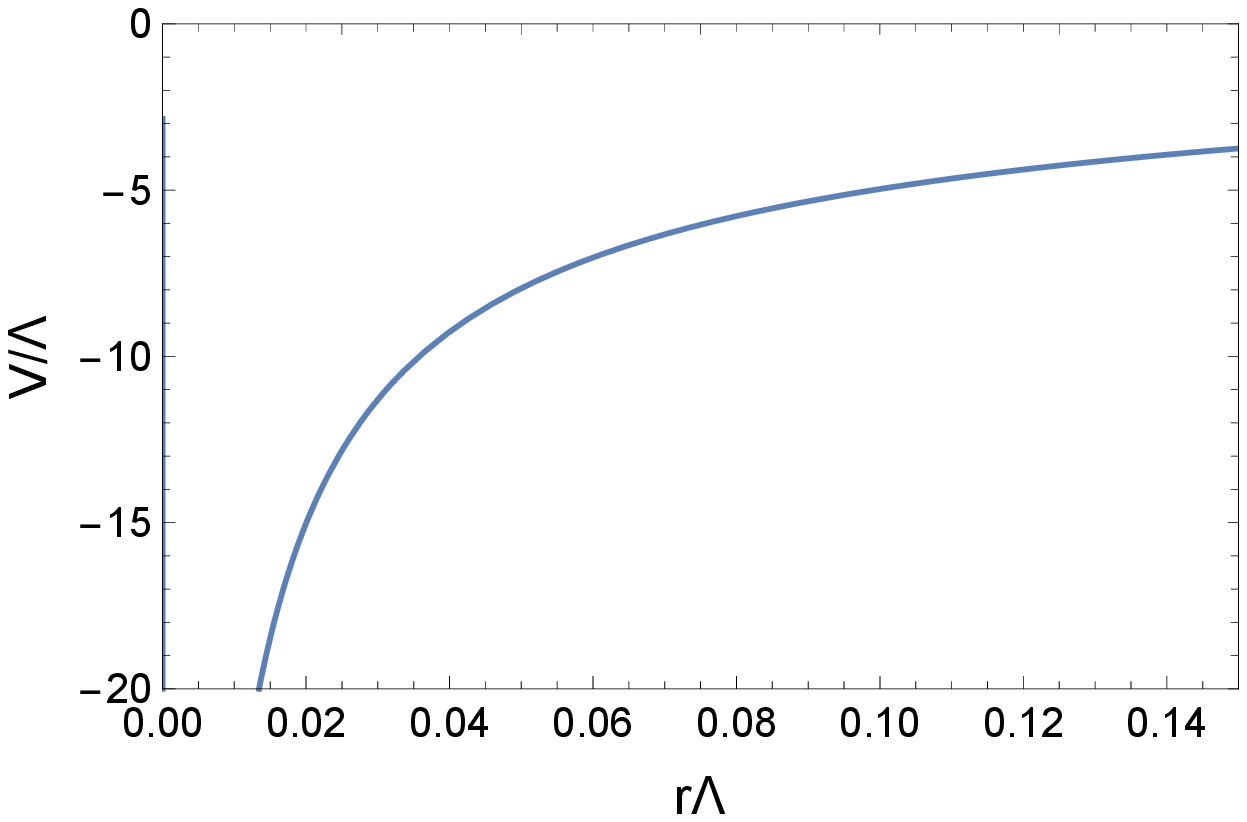}
\end{center}
\end{minipage}
\caption{Renormalon-free prediction $V^{\rm RF}/\LMS$ as a function of $r \LMS$ (left). 
Contributions from $V_{\rm disp}^{\rm RF}/\LMS$ (black dashed) and the $\delta$ part (blue dashed) are also shown separately.
We also show the result obtained based on Ref.~\cite{Sumino:2005cq} (right).
}
\label{fig:RF-whole-LL}
\end{figure}

The total renormalon-free prediction, which is the sum of the $\delta$ part and $v^{\rm RF}_{\rm disp}$, is shown in Fig.~\ref{fig:RF-whole-LL}.
This is compared with a result obtained with the method in Ref.~\cite{Sumino:2005cq} to subtract renormalons (right panel in Fig.~\ref{fig:RF-whole-LL}).
We confirm precise agreement with each other in the examined region.
In fact, the result in the right panel is obtained by adopting the ambiguity function 
of Eq.~\eqref{AmbvLLform} for whole $x$, i.e., $0<x<\infty$.
(In this case, we do not have a $\delta$ part.)
In this sense, we observe consistency among the two different schemes.

\subsection{Static QCD potential with RG method at NLL}
\label{sec:3.3}
\begin{figure}[b]
\begin{center}
\includegraphics[width=10cm]{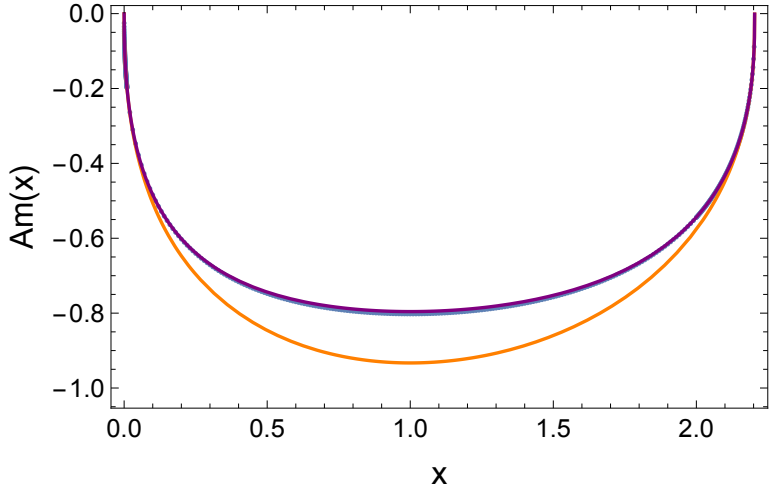}
\end{center}
\caption{Ambiguity function at NLL. The blue points represent numerical results of the integral in Eq.~\eqref{ambNLL}.
The ambiguity function corresponding to the $u=1/2$ renormalon is shown by the orange line, and 
the sum of the $u=1/2$ and $3/2$ ones is shown by the purple line, which almost coincides with the blue points.
We take $n_f=3$.}
\label{fig:am-NLL}
\end{figure}

As an analysis beyond the large-$\beta_0$ or LL approximation,
we extend the analysis in Sec.~\ref{sec:3.2} to the NLL approximation.
The renormalon uncertainty at NLL is obtained as \cite{Sumino:2020mxk}
\begin{align}
{\rm Im} \, v_{+}
&=-\frac{2 C_F}{\pi} \frac{1}{2i} \int_C \frac{d q}{q} \sin(qr) [\alpha_V(q)]_{\text{NLL}} \non
&=-\frac{2 C_F}{\pi} \frac{1}{2i} \int_C \frac{d \hat{q}}{\hat{q}} \sin(\hat{q} \LMS r) [\alpha_V(\hat{q})]_{\text{NLL}} \, ,
\end{align}
where we change the integration variable as $\hat{q}=q/\LMS$.
(Note that $\alpha_s(q)$ in $\alpha_V(q)$ is actually a function of $q/\LMS$.)
Then, we can adopt the ambiguity function as
\be
Am_v(x)=-\frac{2 C_F b_0}{\pi} \frac{1}{2 i} 
\int_C \frac{d \hat{q}}{\hat{q}} 
\sin \lt[\hat{q} x^{1/2} \lt(-\log{x}+\frac{b_1}{b_0^2} \rt)^{\frac{b_1}{2 b_0^2}} \rt] [\alpha_V(\hat{q})]_{\text{NLL}} \quad \text{for $0<x<e^{b_1/b_0^2}$} \label{ambNLL}
\ee
and $0$ for the other region [cf. Sec.~\ref{sec:2.2}].
$x=e^{b_1/b_0^2}$ is a zero of the ambiguity function and this choice corresponds
to the second option~\eqref{secondop}.
The behavior of the ambiguity function is shown in Fig.~\ref{fig:am-NLL}.
Here, we perform the $\hat{q}$-integral along $C$ numerically.
We can compare this result with an asymptotic behavior of the ambiguity function.
The asymptotic behavior at $x \sim 0$ is obtained as
\be
Am_v(x)=b_0 N_{1/2} \lt[x \lt(-\log{x}+\frac{b_1}{b_0^2} \rt)^{b_1/b_0^2} \rt]^{1/2}
+b_0 N_{3/2} \lt[x \lt(-\log{x}+\frac{b_1}{b_0^2} \rt)^{b_1/b_0^2} \rt]^{3/2}+\cdots \, . \label{asymNLL}
\ee
from the $u=1/2$, $3/2$, $\cdots$ renormalons [cf.~Eq.~\eqref{TwoloopAmb}], 
where $N_{1/2}$ and $N_{3/2}$ are defined such that the renormalon uncertainty
is given by
\be
{\rm Im} v_{\pm}=\pm [ N_{1/2} \LMS r+N_{3/2} (\LMS r)^3+\cdots] \, .
\ee
One should note the relation
\be
N_i b_0=\pi u_i^{1+u_i \frac{b_1}{b_0^2}} \frac{K_{u_i} \Gamma(1+\nu_i)}{\Gamma(1+\nu_i)} \, , \label{conventionchange}
\ee
in the convention where one defines parameters in an expansion of the 
Borel transform around $u=u_i$ as 
\be
B(u;u_i)=\frac{K_{u_i} \Gamma(1+\nu_i)}{(1-u/u_i)^{1+\nu_i}} \lt[1+\mathcal{O}(1-u_i/u)\rt] \, 
\ee
with $\nu_i=u_i b_1/b_0^2$.
In Ref.~\cite{Sumino:2020mxk}, the normalization constants $K_{u_i} \Gamma(1+\nu_i)$ 
for $u_i=1/2$ and $3/2$ are explicitly obtained (which are denoted as $N_i$ therein) within the RG method,\fn{
These normalization constants can be accurately obtained within the RG method, 
which allows us to obtain an all-order perturbative series.
This is not always the case when we use fixed order results (Sec.~\ref{sec:4} below). 
}
and should be converted via Eq.~\eqref{conventionchange}.
We have
\be
N_{1/2}=-1.42978, \quad N_{3/2}=0.253216
\ee
for $n_f=3$.
In Fig.~\ref{fig:am-NLL}, we also show the asymptotic form of the ambiguity function \eqref{asymNLL}
with the above normalization constants.
If the ambiguity function up to the $u=3/2$ renormalon is included,
it coincides well with the whole ambiguity function.

\begin{figure}[tbp]
\begin{minipage}{0.5\hsize}
\begin{center}
\includegraphics[width=7cm]{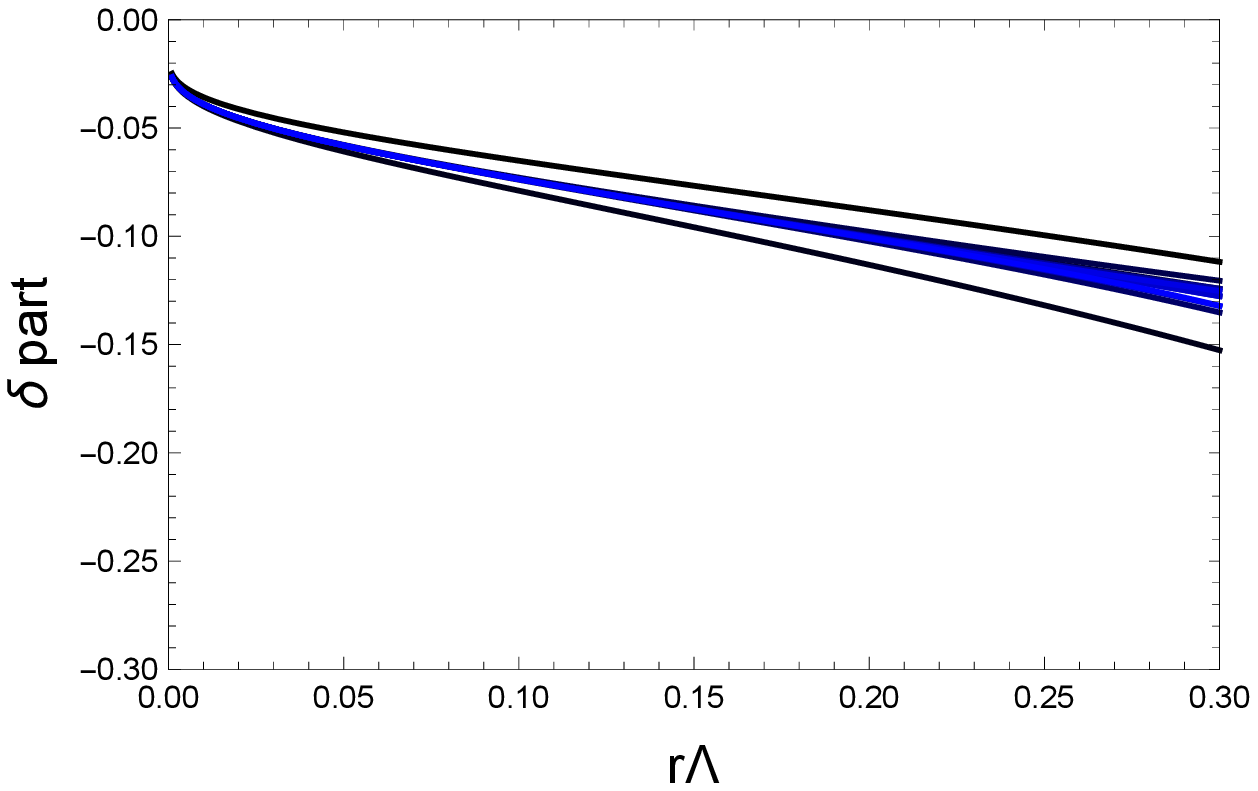}
\end{center}
\end{minipage}
\begin{minipage}{0.5\hsize}
\begin{center}
\includegraphics[width=7cm]{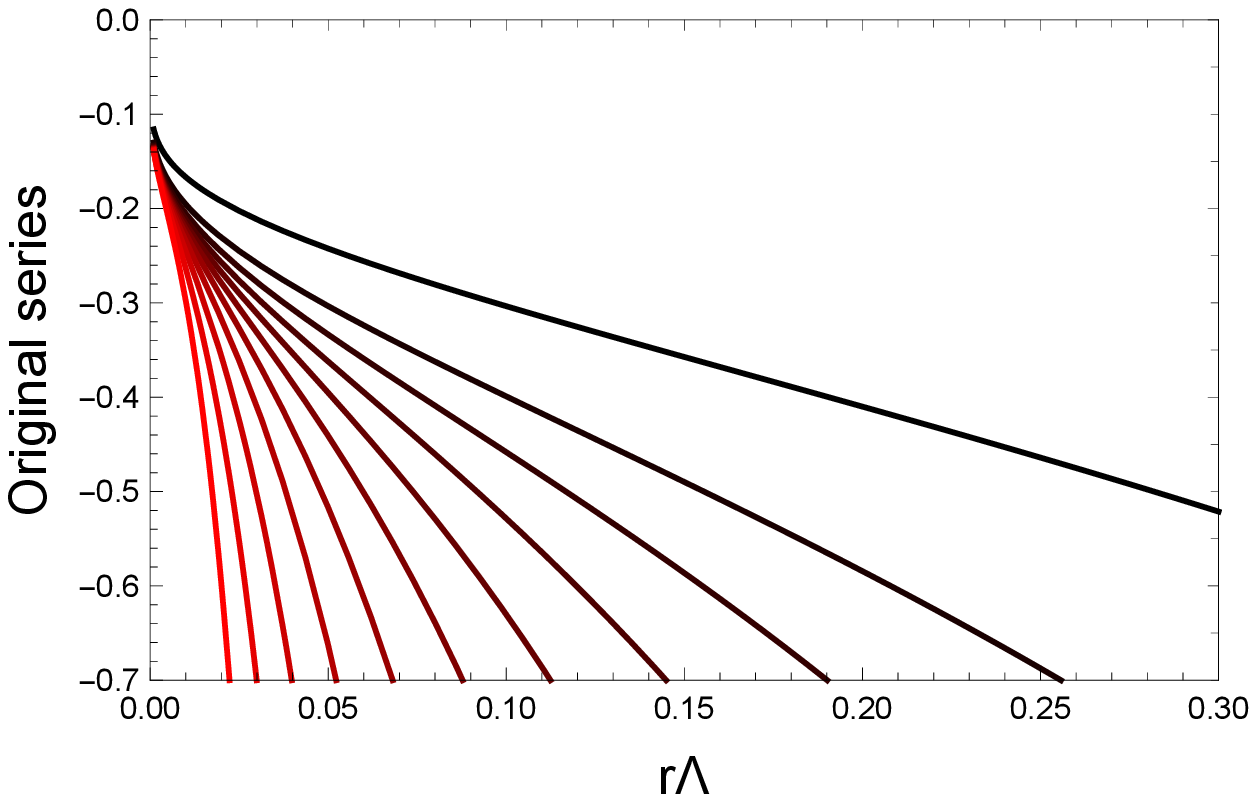}
\end{center}
\end{minipage}
\caption{Perturbative series of the $\delta$ part for the dimensionless potential (left).
This is compared to the original series containing renormalons (right).
Deeper blue (red) line corresponds to higher order result. 
The highest order is $\mathcal{O}(\alpha_s^{11})$.
}
\label{fig:deltaoriginal-NLL}
\end{figure}

\begin{figure}[tbp]
\begin{minipage}{0.5\hsize}
\begin{center}
\includegraphics[width=7cm]{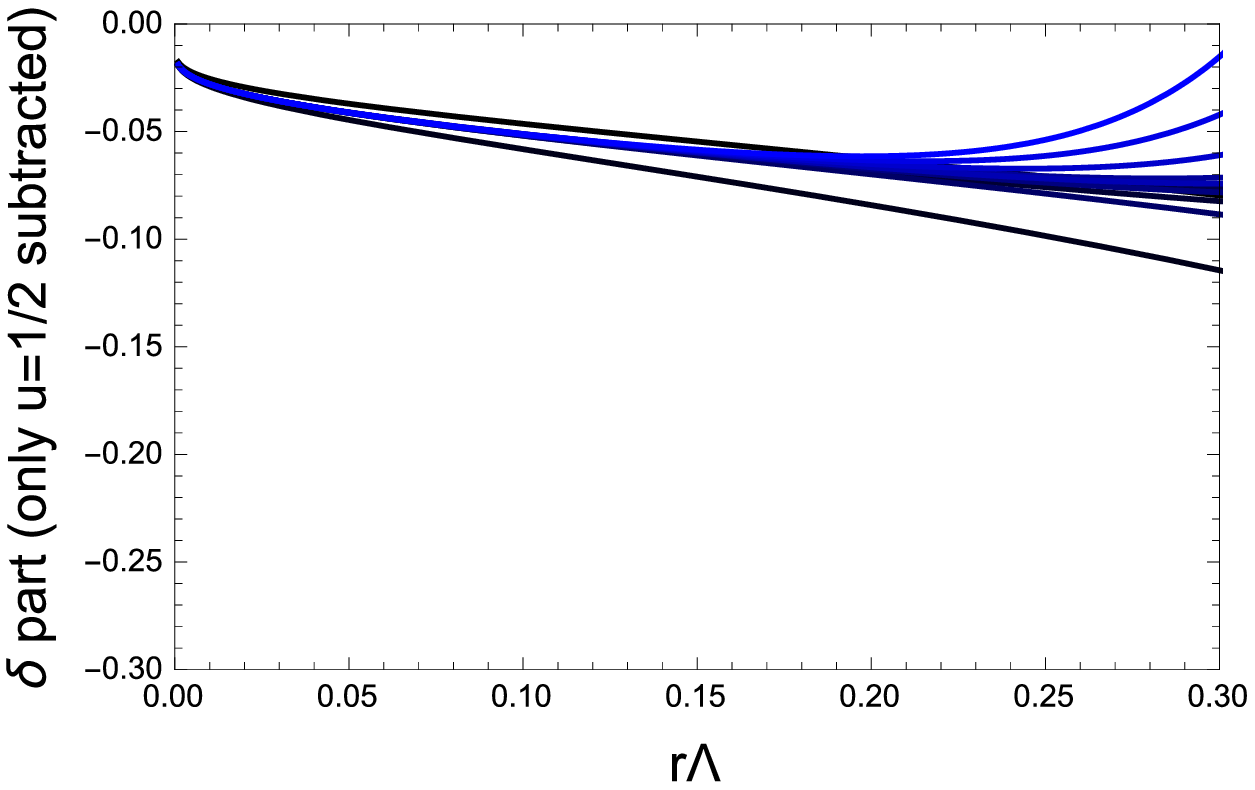}
\end{center}
\end{minipage}
\begin{minipage}{0.5\hsize}
\begin{center}
\includegraphics[width=7cm]{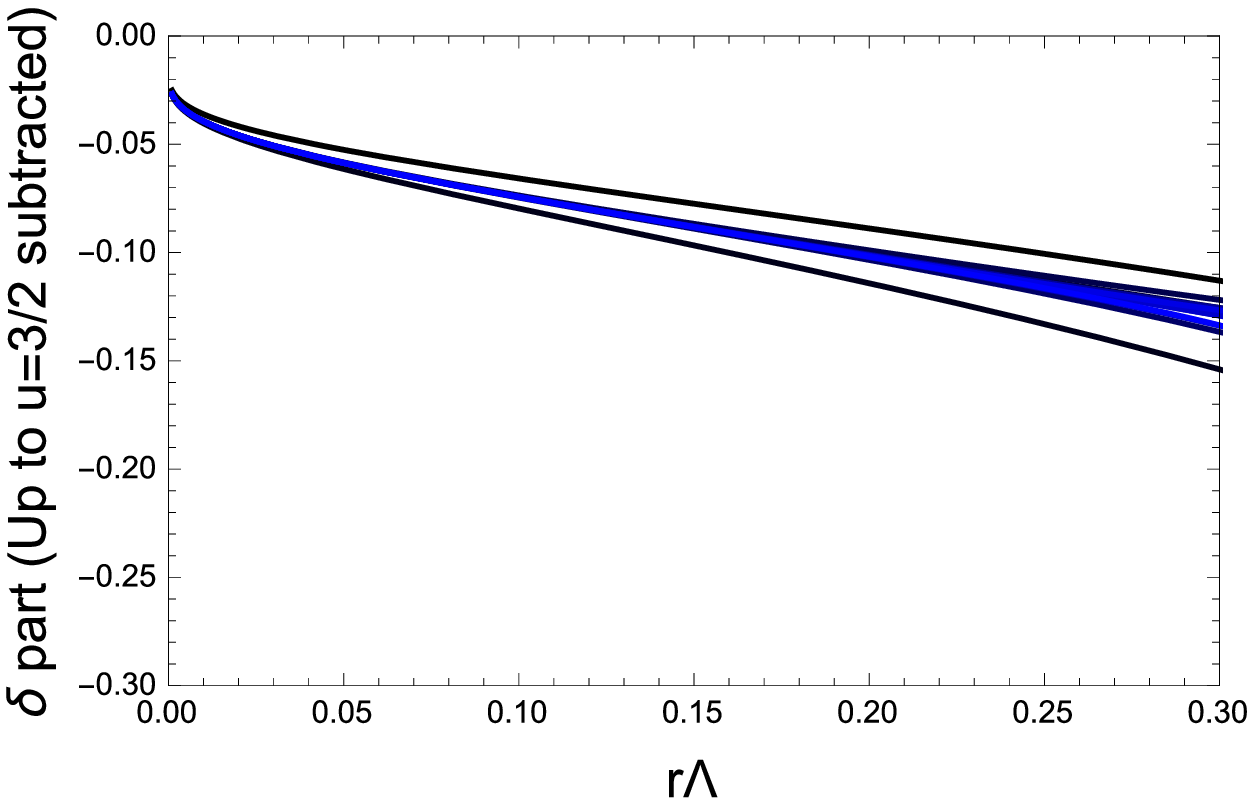}
\end{center}
\end{minipage}
\caption{Perturbative series of the $\delta$ part for the dimensionless potential.
In the left panel, only the $u=1/2$ renormalon is subtracted,
while in the right panel, up to the $u=3/2$ renormalon is subtracted.
Deeper blue line corresponds to higher order result.  
The highest order is $\mathcal{O}(\alpha_s^{11})$.
}
\label{fig:deltavariant-NLL}
\end{figure}

We can obtain $\delta_n$ from the defined ambiguity function.\fn{In this case, we
know the form of $B_v^{\rm sing}(u)$ for each renormalon as in Eq.~\eqref{BorelNLL},
and we may utilize it.} 
($d_n$ is calculated by the $q$-integral of the NLL result for $\alpha_V(q)$.)
The results for $d_n/b_0^n$ and $\delta_n/b_0^{n}$ are given in Table~\ref{tab2}.
\begin{table}[t]
\begin{minipage}{0.5\hsize}
\begin{center}
\begin{tabular}{c|D{.}{.}{10}D{.}{.}{10}}
\hline
\multicolumn{1}{c|}{$n$} & \multicolumn{1}{c}{$d_n/b_0^n$} & \multicolumn{1}{c}{$\delta_n/b_0^n$}  \\ \hline
0 & -1.33333 & -0.286    \\
1 & -2.57628 & -0.373   \\
2 & -9.77401  & 0.928  \\
3 & -7.13349 \times 10^1& 0.430 \\
4 & -6.32924 \times 10^2 & -6.11 \\
5 & -6.71828 \times 10^3 & 1.13 \times 10^1 \\
6 & -8.57255 \times 10^4 & 1.80 \times 10^1 \\
7 & -1.26456 \times 10^{6} & -1.51  \times 10^2 \\
8 & -2.11799 \times 10^{7}   & 3.05  \times 10^2 \\
9 & -3.97236  \times 10^{8}  &   5.98  \times 10^2 \\
10& -8.24602  \times 10^{9} &  -5.93  \times 10^3 \\
\hline
\end{tabular}
\end{center}
\end{minipage}
\begin{minipage}{0.5\hsize}
\begin{center}
\begin{tabular}{c|D{.}{.}{10}D{.}{.}{10}}
\hline
\multicolumn{1}{c|}{$n$} & \multicolumn{1}{c}{$\frac{d_n}{b_0^n} (\Gamma(n+1+\frac{b_1}{2 b_0^2}) 2^n)^{-1} $} & \multicolumn{1}{c}{$\frac{\delta_n}{b_0^n} (n!)^{-1}$}  \\ \hline
0 & -1.33333 & -0.286    \\
1 & -0.923357 & -0.373   \\
2 & -0.731312  & 0.464  \\
3 & -0.786057 & 7.16 \times 10^{-2} \\
4 & -0.793430 & -2.55 \times 10^{-1} \\
5 & -0.780528 & 9.38 \times 10^{-2} \\
6 & -0.778693 & 2.50 \times 10^{-2} \\
7 & -0.776648 & -3.00  \times 10^{-2} \\
8 & -0.774737   & 7.57  \times 10^{-3} \\
9 & -0.773302  &   1.65  \times 10^{-3} \\
10& -0.772126 &  -1.63  \times 10^{-3} \\
\hline
\end{tabular}
\end{center}
\end{minipage}
\caption{\small
Original perturbative coefficient $d_n$ and renormalon subtracted perturbative coefficient $\delta_n$.
In the right panel, we divide $d_n$ by the large order behavior expected from the $u=1/2$ renormalon,
and divide $\delta_n$ by $n!$.
We take $n_f=3$.
\label{tab2}
}
\end{table}
The $\delta$ part [Eq.~\eqref{deltapart}] is shown as a function of $r \LMS$ in Fig.~\ref{fig:deltaoriginal-NLL}.
It is compared with the original series containing renormalon divergences,
and one can see that the $\delta$ part exhibits good convergence also at NLL.
We also study the $\delta$ part when we define it with subtracting only first few renormalons.
In Fig.~\ref{fig:deltavariant-NLL}, we can see that the subtraction up to the $u=3/2$ renormalon is sufficient at the order we work [$\mathcal{O}(\alpha_s^{11})$].
In contrast, only the $u=1/2$ renormalon subtraction seems not satisfatory around this order.

\begin{figure}[t!]
\begin{center}
\includegraphics[width=9cm]{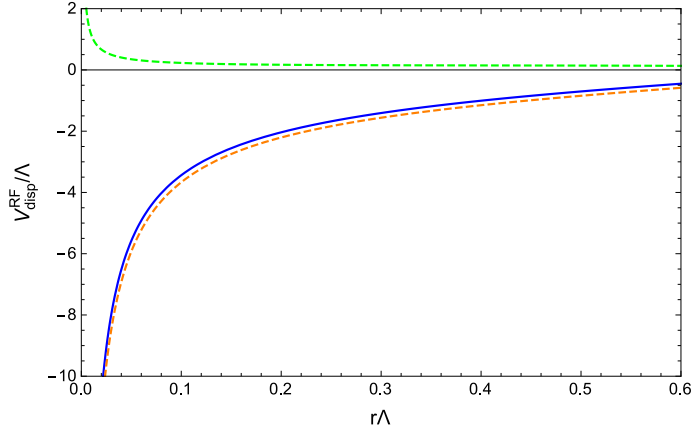}
\end{center}
\caption{$V_{\rm disp}^{\rm RF}/\LMS$ as a function of $r \LMS$.
The red dashed line and green bashed line, respectively, are
the parts extracted from the ambiguity functions corresponding to the $u=1/2$ and $u=3/2$ renormalons.
 The blue line is the sum of them.}
\label{fig:RFdisp-NLL}
\end{figure}

Now we calculate the renormalon-free part corresponding to $X^{\rm RF}_{\rm disp}$ [cf.~\eqref{XRFdisp2}].
Here, we approximate the ambiguity function by the first two terms of Eq.~\eqref{asymNLL} (corresponding to the first two renormalons).
(In this case, strictly speaking, we need to modify the $\delta$ part accordingly
but its modification is small and not significant.)
Then, we obtain a preweight and $V^{\rm RF}_{\rm disp}(r)$ with this ambiguity function.
At NLL, it is difficult to calculate the preweight [Eq.~\eqref{preweight}] analytically and its integral in Eq.~\eqref{XRFdisp2},
and then we perform all the integrals numerically. 
We use convenient formulae collected in Appendix~\ref{app:B}.
$V^{\rm RF}_{\rm disp}(r)$ is shown in Fig.~\ref{fig:RFdisp-NLL}.

\begin{figure}[t!]
\begin{minipage}{0.5\hsize}
\begin{center}
\includegraphics[width=7cm]{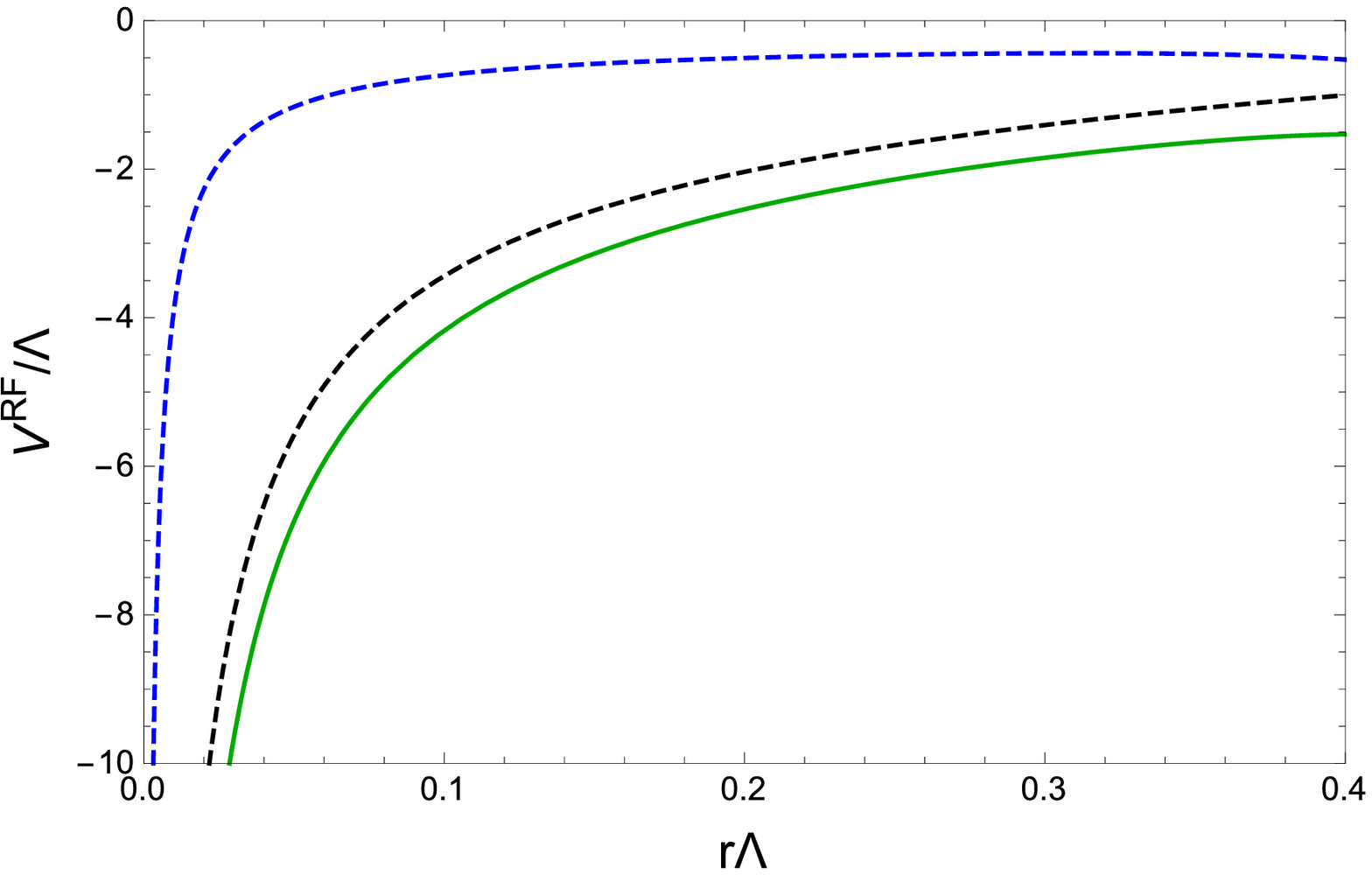}
\end{center}
\end{minipage}
\begin{minipage}{0.5\hsize}
\begin{center}
\includegraphics[width=7cm]{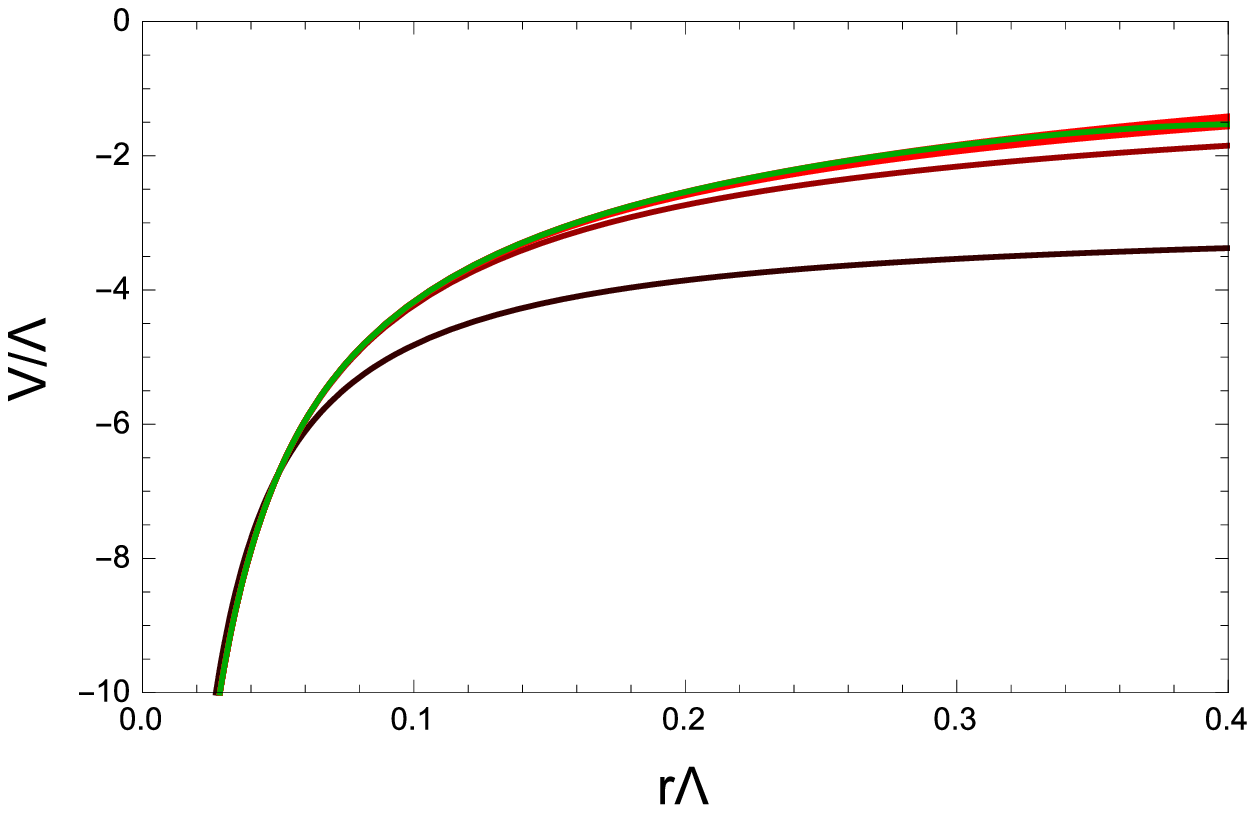}
\end{center}
\end{minipage}
\caption{Renormalon-free prediction for the static QCD potential at NLL (left).
Contributions from $V_{\rm disp}^{\rm RF}/\LMS$ (black dashed) and the $\delta$ part (blue dashed) are also shown separately.
For the $\delta$ part, we use $\mathcal{O}(\alpha_s^{11})$ prediction.
In the right panel, we compare it with fixed order results
at $\mathcal{O}(\alpha_s)$, $\mathcal{O}(\alpha_s^3)$, $\mathcal{O}(\alpha_s^5)$,
$\mathcal{O}(\alpha_s^7)$, $\mathcal{O}(\alpha_s^9)$, and $\mathcal{O}(\alpha_s^{11})$.
Deeper red line corresponds to higher order result.
The renormalization scale is taken as $\Lambda/\mu=0.02$, where $\alpha_s(\mu)=0.144$.
We set $n_f=3$.
}
\label{fig:RF-NLL}
\end{figure}

As a result, we obtain the renormalon-free prediction,
which is the sum of the $\delta$ part and $V^{\rm RF}_{\rm disp}(r)$.
The result is shown in Fig.~\ref{fig:RF-NLL} (left panel).
In the right panel, as a consistency check, we compare the renormalon-free prediction
with fixed-order results.
In plotting the fixed-order results, we adjust the height of the potential at $r\LMS=0.05$.
This adjustment corresponds to subtracting the $u=1/2$ renormalon (whose uncertainty is an $r$-independent constant)
and the perturbative series exhibits convergent behavior. 
This series approaches the renormalon-free prediction (shown by the green line) as the order is raised,
and this shows validity of our renormalon-free prediction.

\newpage
\section{Renormalon-free prediction for static QCD potential at NNNLO}
\label{sec:4}
We apply our formulation to the static QCD potential
starting from its fixed-order result.
Thus, the analysis here does not rely on particular assumptions.
We have the explicit perturbative series to $\mathcal{O}(\alpha_s^4)$ (NNNLO) 
\cite{Appelquist:1977tw,Fischler:1977yf,Peter:1996ig,Peter:1997me,Schroder:1998vy,Smirnov:2008pn,Anzai:2009tm, Smirnov:2009fh, Lee:2016cgz}. 
Let us state the current understanding of the renormalons for this quantity.
The structure of the first IR renormalon at $u=1/2$ was investigated \cite{Pineda:1998id,Hoang:1998nz,Beneke:1998rk} 
(see also Ref.~\cite{Sumino:2020mxk}),
and its uncertainty is exactly proportional to $\LMS$.
This determines the form of the ambiguity function for the $u=1/2$ renormalon.
The overall constant was investigated in Refs.~\cite{Pineda:2001zq,Lee:2002sn}
and the latest result at NNNLO has been obtained in  Refs.~\cite{Ayala:2014yxa,Sumino:2020mxk}
by using the technique developed in Ref.~\cite{Lee:1996yk}.
It was confirmed that the estimate of the normalization constant at $u=1/2$ is stable
against including higher order result and varying the renormalization scale.
This indicates that the normalization constant is obtained with a reasonably small error.
The second IR renormalon at $u=3/2$ has been investigated recently \cite{Sumino:2020mxk},
and its uncertainty takes a form of $\sim \LMS^3 r^2 [1+\mathcal{O}(\alpha_s(1/r))]$.
In Ref.~\cite{Sumino:2020mxk}, however, it was shown that
the normalization constant for the $u=3/2$ renormalon cannot be estimated reliably from the currently available perturbative series.
It may indicate that the $u=3/2$ renormalon does not have a significant effect to the currently available series,
and in this analysis, we only take into account the $u=1/2$ renormalon.

From the above reasoning we consider the ambiguity function corresponding to the $u=1/2$ renormalon
[cf. Eq.~\eqref{AmbLambda}],
\be
Am_v(x)=
\begin{cases}
&b_0 N \lt[x (\log{(1/x)})^{b_1/b_0^2} e^{-\int_0^{-\frac{1}{b_0 \log{x}}} dt \lt(\frac{1}{\beta(t)}+\frac{1}{b_0 t^2}-\frac{b_1}{b_0^2 t} \rt)} \rt]^{1/2} 
\quad \text{for $x<e^{-1}$} \\
& 0 \quad \text{for $x>e^{-1}$} 
\end{cases} \, . \label{NNNLOAm}
\ee
Here we use the four-loop beta function $\beta(\alpha_s)=-\sum_{i=0}^3 b_i \alpha_s^{i+2}$.
We choose the above range $x<e^{-1}$ so that $1/\log{(1/x)}<1$,
which can be regarded as an expansion parameter of the ambiguity function [cf. Eq.~\eqref{ambformula1}].
(This corresponds to $\rho(\mu)=e^{-1}$, where we take $\mu=r^{-1}$.)
In this case, there are no zeros of the ambiguity function and 
we cannot adopt the second option \eqref{secondop}.
The normalization constant for the $u=1/2$ renormalon
has been determined from the NNNLO result as \cite{Sumino:2020mxk}\fn{The relation \eqref{conventionchange} is used
to convert the result in Ref.~\cite{Sumino:2020mxk}.
We note that the normalization constant has an error of about 10 \% \cite{Sumino:2020mxk} due to
higher order uncertainty of the perturbative series.
The error concerning the higher order uncertainty is estimated below.}
\be
b_0 N=-1.63732 \, . \label{normalizationstat}
\ee
We take $n_f=3$ here and hereafter.
Now we have obtained the ambiguity function from Eqs.~\eqref{NNNLOAm} and \eqref{normalizationstat}.
We note that the NNNLO perturbative coefficient contains an IR divergence 
\cite{Appelquist:1977es,Brambilla:1999qa,Kniehl:1999ud,Brambilla:1999xf},
and we remove the pole in $1/\epsilon$  (where the dimension is set as $d=4-2 \epsilon$
in dimensional regularization) and the associated logarithm
in position space. (This scheme is called the scheme A in Ref.~\cite{Sumino:2020mxk}.)

The order of our approximation corresponds to $(n_1, n_2)=(3,3)$ in the notation
introduced in Sec.~\ref{sec:2.7}, that is, the NNNLO perturbative series
and the NNNLO form of the ambiguity function.
We follow the procedure explained in Sec.~\ref{sec:2.7}
to obtain the renormalon-free result at this order.

\begin{figure}[tbp]
\begin{minipage}{0.5\hsize}
\begin{center}
\includegraphics[width=7cm]{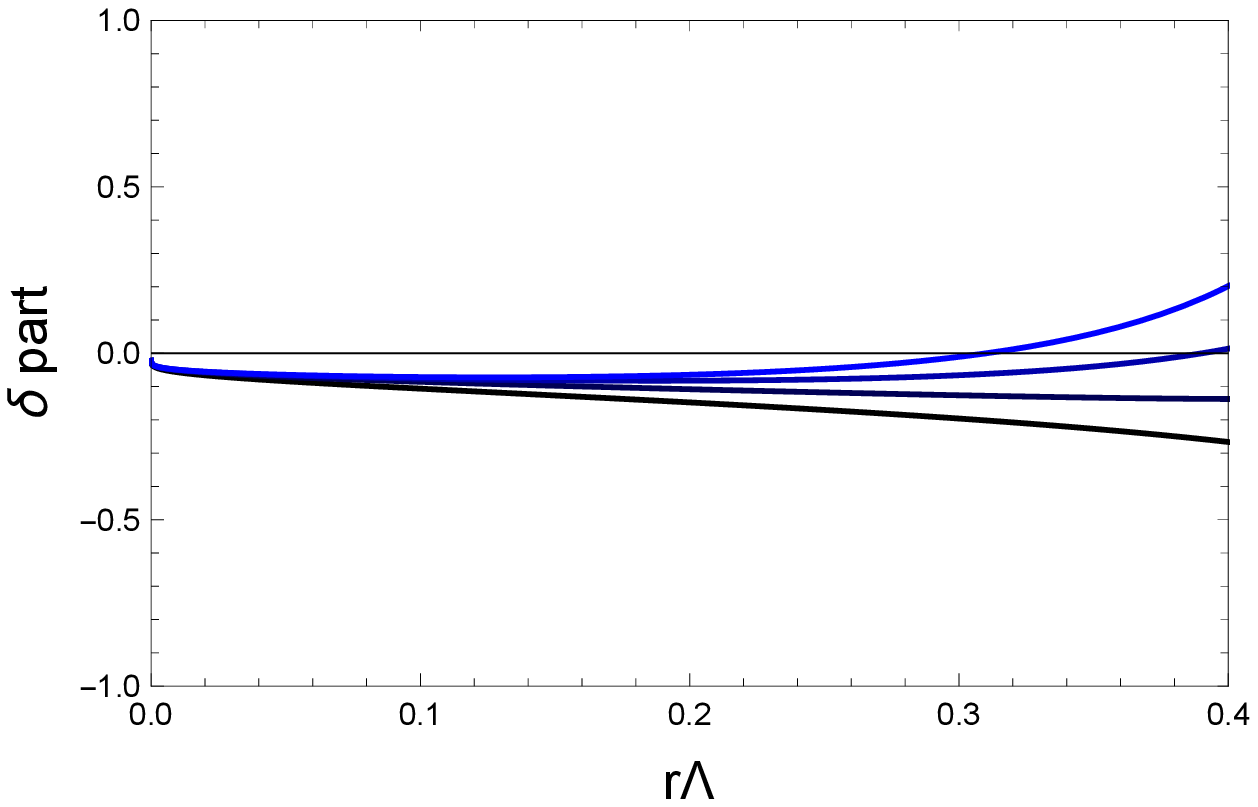}
\end{center}
\end{minipage}
\begin{minipage}{0.5\hsize}
\begin{center}
\includegraphics[width=7cm]{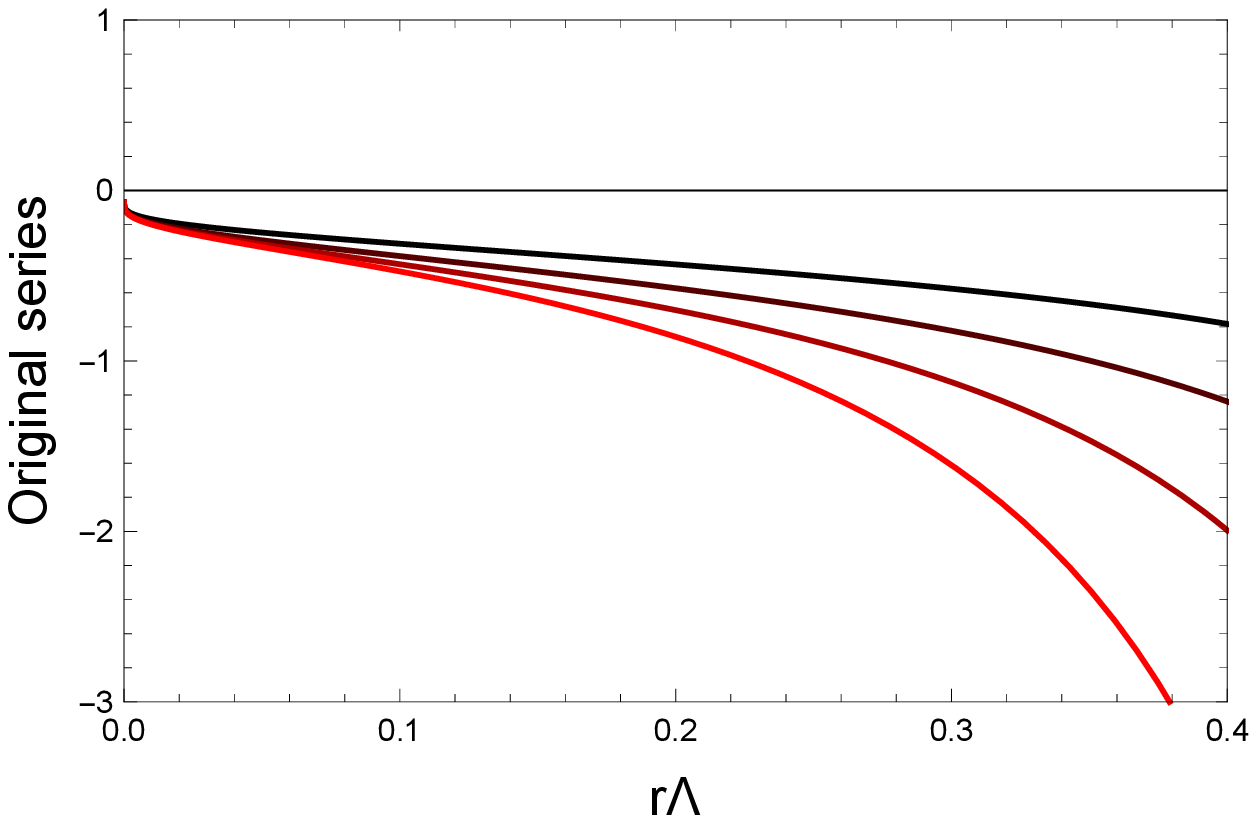}
\end{center}
\end{minipage}
\caption{Perturbative series of the $\delta$ part for the dimensionless potential (left).
This is compared to the original series containing renormalons (right).
Deeper blue (red) line corresponds to higher order result. 
The highest order is $\mathcal{O}(\alpha_s^{4})$.
}
\label{fig:deltaoriginal-NNNLO}
\end{figure}

We present the result of $\delta_n$ in Table~\ref{tab3}.
One can confirm that a large part of $d_n$ is canceled in $\delta_n$.
\begin{table}[t]
\begin{minipage}{0.5\hsize}
\begin{center}
\begin{tabular}{c|D{.}{.}{10}D{.}{.}{10}}
\hline
\multicolumn{1}{c|}{$n$} & \multicolumn{1}{c}{$d_n/b_0^n$} & \multicolumn{1}{c}{$\delta_n/b_0^n$}  \\ \hline
0 & -1.33333 & -0.453887   \\
1 & -2.57628 & 0.523036   \\
2 & -14.1987  & 1.45648 \\
3 & -103.53   & 4.31290 \\
\hline
\end{tabular}
\end{center}
\end{minipage}
\begin{minipage}{0.5\hsize}
\begin{center}
\begin{tabular}{c|D{.}{.}{10}D{.}{.}{10}}
\hline
\multicolumn{1}{c|}{$n$} & \multicolumn{1}{c}{$\frac{d_n}{b_0^n} (\Gamma(n+1+\frac{b_1}{2 b_0^2}) 2^n)^{-1}$} & \multicolumn{1}{c}{$\frac{\delta_n}{b_0^n} (n!)^{-1}$}  \\ \hline
0 & -1.50227 & -0.453887   \\
1 & -1.04035 & 0.523036  \\
2 & -1.19698  & 0.728242 \\
3 & -1.28537   & 0.718816 \\
\hline
\end{tabular}
\end{center}
\end{minipage}
\caption{\small
Original perturbative coefficient $d_n$ and renormalon subtracted perturbative coefficient $\delta_n$.
In the right panel, we divide $d_n$ by the large order behavior expected from the $u=1/2$ renormalon,
and divide $\delta_n$ by $n!$.
We take $n_f=3$.
\label{tab3}
}
\end{table}
We show a behavior of the $\delta$ part [Eq.~\eqref{deltapart}], 
which is compared to that of the original series
in Fig.~\ref{fig:deltaoriginal-NNNLO}.

\begin{figure}
\begin{center}
\includegraphics[width=9cm]{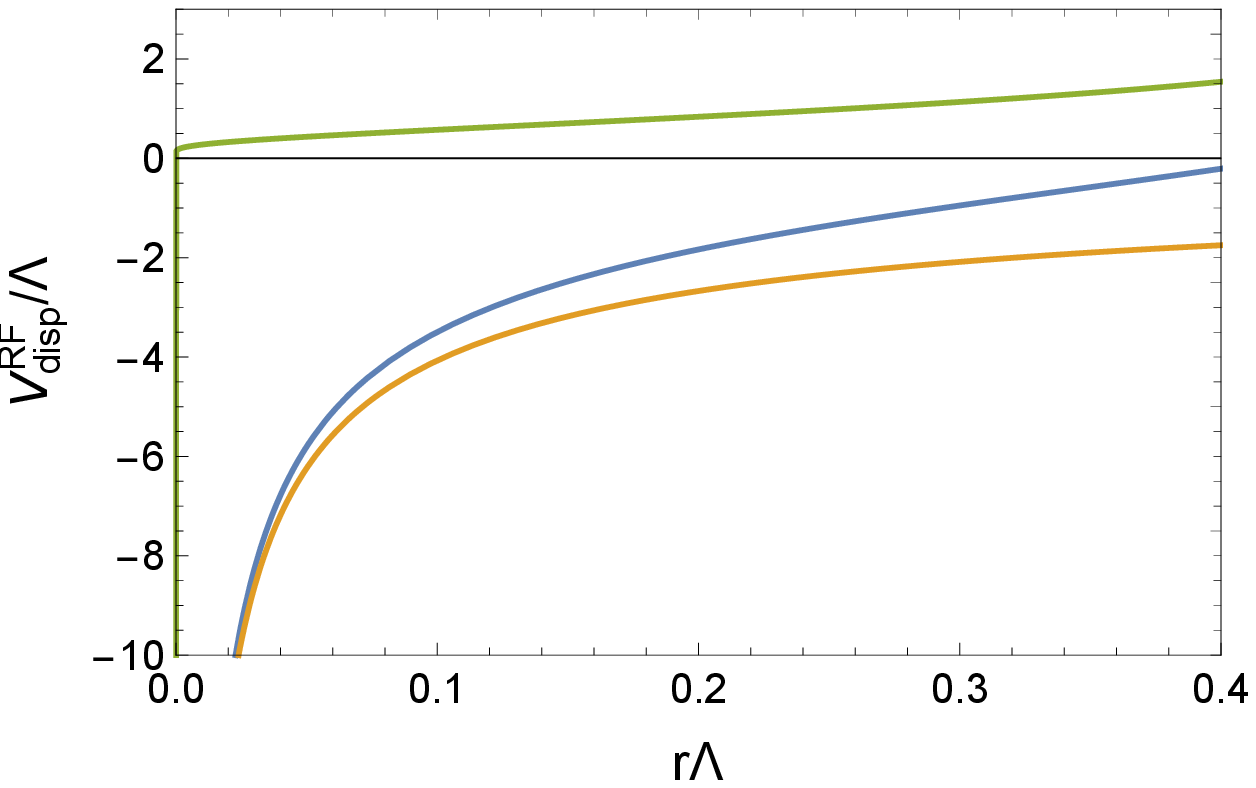}
\end{center}
\caption{$V^{\rm RF}_{\rm disp}/\LMS$ (blue) as a function of $r \LMS$.
Orange line shows the first line of Eq.~\eqref{XRFdisp2}
and green line does the second line of Eq.~\eqref{XRFdisp2}.
}
\label{fig:RF-disp-NNNLO}
\end{figure}

We now give $V^{\rm RF}_{\rm disp}(r)$ [Eq.~\eqref{XRFdisp2}].
We evaluate the running coupling $\alpha_s(1/r)$ in Eq.~\eqref{XRFdisp2} 
with the four-loop beta function.
We show the result in Fig.~\ref{fig:RF-disp-NNNLO}.
We calculate the preweight and its integral in Eq.~\eqref{XRFdisp2} numerically,
where we use the formulae in App.~\ref{app:B}.
The first line of Eq.~\eqref{XRFdisp2} gives a Coulomb-like potential
and the second line of Eq.~\eqref{XRFdisp2} gives a linear-like potential.
We note that such a behavior is obtained as an unambiguous part of the perturbative 
contribution. 
Such a behavior in perturbation theory was first clarified in Ref.~\cite{Sumino:2003yp}.
Using a different formulation, we arrive at a similar conclusion.
We emphasize that this behavior is obtained originally from the 
ambiguity function corresponding to the $u=1/2$ renormalon. 

\begin{figure}
\begin{center}
\includegraphics[width=9cm]{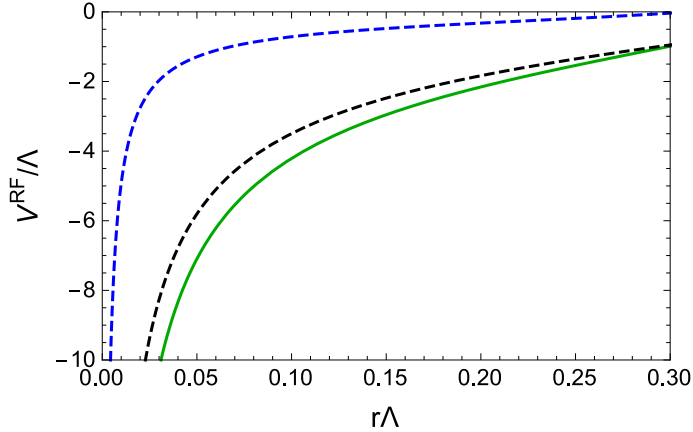}
\end{center}
\caption{Renormalon-free prediction for the static QCD potential $V^{\rm RF}/\LMS$ at NNNLO as a function $\LMS r$.
Contributions from $V_{\rm disp}^{\rm RF}/\LMS$ (black dashed) and the $\delta$ part (blue dashed) are also shown separately.}
\label{fig:RF-NNNLO}
\end{figure}

We finally obtain the NNNLO renormalon-free prediction, 
which is the sum of the $\delta$ part and $V^{\rm RF}_{\rm disp}(r)$.
We show it in Fig.~\ref{fig:RF-NNNLO}.
For the $\delta$ part, we use the highest order $\mathcal{O}(\alpha_s^4)$ result.

\begin{figure}[tbp]
\begin{minipage}{0.5\hsize}
\begin{center}
\includegraphics[width=7.5cm]{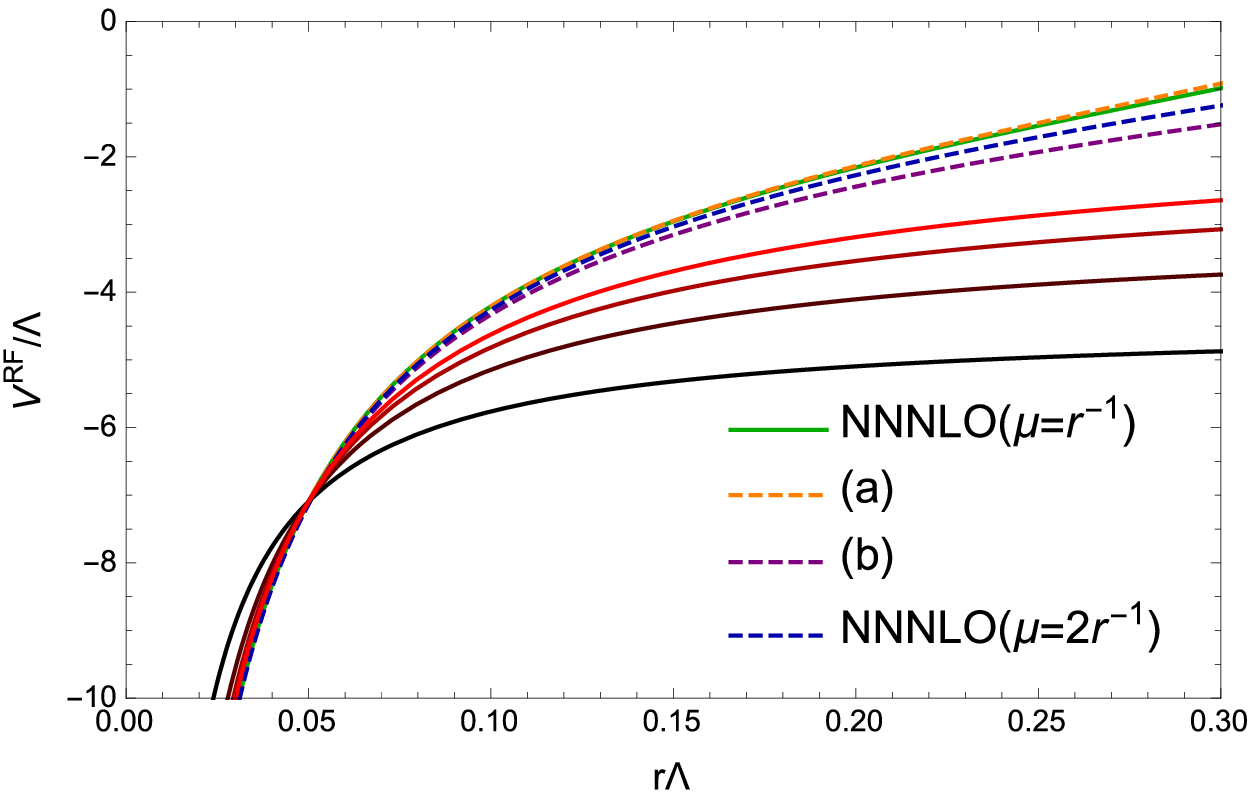}
\end{center}
\end{minipage}
\begin{minipage}{0.5\hsize}
\begin{center}
\includegraphics[width=7.5cm]{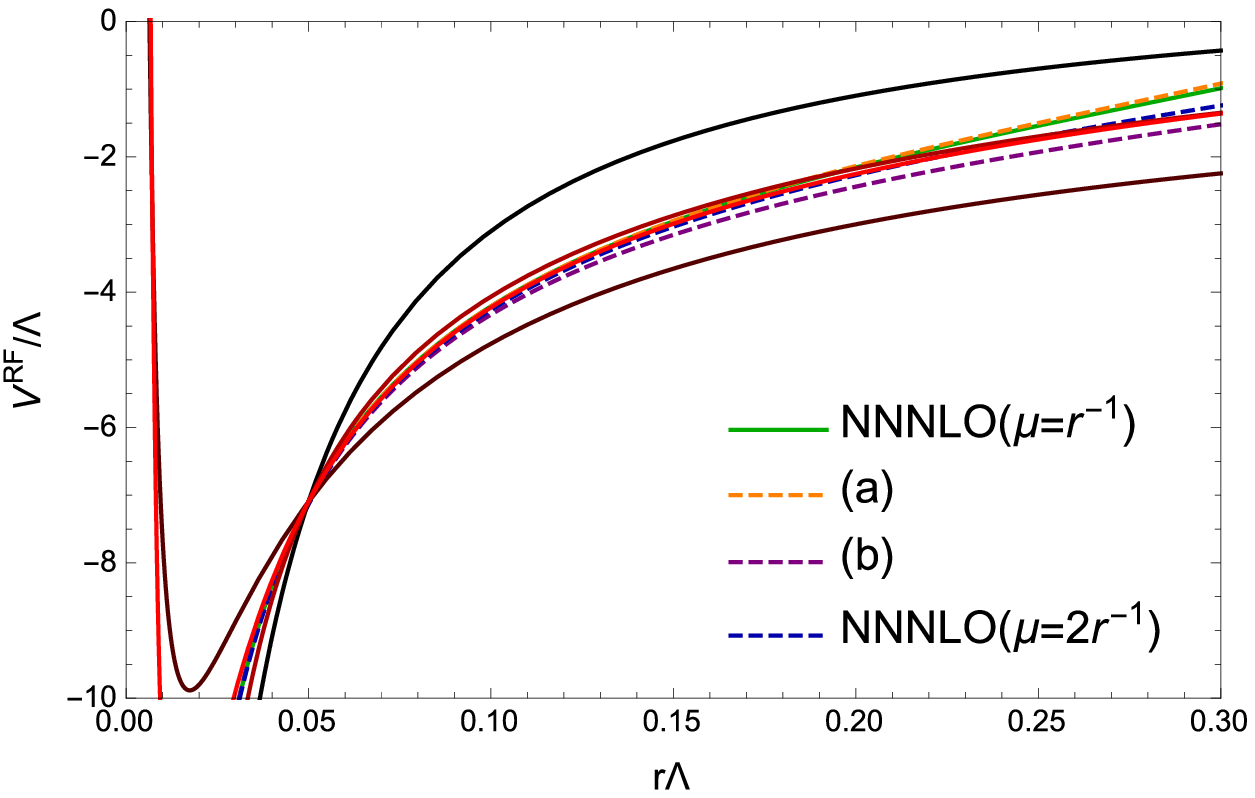}
\end{center}
\end{minipage}
\caption{$V^{\rm RF}/\LMS$ (green), its error, and fixed order results.
The orange dashed line corresponds to prediction (a) 
[NNLO form of the ambiguity function and NNNLO perturbative series] 
and the purple one to prediction (b) [NNNLO form of the ambiguity function
and NNLO perturbative series].
We also show the prediction with the renormalization scale at $\mu=2 r^{-1}$
by the blue dashed line.
We compare them with fixed-order results,
where the black line corresponds to the LO result and deeper red line corresponds to higher order result.
In the left panel, we take the renormalization scale as $\LMS/\mu=0.0026$ such that $\alpha_s(\mu)=0.1$,
and in the right panel, we take it as $\LMS/\mu=0.173$ such that $\alpha_s(\mu)=0.3$.
The heights of the potentials of the fixed-order results and the central value of $V^{\rm RF}/\LMS$
are adjusted at $r \LMS=0.05$.
}
\label{fig:errorNNNLO}
\end{figure}

We discuss the error of this prediction.
We recall that the above prediction is obtained with 
the NNNLO form of the ambiguity function
and the NNNLO perturbative series:  $(n_1,n_2)=(3,3)$.
By the NNNLO form of the ambiguity function, we mean 
that the functional form of the ambiguity function \eqref{NNNLOAm} 
is accurate to $\mathcal{O} ([x\log{(1/x)}^{b_1/b_0^2}]^{1/2} \log^2{(1/x)})$ 
and has an error of $\mathcal{O} ([x\log{(1/x)}^{b_1/b_0^2}]^{1/2} \log^3{(1/x)})$.\fn{
We note that, for the $u=1/2$ renormalon of the static QCD potential, the NNNNLO form
of the ambiguity function is available because 
the renormalon ambiguity is proportional to $\LMS$ and the explicit result of $b_4$ is known 
\cite{Baikov:2016tgj,Herzog:2017ohr,Luthe:2017ttg}.
Here we use the NNNLO form of the ambiguity function just for simplicity.
(From the analysis with (a) below, it is unlikely that neglecting the $b_4$ term induces a significant error.)
}
To estimate the higher order uncertainties concerning 
the form of the ambiguity function and perturbative series, 
we also give renormalon-free predictions with the following inputs:
\begin{description}
\item{(a)}  $(n_1,n_2)=(3,2)$: \\
the NNNLO perturbative series and the NNLO form of the ambiguity function
\item{(b)}  $(n_1, n_2)=(2,3)$:\\
the NNLO perturbative series and the NNNLO form of the ambiguity function
\end{description}
For the NNLO form of the ambiguity function, we set $b_3=b_4=\cdots=0$ in $\beta(t)$ of Eq.~\eqref{NNNLOAm}.
In Fig.~\ref{fig:errorNNNLO}, we give the results of the prediction (a) and (b).
The differences from the $(n_1, n_2)=(3,3)$ result can be regarded as higher order uncertainties.
The higher order uncertainty of the form of the ambiguity function
is small and that of the perturbative series is dominant.
We also examine the remaining renormalization scale dependence.
As we noted in Sec.~\ref{sec:2.6}, the renormalon-free prediction is 
in principle renormalization scale independent.
Hence, remaining sensitivity to a renormalization scale
corresponds to the error of the prediction, and
this analysis provides another error estimate.
We take $\mu=2 r^{-1}$ (while we have taken $\mu=r^{-1}$ so far).\fn{
In this analysis, we evaluate the normalization constant for $\mu=2 r^{-1}$ directly 
from the perturbative series at $\mu=2 r^{-1}$ and 
do not use the exact scaling of the normalization constant
$N \propto (\mu^2 r^2)^{1/2}$.
We note that in the ambiguity function only the normalization constant 
(and only the domain) is changed
and the other parts are independent of the choice of $\mu$.
This is because the $u=1/2$ renormalon ambiguity is proportional to $\LMS$
and in this case $c_k(\mu/Q)$'s in Eq.~\eqref{Borelexp} are independent of $\mu/Q$ ($Q=r^{-1}$ here). 
We also note that if we take $\mu=r^{-1}/2$ the prediction shows a divergent behavior.
This stems from an earlier divergence of the running coupling and
such an analysis does not provide a reasonable error estimate.
}
Based on the argument in Sec.~\ref{sec:2.6},
we should change the domain of the ambiguity function.
From Eq.~\eqref{relationforrho} and the fact that we choose $\rho(\mu=r^{-1})=e^{-1}$, 
we find the proper value to be $\rho(\mu=2 r^{-1}) \simeq 0.0303$.  
The result with $\mu=2 r^{-1}$ is shown in Fig.~\ref{fig:errorNNNLO}.
The largest error is caused by the higher order uncertainty of the perturbative series.
As another systematic error analysis, we change the domain of the ambiguity function
as $\rho(\mu=r^{-1})=e^{-1} \to \rho(\mu=r^{-1})=1$.
We confirmed that the result is hardly changed;
the difference is $\lesssim 0.1 \%$ in the examined distance range.

As a consistency check, we compare our prediction with fixed-order results in Fig.~\ref{fig:errorNNNLO},
where the $u=1/2$ renormalon uncertainty is removed from the fixed order results
by adjusting the $r$-independent constant (as done in Sec.~\ref{sec:3.3}).
For the renormalization scale $\LMS/\mu=0.0026$, we can confirm an agreement at short distances.
On the other hand, for $\LMS/\mu=0.173$, we observe 
an agreement around the region $\LMS r \sim 0.173$.
These are plausible taking into account the fact that fixed order
perturbation theory is reliable around $\mu \sim 1/r$.
We note that $V^{\rm RF}(r)$ contains an all-order perturbative series
in the sense that the expansion of $V^{\rm RF}_{\rm disp}(r)$ in $\alpha_s$ 
gives the infinite series $\sum_{n=0}^{\infty} d_n^{{\rm ren}} \alpha_s^{n+1}$ [cf. Eq.~\eqref{FOfromAmb}].
We also note, however, that the uncertainty coming from this divergent series is removed from $V^{\rm RF}_{\rm disp}(r)$.

Finally, we make comments on relation with Ref.~\cite{Lee:2002sn}.
In the present analysis, we gave the prediction
which is consistent with fixed order perturbation theory
but does not suffer from the $u=1/2$ renormalon uncertainty.
A prediction with these two features has been obtained in Ref.~\cite{Lee:2002sn}.
This was carried out by considering the ``bilocal expansion''
and then a ``Borel resummed'' quantity (the real part of the Borel integral).
Hence, at least numerically, the present result should produce a quite similar result to 
the one which can be obtained with the method of Ref.~\cite{Lee:2002sn}.\fn{
The result in Ref.~\cite{Lee:2002sn} itself was given in quenched QCD ($n_f=0$)
and with the NNLO perturbative series.}
The novel point in the present paper is that
we used a systematic and general method to evaluate the real part (unambiguous part) 
of the Borel integral (which is explained in Sec.~\ref{sec:2}),
and described how the ambiguous part relates to the unambiguous part of the perturbative calculation.

\section{Conclusions and discussion}
\label{sec:5}
In this paper, we presented a formulation to extract an unambiguous 
perturbative prediction from a divergent asymptotic series for a general observable $X(Q^2)$.
We refer to such an unambiguous part as a renormalon-free part.
The renormalon-free part consists of two parts, where
we used a similar idea to Refs.~\cite{Lee:2002sn,Lee:2003hh}.
One is given by series expansion
in $\alpha_s$ which does not contain renormalons ($\delta$ part), 
and the other is the real part of the Borel integral ($X^{\rm RF}_{\rm disp}(Q^2)$)
where the Borel transform possesses renormalons. 
A novel aspect of this paper is that we proposed a systematic method to obtain $X^{\rm RF}_{\rm disp}(Q^2)$
as described below.

To obtain the real part of the Borel integral of the singular Borel tramsform, 
we first introduced an ``ambiguity function,''
as defined in Eq.~\eqref{resumAmb}.
This is inverse Mellin transform of the singular Borel transform
and is deeply connected with renormalon uncertainties.
With the ambiguity function we rewrote the Borel integral by an alternative resummation formula,
which is given by a one-dimensional integral in $x$-plane instead of the Borel $u$-plane. 
In this formula, the integrand of the $x$-integral has only a simple pole as the singularity structure.
This singularity structure is much simpler than that of the Borel integral,
whose integrand has an infinite number of cut singularities.
(Such a transform of singularities itself are rather well known.)
A main advantage in adopting this formula is that 
the structure is common to the resummation formula in the large-$\beta_0$ approximation
and hence it is possible to use the techniques developed there.
We introduced a ``preweight,'' which is given by the dispersive integral
of the ambiguity function, and plays an important role in giving an unambiguous part.  
The main result is given in Eq.~\eqref{XRFdisp2}.
This tells us how the unambiguous part emerges in connection with renormalon ambiguities.
In this method, the ambiguous part, identified as the renormalon uncertainty,
is simultaneously obtained explicitly. 
We also gave detailed RG analyses of the formulation.
Our final result $X^{\rm RF}$ is indeed RG invariant, but $X^{\rm RF}_{\rm disp}$ and the $\delta$ part
are $\mu$ dependent separately. (The sum of them is RG invariant.) 
Nevertheless, the $\mu$ dependence of $X^{\rm RF}_{\rm disp}$
is under good control thanks to the RG equation for the singular Borel transform or
that for the ambiguity function.
We also argued that the present formulation, which generally needs 
a $\delta$ part, is a natural extension
of the formulation in the large-$\beta_0$ approximation from the  viewpoint of RG properties.


We applied this formulation to the Adler function and static QCD potential.
For the Adler function, as a test of the formulation, we considered the large-$\beta_0$ approximation,
where an all-order perturbative series can be obtained. In this case,
we do not have a $\delta$ part and the result completely reduces to the one studied in Refs.~\cite{Ball:1995ni, Mishima:2016xuj}.
We also studied the static QCD potential with the RG method \cite{Sumino:2005cq},
where an approximated all-order perturbative series containing renormalon divergences can be obtained.
We confirmed that the $\delta$ parts exhibit much better convergence than the original perturbative series.\fn{
Nevertheless, the perturbative coefficients in the $\delta$ part still have a growing behavior (see, for instance, Table~\ref{tab1}),
and there is a possibility that the convergence radius is too small for practical use.}
We also confirmed that our renormalon-free predictions are reasonable by comparison with other calculations.
Then we applied the formulation to the fixed-order result for the static QCD potential at NNNLO.
(In Ref.~\cite{Lee:2002sn}, the NNLO result was obtained in quenched QCD.)
The first IR renormalon at $u=1/2$ already has a significant effect to this series,
and we removed this uncertainty and gave a stable result.
We also gave detailed error analyses.

There are useful features of this method.
First our renormalon-free part is consistent with fixed order perturbation theory
(and does no suffer from renormalons), and this is realized 
by a similar idea to the preceding work~\cite{Lee:2002sn,Lee:2003hh}. 
Secondly, this method is compatible with the OPE:
the renormalon uncertainty is consistent with the OPE structure,
and the first Wilson coefficient $C_1$ 
is constructed as a clearly RG invariant quantity.
This is due to the use of the Borel resummation and again common to Refs.~\cite{Lee:2002sn,Lee:2003hh}.
These properties are quite useful to go beyond perturbation theory using the OPE
and are an advantage compared with the truncation regularization of perturbative series.
Thirdly, our formulation can remove subleading renormalons, 
as done in Sec.~\ref{sec:3.2} and Sec.~\ref{sec:3.3}, without difficulties 
(although it is often not an easy task to investigate renormalon structures of the subleading renormalons\fn{
Ref.~\cite{Sumino:2020mxk} clarified the $u=3/2$ renormalon structure for the static QCD potential.}).

It would be possible to apply the present formulation to other observables such as the Adler function (beyond the large-$\beta_0$ approximation). 
The formulation would also be useful to give a clear definition of the gluon condensate 
(see Ref.~\cite{Suzuki:2018vfs} for discussion on this issue within the large-$\beta_0$ approximation) 
and its precise determination as discussed in Sec.~\ref{sec:2.6}.
We would like to discuss these issues in near future.

\section*{Acknowledgements}
The author is very grateful to Sayaka Kawabata for fruitful discussion.
He thanks Yukinari Sumino for encouragement and useful comments on an earlier version
of the manuscript.
He also thanks Hiroshi Suzuki for his encouragement.
This work was supported by JSPS Grant-in-Aid for Scientific Research Grant
Number JP19K14711. 

\newpage
\appendix

\section{RG invariance of Borel integral}
\label{app:A}
We show that the Borel integral is independent of the renormalization scale $\mu$. 
Such an argument has been given in Ref.~\cite{Ayala:2019uaw} and
the present calculation can be regarded as an explicit generalization of the argument of 
Ref.~\cite{Ayala:2019uaw} to all order with respect to $b_i$'s.
Here, it is convenient to adopt the following definitions
\be
\tilde{B}_X(t;Q,\mu)=\sum_{n=0}^{\infty} \frac{d_n(Q,\mu)}{n!} t^n \, , \label{Boreltconv}
\ee
\be
X(Q^2)=\int_0^{\infty} dt \,  \tilde{B}_X(t;Q,\mu) e^{-t/\alpha_s(\mu)} \, ,
\ee
rather than the definition adopted in the main part of this paper for convenience
(Of course one can obtain the same conclusions regardless of chosen conventions).
We regularize the Borel integral by deforming the integration path $0 \to \infty$
as  $0\pm i0 \to \infty \pm i0$ if necessary.
The derivative of the Borel integral with respect to $\mu$ is given by
\begin{align}
&\mu^2 \frac{d}{d \mu^2} \int_0^{\infty} dt \, \tilde{B}_X(t;Q,\mu) e^{-t/\alpha_s(\mu)} \non
&=\int_0^{\infty} dt \, \lt[ \mu^2 \frac{\del}{\del \mu^2} \tilde{B}_X(t;Q,\mu)+ \tilde{B}_X(t;Q,\mu)  \frac{\beta(\alpha_s)}{\alpha_s^2} t  \rt] e^{-t/\alpha_s(\mu)} \, . \label{muderiBorel}
\end{align}
We note that the perturbative coefficients satisfy the RG equation,
\be
\begin{cases}
\mu^2 \frac{\del}{\del \mu^2} d_0=0  \\
\mu^2 \frac{\del}{\del \mu^2} d_n=\sum_{i=0}^{n-1} (n-i) d_{n-(i+1)} b_i \quad \text{for $n \geq 1$} \, .
\end{cases}
\ee
Then, we obtain
\begin{align}
\mu^2 \frac{\del}{\del \mu^2} \tilde{B}_X(t;Q,\mu)
&=\sum_{i=0}^{\infty} b_i \sum_{n=i+1}^{\infty} \frac{n-i}{n!} d_{n-(i+1)} t^n \non
&=\sum_{i=0}^{\infty} b_i  f_i(t) \label{eq:(A5)}
\end{align}
with
\be
f_i(t) := \sum_{n=i+1}^{\infty} \frac{n-i}{n!} d_{n-(i+1)} t^n \, .  \label{eq:(A6)}
\ee
Noting that the $i$th derivative of $f_i(t)$ is given by
\begin{align}
\frac{\del^i f_i}{\del t^i}
=\sum_{n=i+1}^{\infty} \frac{1}{(n-i-1)!} r_{n-i-1} t^{n-i}  
=t \tilde{B}_X(t;Q,\mu) \, , \label{eq:(A7)}
\end{align}
we can rewrite the first term of Eq.~\eqref{muderiBorel} as
\begin{align}
\int_0^{\infty} dt \, \mu^2 \frac{\del}{\del \mu^2} \tilde{B}_X(t;Q,\mu) e^{-t/\alpha_s(\mu)}
&=\sum_{i=0}^{\infty}  b_i  \int_0^{\infty} dt \,  f_i(t) e^{-t/\alpha_s(\mu)} \non
&=\sum_{i=0}^{\infty} b_i  \int_0^{\infty} dt \, f_i(t) (-\alpha_s(\mu))^i \frac{\del^i}{\del t^i} e^{-t/\alpha_s(\mu)} \non
&=\sum_{i=0}^{\infty} b_i   \alpha_s(\mu)^i \int_0^{\infty} dt \,  \frac{\del^i f_i(t)}{\del t^i} e^{-t/\alpha_s(\mu)}  \non
&=-\frac{\beta(\alpha_s)}{\alpha_s^2} \int_0^{\infty} dt \,  t \tilde{B}_X(t;Q,\mu) e^{-t/\alpha_s(\mu)} \, . 
\end{align}
Here, omitting the surface terms we assume $\frac{d^j f_i}{dt^j} e^{-t/\alpha_s(\mu)} \big|_{t=\infty}=0$ for $0 \leq j \leq i-1$
(similarly to Ref.~\cite{Ayala:2019uaw}). On the other hand, $\frac{d^j f_i}{dt^j} e^{-t/\alpha_s(\mu)} \big|_{t=0}=0$
for $0 \leq j \leq i-1$ is ensured from $f_i(t)=\mathcal{O}(t^{i+1})$. (This point is a significant
difference from the argument in Sec.~\ref{sec:2.6}.)
Thus, Eq.~\eqref{muderiBorel} becomes zero, which shows RG invariance of the Borel integral.

\section{Convenient formulae for numerical evaluation}
\label{app:B}
In this appendix, we present convenient formulae for numerical evaluation of the real part
of the preweight [appearing in the second line of Eq.~\eqref{XRFdisp2}] and 
the one-dimensional integral of the preweight in Eq.~\eqref{XRFdisp2}.

The real part of the pre-weight for $z>0$ is evaluated by 
\begin{align}
{\rm Re} \, W(z)
&= {\rm PV} \int_0^{\infty} \frac{dx}{\pi} \frac{Am(x)}{x-z}\non
&= {\rm PV} \int_0^{c} \frac{dx}{\pi} \frac{Am(x)}{x-z}+\int_{c}^{\infty} \frac{dx}{\pi} \frac{Am(x)}{x-z}
\end{align}
where $c$ is taken as $c>z$. 
The first integral can be rewritten in the following form,
which is convenient for numerical integral:
\begin{align}
{\rm PV} \int_0^{c} \frac{dx}{\pi} \frac{Am(x)}{x-z}
&=\lt(\int_0^{z-\epsilon}+\int_{z+\epsilon}^{c} \rt) \frac{dx}{\pi} \frac{Am(x)}{x-z} \non
&=\lt(\int_0^{z-\epsilon}+\int_{z+\epsilon}^{c} \rt) \frac{dx}{\pi} \lt[ \frac{Am(x)-Am(z)}{x-z} +\frac{Am(z)}{x-z} \rt] \non
&=\int_0^c \frac{dx}{\pi} \frac{Am(x)-Am(z)}{x-z}+\frac{1}{\pi} Am(z) \log\lt| \frac{c-z}{z} \rt| \, .
\end{align}

Now we consider the one-dimensional integral in Eq.~\eqref{XRFdisp2}:
\be
\frac{1}{b_0} \int_0^{\infty} \frac{dz}{\pi z} W_{X+}(z) \frac{- \pi}{\lt(\log{z}+\frac{1}{b_0 \alpha_s} \rt)^2+\pi^2} \, . \label{intfocus}
\ee
We can use
\be
-\frac{d}{dz}\arctan\lt[\frac{\log \lt(z e^{\frac{1}{b_0 \alpha_s(Q)}} \rt)}{\pi} \rt]
=\frac{1}{z} \frac{-\pi}{\pi^2+\log^2(z e^{\frac{1}{b_0 \alpha_s(Q)}})} \, , 
\ee
to rewrite this integral. We present two methods.
\subsection*{Method I}
With a constant $c>0$, we can rewrite Eq.~\eqref{intfocus} as
\begin{align}
&\frac{1}{b_0} \int_0^{\infty} \frac{dz}{\pi z} W_{X+}(z) \frac{- \pi}{\lt(\log{z}+\frac{1}{b_0 \alpha_s} \rt)^2+\pi^2} \non
&=\frac{1}{b_0} \lt(\int_0^{c}+\int_c^{\infty} \rt) \frac{dz}{\pi z} W_{X+}(z) \frac{- \pi}{\lt(\log{z}+\frac{1}{b_0 \alpha_s} \rt)^2+\pi^2} \non
&=\frac{1}{b_0} \int_0^{c} \frac{dz}{\pi z} [W_{X+}(z) -W_{X+}(0)] \frac{- \pi}{\lt(\log{z}+\frac{1}{b_0 \alpha_s} \rt)^2+\pi^2} \non
&\quad+\frac{1}{b_0} W_{X+}(0)  \int_0^{c} \frac{dz}{\pi z} \frac{- \pi}{\lt(\log{z}+\frac{1}{b_0 \alpha_s} \rt)^2+\pi^2} \non
&\quad+\frac{1}{b_0} \int_c^{\infty} \frac{dz}{\pi z} W_{X+}(z) \frac{- \pi}{\lt(\log{z}+\frac{1}{b_0 \alpha_s} \rt)^2+\pi^2} \non
&=\frac{1}{b_0} \int_0^{c} \frac{dz}{\pi z} [W_{X+}(z) -W_{X+}(0)] \frac{- \pi}{\lt(\log{z}+\frac{1}{b_0 \alpha_s} \rt)^2+\pi^2} \non
&\quad+\frac{1}{b_0} \frac{1}{\pi} W_{X+}(0)  \lt(-\arctan\lt[\frac{\log{c}+\frac{1}{b_0 \alpha_s} }{\pi}\rt] -\frac{\pi}{2} \rt)\non
&\quad+\frac{1}{b_0} \int_c^{\infty} \frac{dz}{\pi z} W_{X+}(z) \frac{- \pi}{\lt(\log{z}+\frac{1}{b_0 \alpha_s} \rt)^2+\pi^2}  \, .
\end{align}
\subsection*{Method II}
We can also rewrite Eq.~\eqref{intfocus} as
\begin{align}
&\frac{1}{b_0} \int_0^{\infty} \frac{dz}{\pi z} W_{X+}(z) \frac{- \pi}{\lt(\log{z}+\frac{1}{b_0 \alpha_s} \rt)^2+\pi^2} \non
&=\frac{1}{b_0} \int_0^{\infty} \frac{dz}{\pi} W_{X+}(z)  \lt\{-\arctan\lt[\frac{\log \lt(z e^{\frac{1}{b_0 \alpha_s(Q)}} \rt)}{\pi} \rt] \rt\}' \non
&~~+\frac{1}{b_0} \int_0^{\infty} \frac{dz}{\pi} W'_{X+}(z) \arctan\lt[\frac{\log \lt(z e^{\frac{1}{b_0 \alpha_s(Q)}} \rt)}{\pi} \rt] \non
&=-\frac{1}{b_0} \frac{W_{X+}(0)}{2}
+\frac{1}{b_0} \int_0^{\infty} \frac{dz}{\pi} W'_{X+}(z) \arctan\lt[\frac{\log \lt(z e^{\frac{1}{b_0 \alpha_s(Q)}} \rt)}{\pi} \rt] \, .
\end{align}

\bibliographystyle{utphys}
\bibliography{BibQCD}

\end{document}